\documentclass[preprint2]{aastex}
\usepackage{amsmath,amsthm}
\shorttitle{A panchromatic study of BLAST galaxies}

\shortauthors{Moncelsi et al.}

\begin{document}
\title{A panchromatic study of BLAST counterparts: total star-formation rate, morphology, AGN fraction and stellar mass}
%\title{Total star-formation rates in BLAST galaxies}

\author{
Lorenzo Moncelsi\altaffilmark{1}, Peter A. R. Ade\altaffilmark{1},
Edward L. Chapin\altaffilmark{2}, Luca Cortese\altaffilmark{1,3},
Mark J. Devlin\altaffilmark{4}, Simon Dye\altaffilmark{1}, Stephen
Eales\altaffilmark{1}, Matthew Griffin\altaffilmark{1}, Mark
Halpern\altaffilmark{2}, Peter C. Hargrave\altaffilmark{1}, Gaelen
Marsden\altaffilmark{2}, Philip Mauskopf\altaffilmark{1}, Calvin
B. Netterfield\altaffilmark{5}, Enzo Pascale\altaffilmark{1},
Douglas Scott\altaffilmark{2}, Matthew D. P.
Truch\altaffilmark{4}, Carole Tucker\altaffilmark{1}, Marco P.
Viero\altaffilmark{5}, Donald V. Wiebe\altaffilmark{2}
\email{lorenzo.moncelsi@astro.cf.ac.uk} } \altaffiltext{1}{Cardiff
University, School of Physics \& Astronomy, Queens Buildings, The
Parade, Cardiff, CF24 3AA, U.K.} \altaffiltext{2}{Department of
Physics \& Astronomy, University of British Columbia, 6224
Agricultural Road, Vancouver, BC V6T 1Z1, Canada}
\altaffiltext{3}{European Southern Observatory,
Karl-Schwarzschild-Str. 2, D-85748, Garching, Germany}
\altaffiltext{4}{Department of Physics \& Astronomy, University of
Pennsylvania, 209 South 33rd Street, Philadelphia, PA, 19104,
U.S.A.} \altaffiltext{5}{Department of Astronomy \& Astrophysics,
University of Toronto, 50 St. George Street Toronto, ON M5S 3H4,
Canada}

\begin{abstract}
We carry out a multi-wavelength study of individual galaxies
detected by the Balloon-borne Large Aperture Submillimeter
Telescope (BLAST) and identified at other wavelengths, using data
spanning the radio to the ultraviolet (UV). We develop a Monte
Carlo method to account for flux boosting, source blending, and
correlations among bands, which we use to derive deboosted
far-infrared (FIR) luminosities for our sample. We estimate total
star-formation rates for BLAST counterparts with $z\leq0.9$ by
combining their FIR and UV luminosities. Star formation is heavily
obscured at $L_{\rm FIR} \gtrsim 10^{11}$\,L$_{\sun}$, $z \gtrsim
0.5$, but the contribution from unobscured starlight cannot be
neglected at $L_{\rm FIR} \lesssim 10^{11}$\,L$_{\sun}$, $z
\lesssim 0.25$. We assess that about 20\% of the galaxies in our
sample show indication of a type-1 active galactic nucleus (AGN),
but their submillimeter emission is mainly due to star formation
in the host galaxy. We compute stellar masses for a subset of 92
BLAST counterparts; these are relatively massive objects, with a
median mass of $\sim10^{11}$\,M$_{\sun}$, which seem to link the
24\,$\mu$m and SCUBA populations, in terms of both stellar mass
and star-formation activity. The bulk of the BLAST counterparts at
$z\lesssim1$ appear to be run-of-the-mill star-forming galaxies,
typically spiral in shape, with intermediate stellar masses and
practically constant specific star-formation rates. On the other
hand, the high-$z$ tail of the BLAST counterparts significantly
overlaps with the SCUBA population, in terms of both
star-formation rates and stellar masses, with observed trends of
specific star-formation rate that support strong evolution and
downsizing.
\end{abstract}


\section{Introduction}\label{sec:intro}
The physical processes associated with the evolution of the
Universe have left an imprint in the extragalactic background
light. The far-infrared (FIR) portion of the background is
associated with forming galaxies in which the ultraviolet (UV)
photons emitted by new born stars are absorbed and re-radiated by
dust in the IR. Roughly half of the energy content of the
starlight integrated over the age of the Universe is stored in the
Cosmic Infrared Background (CIB), glowing with a broad peak at
around 200\,$\mu$m \citep{puget96,Fixsen98,Dwek1998}. The tight
connection between star formation and FIR luminosity provides a
route to understanding the history of star formation in the
Universe, by means of studying the CIB at wavelengths close to its
peak \citep{Gispert2000,RR01,Chary2001,Hauser2001}.

The first leg on this route is to identify the sources
contributing to the CIB. Ground-based surveys with the
Submillimetre Common-User Bolometer Array (SCUBA) have revealed
the existence of a population of distant, highly dust-obscured
galaxies, similar to the Ultra Luminous Infrared Galaxies (ULIRGs)
detected by {\sl IRAS} \citep{Smail1997, Hughes1998, Barger1998},
which make up all the background at 850\,$\mu$m \citep{Blain1999}.
However, at these wavelengths the energy in the CIB is only
one-thirtieth of the value at its peak, and the SCUBA population
only contributes 20--30\% to the CIB at its peak
\citep{Coppin2006,Dye2007}.

Recent progress has been made through new observations obtained at
24, 70, and 160\,$\mu$m by the MIPS instrument aboard the {\sl
Spitzer Space Telescope} \citep{Rieke2004}, and at 250, 350 and
500\,$\mu$m by the Balloon-borne Large Aperture Submillimeter
Telescope \citep[BLAST,][]{devlin04,Pascale2008}, a forerunner of
the SPIRE instrument \citep{Griffin2010} on the {\sl Herschel
Space Observatory} \citep{Pilbratt2010}. These wavelengths bracket
the CIB peak; several authors have shown through stacking analyses
that 24\,$\mu$m-selected galaxies resolve the CIB background, both
on the short-wavelength side of the peak \citep{Dole2006} and on
its long-wavelength side \citep{Devlin2009, Marsden2009}.

Sources identified at 24\,$\mu$m are mostly unresolved in the FIR,
and have a redshift distribution with median of 0.9
\citep{Pascale2009}. A detailed multiwavelength study of these
sources is the necessary next step. Starting from a catalog of
$\geq5\sigma$ BLAST sources, \citet[][hereafter D09]{Dye2009} have
identified counterparts in 24\,$\mu$m and radio catalogs (BLAST
IDs). These tend to be relatively nearby sources (median $z$ of
0.6, inter-quartile range of 0.2--1.0), with a median dust
temperature of 26 K and a median bolometric FIR luminosity of $4
\times 10^{11}$\,L$_{\sun}$, which contribute 20\% to the CIB at
250\,$\mu$m. Identified BLAST sources typically lie at lower
redshifts and have lower rest-frame dust temperatures compared to
submillimeter (submm) sources detected in surveys conducted with
SCUBA \citep{Chapman2005,Pope2005}. However, D09 also note that
the $\sim40\%$ of BLAST sources without identified counterparts
probably lie at higher redshifts on average. Finally, D09
illustrate how the apparent increase in dust temperature and FIR
luminosity with redshift occurs as a result of selection effects.

We also note that three other multi-wavelength studies of fainter
BLAST sources discovered in the deepest part of the map have been
undertaken. \citet{Dunlop2010} concentrate on 250\,$\mu$m
radio-identified sources within GOODS-S \citep[][see Section
\ref{subsec:BLAST}]{Dickinson2003} where the deepest ancillary
data coincide. \citet{Chapin2010} use overlapping BLAST
250--500\,$\mu$m and LABOCA 870\,$\mu$m \citep{Weiss2009} data in
the larger Extended Chandra Deep-Field South (ECDFS) to constrain
the Rayleigh-Jeans tail more accurately than was possible in D09.
Finally, \citet{Ivison2010} study the FIR/radio correlation for a
catalog of BLAST 250\,$\mu$m-selected galaxies in the ECDFS; this
sample is deeper than the D09 one, and yet slightly shallower than
the selection in \citet{Dunlop2010}. There is little overlap
between the sources used in these studies and the
shallower/wider-area sample from D09.

The basis of our present study is the D09 sample as its brighter,
and lower-redshift objects were most easily followed-up in the
optical and UV. However, we first extend the submm analysis of D09
by accounting for flux boosting, source blending and correlations
among BLAST bands that inevitably arise in IR surveys as a
consequence of finite instrumental angular resolution and source
confusion \citep{Coppin2005}. We then identify counterparts to the
BLAST IDs in the near- and far- UV {\sl GALEX} maps, in order to
quantify the total dust-obscured and unobscured star formation, as
described by several authors
\citep{Bell2003,Hirashita2003,Iglesias-Paramo2006,Buat2007}. We
also extend the analysis of \citet[][hereafter E09]{Eales2009} to
combine spectroscopic data of BLAST IDs with optical, near-IR
(NIR) and mid-IR (MIR) photometry in order to place firmer
constraints on source redshifts, morphology, AGN fraction, and
stellar masses.

We are able to assign spectroscopic and photometric redshifts to
$\sim$62\% of the BLAST IDs. We use this information to estimate
the rest-frame total FIR luminosity from the combined BLAST and
MIPS photometry. We compare our FIR luminosities with those
obtained from MIPS photometry only, finding a significant
discrepancy for high luminosity sources ($L_{\rm FIR} \gtrsim 5
\times 10^{11}$\, L$_{\sun}$) at $z\gtrsim0.5$. The BLAST and
SPIRE wavebands are therefore fundamental in constraining the peak
of hidden star formation at high redshift \citep[see also
e.g.][]{Schulz2010,Elbaz2010}.

In addition, UV counterparts are found for about 60\% of the BLAST
IDs. This allows us to estimate the fraction of UV photons that
manage to escape the dust shroud, which is then combined with FIR
data to build an estimator of the total star-formation rate
(SFR$_{\rm tot}$) ongoing in these sources. Recent observations at
the same wavelengths \citep{Rodighiero2010} delineate the UV
contribution as marginal at all redshifts. We find that star
formation is heavily obscured at $L_{\rm FIR} \gtrsim
10^{11}$\,L$_{\sun}$, $z \gtrsim 0.5$, but unobscured starlight
plays an important role in low redshift, low FIR luminosity
sources ($z \lesssim 0.25$, $L_{\rm FIR} \lesssim
10^{11}$\,L$_{\sun}$), in agreement with \citet{Buat2010}.

We reanalyze the optical spectroscopy data from the AAOmega survey
presented in E09 to obtain H$\alpha$ equivalent widths and
[NII]/H$\alpha$ line ratios. This spectral analysis, combined with
a qualitative study of the radio, MIR and optical emission, allows
us to assess whether or not a BLAST galaxy is hosting an active
nucleus: roughly 20\% of the objects in our sample show evidence
of AGN presence. Recent observations of FIR-selected samples
\citep{Wiebe2009,Coppin2010,Muzzin2010,Hatziminaoglou2010,Shao2010,Elbaz2010}
show that the submm emission of such objects is mainly due to star
formation ongoing in the host galaxy, rather than due to the AGN.
Therefore we do not to explicitly exclude AGN from our analysis,
unlike other authors
\citep{Bell2003,Iglesias-Paramo2006,Buat2007}, but rather flag
them as such. Visual examination of BLAST IDs in UV, optical and
MIR images (see Appendix A) is used to derive a broad
morphological classification of these objects: at low redshift we
find predominantly spirals, whereas most of the BLAST sources
identified at high redshift are compact and show AGN signatures.
This is probably a selection bias, as the fraction of submm
sources identified at other wavelengths gradually decreases with
$z$ (see D09), and the farthest objects can often be identified
only if they are particularly bright in the radio or in the
optical, frequently an indication of AGN presence. As a matter of
fact, the analysis carried out by \citet{Dunlop2010} shows that a
deep survey at 250\,$\mu$m not only contains low-$z$ spirals, but
also extreme dust-enshrouded starburst galaxies at $z\sim 2$. Our
analysis tends to miss the latter because they are typically
extremely faint in the optical/UV, unless they also host an AGN.

Finally, stellar masses ($M_{\star}$) are estimated using the
method detailed in \citet{Dye2008b}, in order to study whether or
not specific star-formation rates (SSFR $\equiv$ SFR/$M_{\star}$)
depend on stellar mass and $L_{\rm FIR}$. The SSFR plays an
important role as it measures the time scale of recent star
formation in a galaxy, as compared to the star-formation rate
integrated over the galaxy's history. Several studies \citep[][and
references therein]{Santini2009,Rodighiero2010} report that the
SSFR increases with redshift at all masses, whereas the dependence
of SSFR on mass is one of the most debated questions. In
particular, we aim to understand whether or not sources selected
at wavelengths longward of 200\,$\mu$m are experiencing a major
episode of star formation, forming stars more actively than in
their recent past and building up a substantial fraction of their
final stellar mass. We highlight a dichotomy in the BLAST
population: sources at $z\lesssim1$ appear to be run-of-the-mill
star-forming galaxies with intermediate stellar masses (median
$M_{\star}\sim7\times10^{10}$\,M$_{\sun}$) and approximately
constant SSFRs, whereas the high-$z$ tail of the BLAST
counterparts significantly encroaches on the SCUBA population
detected in the SHADES survey \citep{Dye2008a}, in terms of both
stellar masses and SSFRs. This is expected since there is good
overlap between fainter BLAST sources and 870\,$\mu$m-selected
galaxies \citep{Dunlop2010,Chapin2010}, but it is also important
to establish an additional link with a shallower BLAST sample,
using a methodology equivalent to that of SHADES. In addition,
since the more massive BLAST galaxies at intermediate redshifts
($0 < z < 1$) seem to form stars more vividly than the equally
massive and aged 24\,$\mu$m sources detected in the GOODS survey,
we suggest that the BLAST counterparts may act as linking
population between the 24\,$\mu$m-selected sources and the SCUBA
starbursts.

The layout of this paper is as follows. In Section \ref{sec:data}
we describe in detail the maps, images and catalogs used
throughout this paper. Section \ref{sec:FIR_lum_SFR} and Section
\ref{sec:UV_lum_SFR} are concerned with luminosities and
star-formation rates in the FIR and UV, respectively. In Section
\ref{sec:tot_SFR} we build a unified estimator of total star
formation and discuss the first results. In Section \ref{sec:AGN}
we estimate the AGN content of our sample, while in Section
\ref{sec:morph} we outline a broad morphological scheme for our
sources. In Section \ref{sec:masses} we compute the stellar masses
and present the main results. Section \ref{sec:summary} contains
our conclusions. Throughout this paper we assume the standard
concordance cosmology: $\rm \Omega_M=0.274,\
\Omega_{\Lambda}=0.726,\ H_0 = 70.5\ km\ s^{-1}\ Mpc^{-1}$
\citep{Hinshaw2009}.

\section{Data}\label{sec:data}
This Section describes the data sets used for our analysis,
spanning from the UV to the submillimeter.

\subsection{Submillimeter data}\label{subsec:BLAST}
We use data from the wide-area extragalactic survey of BLAST
described by \citet{Devlin2009}, and centered on the Great
Observatory Origins Deep Survey-South \citep[GOODS-S,][which in
turn is centered on the Chandra Deep-Field South,
CDFS]{Dickinson2003} region. The maps cover an area of
8.7\,deg$^2$ with a $1\sigma$ depth of 36, 31 and 20\,mJy at 250,
350, and 500\,$\mu$m, respectively. We refer to this region as the
BLAST GOODS South Wide (BGS-Wide). A smaller region of
0.8\,deg$^2$ (BGS-Deep) nested inside BGS-Wide has a $1\sigma$
depth of 11, 9 and 6\,mJy at 250, 350, and 500\,$\mu$m,
respectively. These depths account for the instrumental noise
only, and do not include confusion. \citet{Marsden2009} estimate
that fluctuations arising from unresolved sources are a factor two
larger than instrumental noise at 500\,$\mu$m, in BGS-Deep.
Catalogs of sources detected at each wavelength in BGS-Deep and
BGS-Wide are presented by \citet{Devlin2009}.

D09 combine these single-wavelength catalogs by selecting sources
with a $\geq5\sigma$ (instrumental only, no confusion noise)
significance in at least one of the bands. They use this
multi-band catalog to identify counterparts (BLAST primary IDs) in
deep radio \citep[ACTA and VLA,][]{Norris2006,Miller2008} and
24\,$\mu$m \citep[SWIRE and
FIDEL,][]{Lonsdale2004,Dickinson2007,Magnelli2009} surveys. The
BLAST primary IDs all have $\leq5\%$ probability of being a chance
alignment. They also compile a list of secondary IDs, with
different counterparts associated to the same BLAST source as the
primary ID, but with larger probability of being a chance
alignment.

In this work we present an extended version of the D09 catalog of
BLAST primary IDs which contains 227 BLAST sources. In the
following sections we update this list to include UV data, recent
redshifts, corrections for submm flux boosting and blending,
morphology, AGN features, and star-formation rates (see Appendix B
for data tables). The list of secondary IDs is extensively
discussed in E09 and we do not investigate them further.

We emphasize again that the sample studied in this work comprises
the subset of BLAST-selected bright sources for which optical
spectroscopy/photometry is available, and/or for which we find a
clear counterpart in the UV. Naturally, this is only a fraction of
sources that would be in a purely BLAST-selected catalog, skewed
towards lower redshifts and strong optical/UV fluxes.

\subsection{Optical spectroscopy}\label{subsec:AAOmega}

A spectroscopic follow-up of the BLAST IDs was carried out with
the AAOmega optical spectrograph at the Anglo-Australian
Telescope. The BLAST spectroscopic redshift survey is discussed in
E09, as well as the reduction of the spectral data; here we extend
their analysis and results (see Sections \ref{sec:AGN},
\ref{sec:morph} and Tables \ref{tab:morphology},
\ref{tab:catalog1}).

AAOmega \citep[AAO,][]{Sharp2006} consists of 392, 2\arcsec-wide
fibers feeding light from targets within a 2$^\circ$
field-of-view; the configuration of diffraction gratings was
chosen to yield a wavelength coverage from 370 to 880\,nm, with
spectral resolution $\lambda/\delta\lambda \simeq 1300$. At
redshifts lower than 1, this allows us to detect two or more of
the following lines: [OII] 372.7, Calcium H and K, H$\beta$,[OIII]
495.9 and 500.7, H$\alpha$, [NII] 658.3 and [SII] 671.6 and 673.1.
At redshifts greater than 1, we only rely on broad emission lines,
such as Lyman $\alpha$, SiIV 140.3, CIII] 190.9 and CIV 154.9.

We have produced two prioritized lists of targets. The first list
comprises $\geq 3.5\sigma$ BLAST sources with primary radio or
24\,$\mu$m counterparts\footnote{If only the 24\,$\mu$m
counterpart is present, we refine the position of the source by
matching it with optical or IRAC 3.6\,$\mu$m coordinates.}.
Sources selected at 24\,$\mu$m are also included in the target
list to use all the available fibers. The second list contains the
secondary BLAST IDs, plus 24 $\mu$m sources.

The net observing time for the list of primary targets was 7\,hr,
obtaining spectra for 669 sources (316 BLAST IDs, and 356 SWIRE
sources). The list of secondary targets was observed for only
1\,hr (due to poor weather), obtaining 335 spectra (77 BLAST IDs,
and 258 SWIRE sources). Spectroscopic redshifts were consequently
obtained by E09 for 212 BLAST IDs in the primary list, 193 of
which have $\geq75\%$ confidence level (c.l.), and for 11 BLAST
IDs in the secondary list (all with $\geq75\%$ c.l.).

It is important to clarify here that the two lists used for the
AAO observations are not fully coincident with the D09 list
discussed in the previous section and used in this work. However,
a large overlap among sources in these lists is present and 82
sources from the D09 catalog of BLAST IDs have AAO redshifts, all
with $\geq95\%$ c.l. (see Table~\ref{tab:catalog1}).

Using the available spectra we estimate H$\alpha$ Equivalent
Widths (EWs) and [NII]/H$\alpha$ line ratios for 56 of these 82
sources. The remaining 26 sources either are at too high redshift
for the H$\alpha$ line to fall in our spectral coverage ($z\gtrsim
0.33$), or have spectra with poor signal-to-noise ratio. We
calculate the uncertainties on the EWs as quadrature sum of the
measurement error, estimated with a bootstrapping technique
applied to the individual spectra, and the Poisson noise,
estimated following \citet[][Equation 7]{Vollmann2006}.

We list the rest-frame EWs, EW$_{\rm rf}$ = EW$/(1+z)$, in
Table~\ref{tab:catalog1}, along with their uncertainties and the
[NII]/H$\alpha$ line ratios. Note that we have applied a
1\,\AA~correction to the H$\alpha$ EW$_{\rm rf}$ for underlying
stellar absorption \citep{Hopkins2003,Balogh2004}.

\subsection{UV data}\label{subsec:GALEX}
We identify near-UV (NUV) and far-UV (FUV) counterparts to BLAST
IDs by searching for {\sl GALEX} sources in the Deep Imaging
Survey \citep[DIS,][data release GR--4/5]{Martin2005} within
6\arcsec~of the radio or 24\,$\mu$m counterpart\footnote{If both
counterparts are present, we use the arithmetic mean between the
two sets of coordinates: [$\alpha_{\rm BLAST}$, $\delta_{\rm
BLAST}$].}, a separation just slightly larger than the {\sl GALEX}
PSF FWHM \citep{Morrissey2007}. This choice is justified by the
presence of a few extended objects, unresolved by the
submillimetric beam, that contribute to the same BLAST source (see
Section \ref{sec:morph}). After visual inspection of the UV
images, we have added one additional interacting system extending
beyond 6\arcsec~from the BLAST ID (\#2); in this case we integrate
the UV magnitude from both the interacting objects, because they
fall within the same BLAST beam. We estimate FUV and NUV
magnitudes using the standard {\sl GALEX} pipeline
\citep{Morrissey2007} for most IDs, whereas we perform aperture
photometry on 13 extended objects. A magnitude is considered to be
unreliable if the source is either confused or blended with a
star.

We find that 144 BLAST IDs have a NUV counterpart (136 with
reliable magnitude), and 113 have a FUV counterpart (107 with
reliable magnitude). Three sources are outside the area covered by
the DIS, and the remaining 80 BLAST IDs have no obvious
counterpart. By comparing the flux estimates for objects detected
in more than one {\sl GALEX} tile (pointing), we find that the
average uncertainty associated with the reproducibility of the
measurement is 0.06 and 0.11\,mag in NUV and FUV, respectively.
For bright galaxies, these values are larger than the uncertainty
in the calibration \citep[0.03 and 0.05\,mag in the NUV and FUV,
respectively;][]{Morrissey2007}, and in the source extraction
procedure ($\leq 0.02$ mag). The uncertainty on a quoted UV
magnitude is therefore the sum in quadrature of these three terms,
and it lies in the $1\sigma$ range of 0.07--0.25\,mag and
0.12--0.5\,mag in NUV and FUV, respectively.

{\sl GALEX} postage-stamp images, $2\arcmin\times 2\arcmin$ wide,
are used to study the UV-morphology of the BLAST IDs; a
selection\footnote[3]{The complete set of full-color cut-outs can
be found at http://blastexperiment.info/results\_images/moncelsi/}
of these is shown in Figure~\ref{fig:postage_stamps}. UV
magnitudes and uncertainties are listed in
Table~\ref{tab:catalog2}.

\subsection{SWIRE 70 and 160\,$\mu$m MIPS maps}\label{subsec:MIPS}
We use 70 and 160\,$\mu$m fluxes extracted from SWIRE maps
\citep{Lonsdale2004} at positions [$\alpha_{\rm BLAST}$,
$\delta_{\rm BLAST}$] to constrain the SED of each BLAST source at
wavelengths shorter than the emission peak (see Section
\ref{subsec:FIR_lum}). These maps overlap almost completely with
BGS-Wide, and all the $\geq5\sigma$ BLAST sources investigated in
this work lie within them. The $1\sigma$ depth of the maps is 3.6 and 20.8\,mJy at 70 and 160\,$\mu$m, respectively.

\subsection{MIR/NIR/optical images and catalogs}\label{subsec:IRAC_optical}
In addition to the aforementioned UV {\sl GALEX} images, we
investigate BLAST source morphology using optical and IR images.
The latter are 3.6, 4.5, 5.8, and 8\,$\mu$m IRAC \citep{Fazio2004}
images from the SWIRE survey. In the optical, we examine
($U\,g\,r$)-band images, acquired with the 4m Cerro Tololo
Inter-American Observatory (CTIO) as part of the SWIRE survey, and
$R$-band images from the COMBO--17 survey \citep{wolf04,wolf08}.
In Figure~\ref{fig:postage_stamps} we show $2\arcmin\times
2\arcmin$ cut-outs for a selection\footnotemark[3] of BLAST IDs.

\setcounter{footnote}{3}

For the purpose of studying the morphology, AGN fraction and
stellar mass, we have also matched, using a search radius of
3\arcsec~as in D09, the catalog of BLAST IDs to the following
catalogs:
\begin{enumerate}\addtolength{\itemsep}{-0.5\baselineskip}
    \item the SWIRE band-merged catalog consisting of optical ($U\,g\,r\,i\,z$) and MIR IRAC fluxes\footnote{The lower limits for inclusion in the catalog are 7 ($10\sigma$), 7 ($5\sigma$), 41.8 ($5\sigma$) and 48.6\,$\mu$Jy ($5\sigma$) at 3.6, 4.5, 5.8 and 8\,$\mu$m, respectively.} \citep{Surace2005};
    \item the 17 band COMBO--17 optical catalog \citep{wolf04,wolf08};
    \item the Multi-wavelength Survey by Yale-Chile \citep[MUSYC;][]{Gawiser2006} catalog for NIR photometry ($J$- and $K$-band).
\end{enumerate}
As a result of this analysis, out of 227 BLAST IDs:
\begin{itemize}\addtolength{\itemsep}{-0.5\baselineskip}
     \item 205 (90\%) have an IRAC counterpart from the SWIRE survey;
     \item 114 (50\%) have an optical (SWIRE and/or COMBO--17), \emph{and} either a NIR (MUSYC) or
MIR (3.6 or 4.5\,$\mu$m, IRAC) counterpart\footnote{We note that
the sky overlap among BGS, SWIRE, COMBO--17 and MUSYC is limited
to a $\sim4.15$\,deg$^2$ region.};
    \item 102 of the above 114 are detected in a
    minimum of 5 bands (optical, NIR and MIR);
    \item 52 of the above 102 have $J$- and $K$-band photometry
    from MUSYC.
\end{itemize}
We use the wealth of ancillary information for a variety of
purposes: we refer to Sections \ref{sec:AGN}, \ref{sec:morph} and
\ref{sec:masses} for discussions on AGN fraction, morphology and
stellar masses.

\subsection{Redshifts}\label{subsec:redshifts}
In addition to the 82 spectroscopic redshifts obtained with AAO
for the BLAST primary IDs, we have found 5 additional
spectroscopic redshifts by exploring the NASA/IPAC Extragalactic
Database (NED) with a 1\arcsec\ search radius around each ID. For
the other sources, we use photometric redshifts from the
MUSYC-EAZY \citep{Taylor2009}, COMBO--17 \citep[][only sources
with $R\leq24$]{wolf04,wolf08} and \citet[][RR08]{RR08} catalogs,
using again a 1\arcsec\ search radius. We carefully inspect each
individual alignment by taking into account the imaging data in
Figure~\ref{fig:postage_stamps}, the UV photometry, the SED in the
FIR/submm, and any additional information available from NED. In
cases of BLAST IDs with more than one associated photometric
redshift, priority is given in the order: EAZY, COMBO--17, and
RR08. We have thereby acquired 53 additional photometric
redshifts, of which 20 are from EAZY, 6 from COMBO--17 and 27 from
RR08.

\begin{figure}[htbp]
\epsscale{1.} \plotone{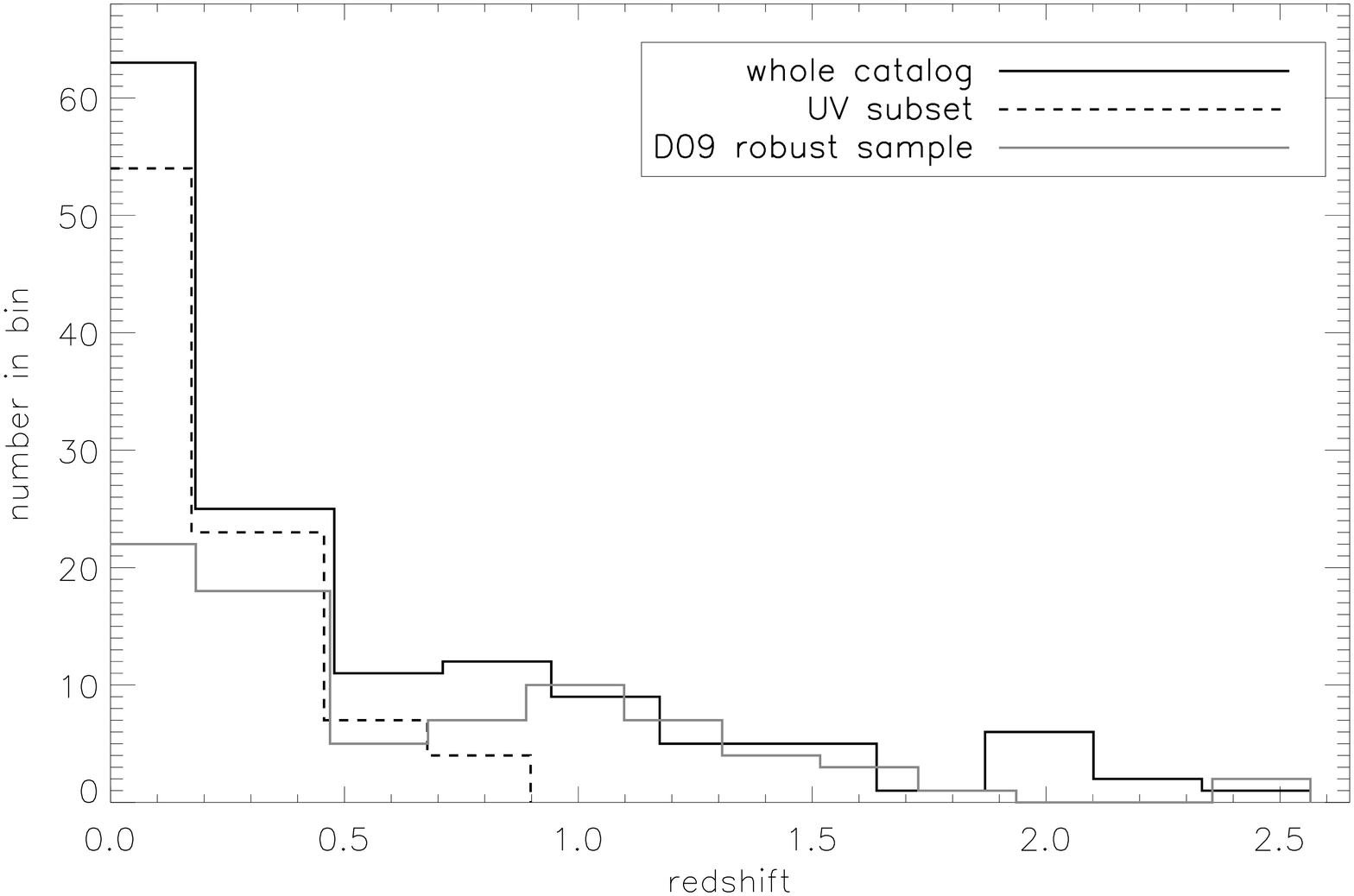} \caption{Redshift
distributions for the whole catalog of BLAST IDs and for the
subsample with UV data. The former has a median of 0.29 and an
inter-quartile range of 0.12--0.84; the latter has a median of
0.18 and an inter-quartile range of 0.10--0.34. We also show the
redshift distribution for the robust sample of D09, with median of
0.6 and an inter-quartile range of
0.2--1.0.}\label{fig:redshift_distr_NUV_FUV}
\end{figure}

We have succeeded in assigning 140 redshifts out of 227 ($\sim
62\%$) objects in our sample. The redshifts are listed in
Table~\ref{tab:catalog1}, along with their provenance.
Figure~\ref{fig:redshift_distr_NUV_FUV} shows the redshift
distribution of the whole BLAST ID catalog, and of the UV subset
used in Section \ref{sec:tot_SFR} for discussion of the total
star-formation rates. The number of sources with redshift is
doubled with respect to the robust sample of D09\footnote{The
robustness of a source is assessed by D09 based solely on the
goodness of the SED fit.}, but the median redshift is roughly
halved. This apparent pronounced discrepancy, limited to the $z
\lesssim 0.2$ bin, amounts to 40 sources and is due to the
combination of two selection effects. First, roughly 15 sources in
D09 with $z \lesssim 0.2$ (mostly from RR08) do not make it into
the robust sample, mainly because the photometric redshift is
intrinsically unreliable or, in a handful of cases, because the
BLAST source has been spuriously identified with the counterpart.
Second, 27 other sources with redshifts estimated in this work
have no redshift in D09, because they have neither sky coverage
from COMBO--17 nor from RR08; of these 27, 21 are from AAO, and 24
have $z \lesssim 0.25$. Therefore the apparent excess of low-$z$
sources with respect to D09 partly reflects the inclusion of the
AAO spectroscopic redshifts (naturally skewed towards low-$z$) and
partly lies in the intrinsic robustness in D09 of either the
photometric redshift or the counterpart itself.

It is worth noting here that this study misses a large fraction of
the high-$z$ BLAST sources that are known to constitute an
important part of the BLAST population
\citep{Devlin2009,Marsden2009,Pascale2009}. This is again due to
the combination of two factors. First, $\sim38\%$ of the BLAST IDs
presented in this paper do not have a redshift estimate; using
information about the UV identification rate (similarly to D09),
we can argue that more than half of the sources without a redshift
estimate lie at $z\gtrsim0.7$. In fact, 90 out of 99 (91\%)
sources at $z\leq0.7$ (and 96 out of 115, 83\%, sources at
$z\leq1$) have a {\sl GALEX} counterpart; now, of the 87 sources
with no redshift estimate, 57 (66\%) do not have a {\sl GALEX}
counterpart. Under the assumption that the UV identification rate
is a reasonable (if coarse) estimator of redshift, arguably more
than half of the sources without a redshift estimate lie at
$z\gtrsim0.7$ and roughly half lie at $z\gtrsim1$. Secondly, D09
starts with a catalog composed of bright, $\geq5\sigma$ sources
with flux densities $\geq33$\,mJy at 250\,$\mu$m, $\geq27$\,mJy at
350\,$\mu$m, and $\geq19$\,mJy at 500\,$\mu$m; \citet{Dunlop2010}
and \citet{Chapin2010} clearly show the necessity of digging
deeper into the BLAST maps, with the aid of the deepest available
multi-wavelength data, in order to identify the faintest, high-$z$
BLAST galaxies. Of course, this is done at the expense of the size
of the submm sample, which inevitably drops to a few tens of
sources.

Nonetheless, the present study is still unique in terms of size of
the sample, wavelength coverage, depth and quality of the
ancillary data. Indeed, {\sl IRAS} sources have been studied at
many wavelengths \citep[e.g.][]{DellaValle2006,Mazzei2007}, but
with little knowledge of the details of the cold dust emission
from which the FIR star-formation rate estimates come. Some
improvements have been made with the SCUBA Local Universe and
Galaxy Survey \citep[SLUGS;][]{Dunne2000,Vlahakis2005}, but still
with limited ability to estimate the bolometric FIR luminosity.
The results in this paper probably will not be immediately
replaced by deeper surveys undertaken by {\sl Herschel}; in fact,
even the much more sensitive observations carried out with SPIRE
will have to face the lack of deeper ancillary data. This is
especially true in the optical/NIR, where most of the $z>2$ submm
galaxies are much too faint to be detected by instruments like
AAOmega, and in the radio, where the identification rate of the
faintest $z>2$ sources drops drastically, even when using the
deepest available data (VLA).

\section{FIR luminosities and SFRs}\label{sec:FIR_lum_SFR}

\subsection{Deboosting the BLAST fluxes}\label{subsec:deboost}
The sources in the BLAST catalog used by D09 to identify
counterparts in the radio and 24\,$\mu$m were detected directly
from the maps of BGS-Deep and BGS-Wide. While the details of the
catalog are discussed there, it is useful to summarize here the
procedure to clarify what are the potential biases.

First, a catalog of BLAST sources with detection significance
higher than 3$\sigma$ is made at each wavelength, independently.
Each entry in the catalog is then positionally matched across the
three bands, with the requirement of a 5$\sigma$ detection in at
least one band. The significance here is relative to instrumental
noise, and does not include confusion noise. A new position is
assigned to the source by averaging its positions in the original
single-wavelength catalogs, with weights estimated by taking into
account the beam sizes and the signal-to-noise ratios (SNRs) of
the detections at each wavelength. This combined catalog is then
used to identify counterparts in the radio and at 24\,$\mu$m, and
a new flux density is measured from the 70 to 500\,$\mu$m maps at
the accurate position of the counterpart.

The BLAST differential source counts fall very rapidly with flux
density \citep[approximately following $dN/dS \propto
S^{-3}$,][]{Patanchon2009}, thus Eddington bias as well as source
confusion will cause the fluxes to be boosted. This effect has to
be estimated to properly compute the FIR luminosity of each
source. \citet{Coppin2005} have proposed a Bayesian approach that
can be applied to estimate the most likely flux distribution when
the noise properties of the detection and the underlying source
distribution are known. Their method is derived under the
assumption that the flux density comes from just one source, plus
noise. This cannot be applied to BLAST sources because of
blending: the measured flux density can either come from just one
source, or more likely from several sources blended together by
the beam, which then appear as one single source of larger flux
density.

\begin{figure}[htbp]
\epsscale{0.9} \plotone{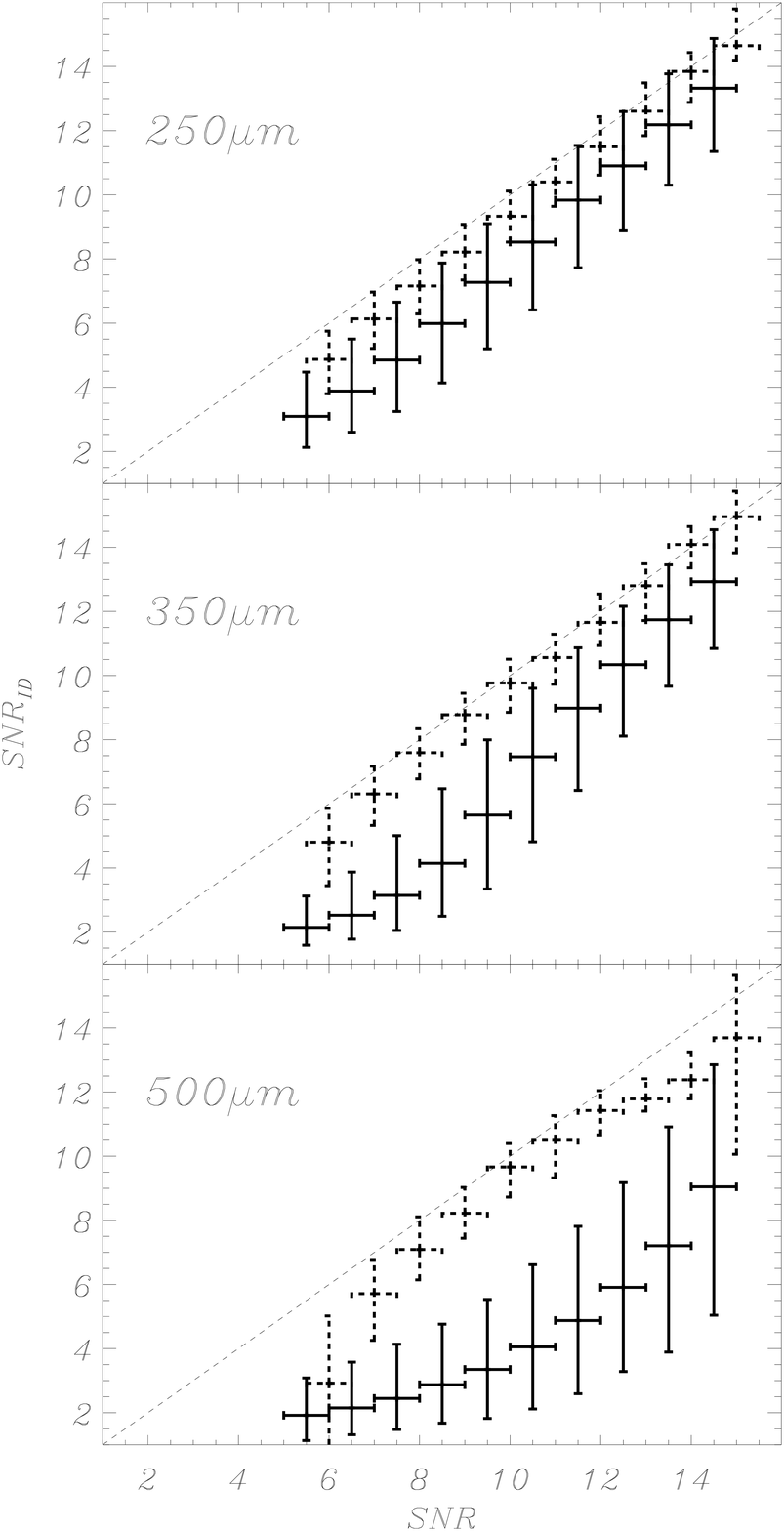} \caption{Effects of flux
boosting, and source blending at BLAST wavelengths in BGS-Deep
(solid error bars) and in BGS-Wide (dashed bars). For a source
with a measured SNR at a given wavelength, the points show the
distribution of the SNR$_{\rm ID}$ retrieved from simulations,
binned in 1-SNR wide bins. Each point indicates the median value
of the distribution in each bin, and the low and high error bars
are the first and third interquartiles, respectively. The dashed
line indicates where the points would lie in the absence of
biases. The effects are mild in the wide region, where
instrumental noise dominates, and get more severe in BGS-Deep,
where confusion noise dominates, and source blending becomes more
important. At the longest wavelength, the beam size blends fluxes
from many adjacent sources, giving a strong bias. This is not a
major problem for our analysis, which deals with sources
identified at low, or moderate redshifts.}\label{fig:deboost}
\end{figure}

We have developed a different method to account for boosting of
BLAST fluxes, which is entirely based on Monte Carlo simulations.
We generate 100 noise-less sky maps using the BLAST measured count
models \citep{Patanchon2009}, and no clustering. Noise is added to
each simulated map to a realistic level for the BGS-Deep and
BGS-Wide regions. Sources are then retrieved with the same method
used on the real maps \citep{Devlin2009}. Considering all the
input components within a FWHM beam distance from each retrieved
source, we stipulate that the input component with largest flux
density is the actual counterpart\footnote{We know that this
assumption is always verified in BGS-Wide but less so in BGS-Deep,
where, in 21\% of the cases, the second brightest component
contributes to more than 50\% of the retrieved flux (see E09,
Appendix B).} (ID). The source flux density is then remeasured at
the position of the ID. Finally, we compare this flux density with
that of the input source. By repeating this for each source
detected in each simulation, we generate distributions of
input/output SNR, where the relevant noise is the instrumental
noise at the position of the ID. These simulations are similar to
those used in \citet{Chapin2010} to study the effects of confusion
for their deeper sample.

Figure~\ref{fig:deboost} shows the result of this analysis. In
each bin we display the median of the distribution of input SNR
(labeled SNR$_{\rm ID}$) corresponding to the measured SNR. The
error bars define the first and third interquartiles. To obtain
the deboosted flux density likelihood, it suffices to multiply the
$y$-axis by the corresponding instrumental noise. It is clear from
this figure that sources in the BGS-Wide region are only
moderately affected by boosting. The situation is substantially
different for BGS-Deep, and the effect of boosting increases with
wavelength, as expected, due to the telescope PSF becoming larger.
At the longest BLAST wavelength, the fluxes are severely affected
by boosting: a source detected even with a 10$\sigma$ significance
level has a deboosted flux only about half of what is measured
directly from the map. By comparing the deboosted values for
BGS-Wide at 250 and 350\,$\mu$m, we notice that the longer
wavelength appears to be slightly less biased. This arises from
the fact that the two PSFs are not very different in size (36 and
42\arcsec, respectively), but the 250\,$\mu$m PSF has larger
sidelobes \citep{Truch2009}.

\subsection{SED fitting and FIR luminosities}\label{subsec:FIR_lum}

In order to estimate the rest-frame FIR luminosity ($L_{\rm FIR}$)
of each BLAST source in our sample, we perform SED fitting using
the MIPS flux densities (70 and 160\,$\mu$m only) and the
deboosted BLAST flux densities; the model template is a modified
blackbody spectrum \citep[with spectral index $\beta =
1.5$,][]{Hildebrand1983}, with a power law $\nu^{-\alpha}$
replacing the Wien part of the spectrum, to account for the
variability of dust-temperatures within a galaxy \citep[we choose
$\alpha=2$,][]{Blain2003,Viero2010}. \citet{Pascale2009} have
shown that the estimated FIR luminosities depend weakly on the
choice of $\alpha$, whereas the estimated dust temperatures are
more sensitive to the template used. Since our analysis does not
employ temperature measurements, the value of $\alpha$ we adopt is
not critical. We also note here that the SED template chosen is
the one that best performs in fitting the spectrum of two
often-used IR-luminous local galaxies, Arp 220 and M82; by
sampling their SEDs at the five observed wavelengths in question,
the nominal FIR luminosities and dust temperatures are correctly
retrieved (within uncertainties) not only at $z\sim0$, but also
when their spectra are redshifted up to $z=2$.

The way each BLAST flux density is deboosted depends on its SNR.
If this is larger than 15, no correction is applied. If the
measured flux density is smaller than twice the square root of the
sum in quadrature of instrumental and confusion noise \citep[as
reported in][]{Marsden2009}, the detection is treated as an upper
limit. In all other cases, the above deboosting distributions are
used. For sources in BGS-Deep the deboosting likelihood
distribution is well approximated by a Gaussian function, but this
is less true in BGS-Wide (especially at low SNR). Therefore, we
use the sampled distribution for sources in BGS-Wide, and a
Gaussian approximation in BGS-Deep.

The portion of noise arising from confusion is highly correlated
among bands. The Pearson coefficients of the correlation matrix
are listed in Table~\ref{tab:Pearson}, and have been estimated
from the (beam-convolved) BGS-Deep and BGS-Wide maps. As expected,
the correlation effects are more important for sources in
BGS-Deep, and we do take this into account in the SED fitting
algorithm, whereas no correlations among bands are considered for
sources in BGS-Wide. This turns out to be convenient, as in
BGS-Deep the distributions are Gaussian, and a correlation
analysis is relatively straightforward. This would not be the case
for the sources in BGS-Wide.

\begin{deluxetable}{ccccccc}
\tablewidth{0pt} \small \tablecaption{Correlations among BLAST
bands\label{tab:Pearson}}
\tablehead{\colhead{Band} & \multicolumn{6}{c}{Pearson correlation matrix} \\
    &\multicolumn{3}{c}{BGS-Deep} & \multicolumn{3}{c}{BGS-Wide}\\
\colhead{[$\mu$m]} & \colhead{250\,$\mu$m} & \colhead{350\,$\mu$m}
& \colhead{500\,$\mu$m} & \colhead{250\,$\mu$m} & \colhead{350
$\mu$m} & \colhead{500\,$\mu$m}}\startdata
250 & 1 &0.68 & 0.66 & 1 & 0.26 & 0.29\\
350 &   & 1   & 0.69 &   &  1   & 0.29\\
500 &   &     & 1    &   &      & 1\\
\enddata
\end{deluxetable}

MIPS fluxes at 70 and 160\,$\mu$m were also used in the fitting
routine to constrain the SED at wavelengths shorter than the
emission peak. Deboosting these bands is beyond the scope of this
paper, and it is less necessary because the source counts are
shallower than the BLAST ones
\citep[see][]{Frayer2009,Bethermin2010}. The SED fitting procedure
copes with the size of the photometric bands (color-correction),
and the instrumental plus photometric uncertainties
\citep{Truch2009}. Correlations are properly taken into account
via a Monte Carlo procedure.

\begin{figure}[htbp]
\epsscale{0.69} \plotone{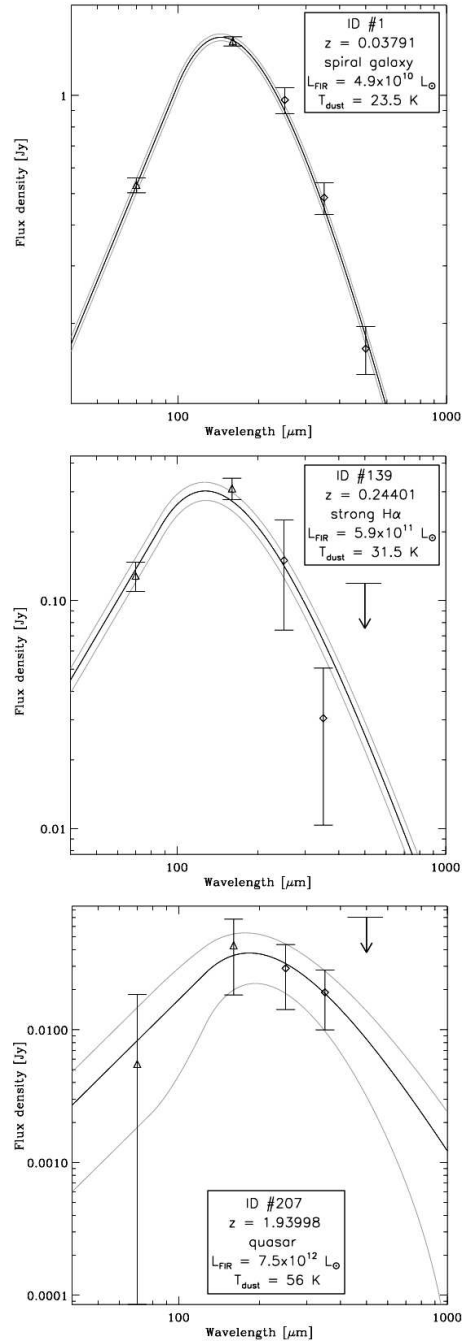} \caption{SED fitting of
the FIR flux densities for three representative objects in our
sample. Points with error bars are from BLAST (deboosted,
color-corrected 250, 350, and 500\,$\mu$m) and MIPS (70 and
160\,$\mu$m); arrows indicate upper limits (see text). Black solid
lines show the best-fit curves, with 68\% confidence levels
displayed as gray solid lines. The fitting routine accounts for
the finite BLAST bandwidths and for the correlated calibration
uncertainties. The model template is a modified graybody with an
emissivity law $\beta=1.5$ \citep{Hildebrand1983} and a power law
$\nu^{-\alpha}$ replacing the Wien part of the spectrum
\citep[$\alpha=2$,][]{Blain2003,Viero2010}.}\label{fig:SEDs}
\end{figure}

In Figure~\ref{fig:SEDs} we show the fitted FIR SED for three representative objects in our sample: a low-redshift spiral galaxy; a mid-redshift strong H$\alpha$ emitter; and a high-$z$ quasar. The resulting FIR luminosities, listed in Table~\ref{tab:catalog2}, are the rest-frame SED integral between 8 and 1000\,$\mu$m \citep{Kennicutt98}.

\begin{figure}[htbp]
\epsscale{1.1} \plotone{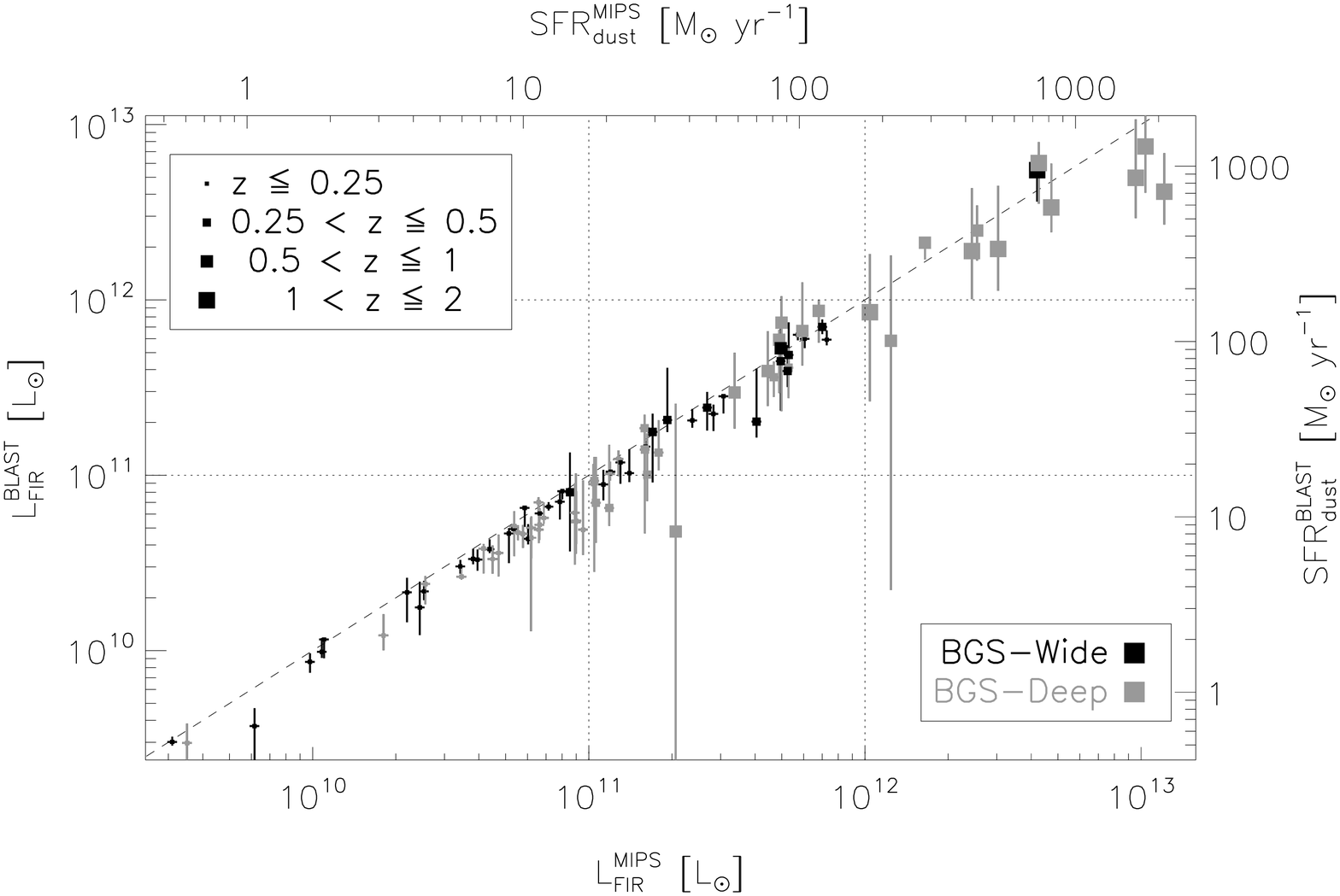}
\caption{Comparison of estimates of total FIR luminosity for a
$z\leq 2$ subset of 93 sources with 24\,$\mu$m counterpart. On the
$x$-axis we used the prescription of \citet[][Equation
4]{DaleHelou2002} based on 24, 70 and 160\,$\mu$m MIPS fluxes; the
error bars are set to 4\%, which represents the mean discrepancy
between their prescription and their model bolometric IR
luminosities. On the $y$-axis we used the FIR luminosity estimates
and uncertainties described in Section \ref{subsec:FIR_lum}.
Sources lying in the BGS-Wide region are in black and sources in
BGS-Deep are in gray. Symbol sizes increase with redshift as shown
in the legend. The secondary axes are both calculated using
Equation~\ref{eq:SFR_FIR}. The dashed line shows $y=x$, for
reference.}\label{fig:L_FIR_MIPS_vs_BLAST}
\end{figure}

In Figure~\ref{fig:L_FIR_MIPS_vs_BLAST} we compare our estimates
of rest-frame FIR luminosity with those obtained using only MIR
flux densities, to investigate the level of uncertainty when data
are not available in the submm. Following the prescription of
\citet{DaleHelou2002}, we calculate the FIR luminosities using
only MIPS flux densities (24, 70 and 160\,$\mu$m) for a $z\leq 2$
subset of 93 sources with 24\,$\mu$m counterpart. There is
considerable agreement up to $L_{\rm FIR} \lesssim 5 \times
10^{11}$\,L$_{\sun}$ and $z\lesssim0.5$. At higher redshifts (and
luminosities) we find a poorer concordance; the MIPS-only
estimates tend to overestimate the FIR luminosity, by as much as a
factor of two in some cases. Other authors
\citep{Pope2006,Papovich2007,Kriek2008,Murphy2009,Muzzin2010,Elbaz2010,Nordon2010}
find similar trends; this is expected as the MIPS bands sample the
SED peak progressively less and less as redshift increases, thus
pulling the SED toward shorter wavelengths, and resulting in a
higher $L_{\rm FIR}$. This emphasizes how essential the BLAST and
SPIRE wavebands are to constrain the IR emission peak of
star-forming galaxies at high redshift \citep[see also
e.g.][]{Schulz2010,Elbaz2010}.

\subsection{FIR Star-Formation Rates}\label{subsec:SFR_FIR}
The FIR luminosities are a sensitive tracer of the young stellar
population and, under some reasonable assumption, can be directly
associated to the Star-Formation Rates (SFRs). This is
particularly true for dusty starburst galaxies, because the
optically thick dust surrounding star forming regions is very
effective in absorbing the UV photons emitted by young, massive
stars and converting this energy into IR emission.

\begin{figure}[htbp]
\epsscale{1.1} \plotone{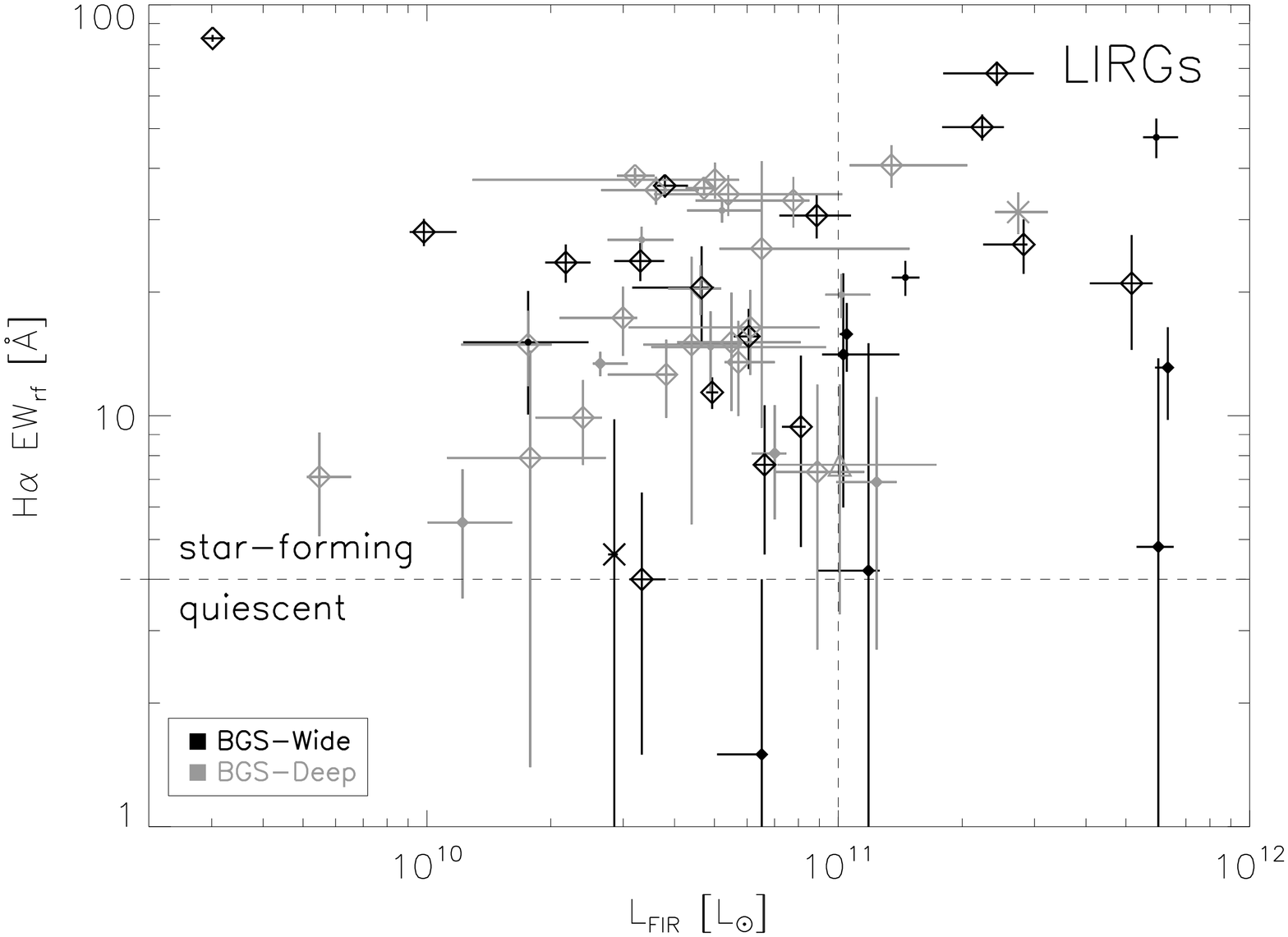} \caption{H$\alpha$
rest-frame Equivalent Widths (EW$_{\rm rf}$) as a function of the
FIR luminosity for the subset of 56 $z\lesssim 0.33$ sources
described in Section \ref{subsec:AAOmega}. Note that we applied a
1\,\AA~correction to the H$\alpha$ EW$_{\rm rf}$ for underlying
stellar absorption \citep{Hopkins2003}. Sources lying in the
BGS-Wide region are in black, sources in BGS-Deep are in gray. We
also encode here the morphological information discussed in
Section \ref{sec:morph}: spiral galaxies are indicated with empty
diamonds; compact objects with empty squares; ellipticals with
triangles; interacting systems with crosses; Seyfert galaxies with
filled diamonds; and objects without morphological classification
with filled circles. The horizontal dashed line at
4\,\AA~separates galaxies with ongoing star formation from
quiescent ones \citep{Balogh2004}. Clearly all galaxies in our
sample but one are compatible with being actively forming
stars.}\label{fig:LFIR_vs_HaREW}
\end{figure}

Under the assumption that the above is the only physical process
heating up the dust, \citet{Kennicutt98} has derived the following
relation between SFR and bolometric FIR luminosity:
\begin{equation}\label{eq:SFR_FIR}
{\rm SFR}_{\rm dust}\left[\frac{{\rm M}_{\sun}}{\rm yr}\right] =
1.73 \times 10^{-10} \times L_{\rm FIR}[{\rm L}_{\sun}].
\end{equation}
Our sample includes sources with a wide range of FIR luminosities.
On one end, the FIR energy output is similar to the one found in
Luminous IR galaxies (LIRGs, $L_{\rm FIR} > 10^{11}$\,L$_{\sun}$),
and Ultra Luminous IR galaxies (ULIRGs, $L_{\rm FIR} >
10^{12}$\,L$_{\sun}$). In this type of source, AGN can play an
important role in heating up the dust, resulting in a bias in the
SFR calculation (an effect discussed further in Section
\ref{sec:AGN}). At lower FIR luminosities, we have strong
additional evidence indicating that most of the galaxies sampled
by BLAST are actively star-forming. This is shown in
Figure~\ref{fig:LFIR_vs_HaREW}: available H$\alpha$ rest-frame
Equivalent Widths (EW$_{\rm rf}$) are plotted against FIR
luminosity for 56 sources at $z\lesssim 0.33$ (see Section
\ref{subsec:AAOmega}). The horizontal dashed line at
4\,\AA~separates galaxies with ongoing star formation from
quiescent ones \citep{Balogh2004}. All sources but one have
H$\alpha$ signature of ongoing star formation. It is highly
unlikely that, despite the poor statistics of this plot, we could
be missing a population of quiescent objects with $L_{\rm FIR}
\lesssim 10^{10}$\,L$_{\sun}$, whose FIR emission is due to a
different physical process than the one described above.

Nonetheless, as the FIR luminosity decreases, our sources approach
more normal star forming galaxies. In this type of source a
non-negligible contribution to dust heating comes from older
stellar populations, which would bias the SFR estimate high
\citep{Bell2003,Hirashita2003,Iglesias-Paramo2004,Iglesias-Paramo2006}.
The reduced optical depth of dust also needs to be taken into
account or it would result in a lower estimate of SFRs
\citep{Inoue2002}. Both these effects are considered in the
following discussion (Section \ref{sec:tot_SFR}) on the total SFR
in our sample.

\section{UV luminosities and SFRs}\label{sec:UV_lum_SFR}

\subsection{UV fluxes and rest-frame luminosities}\label{subsec:UV_lum}
The amount of unobscured star formation ongoing in each galaxy of
our sample can be estimated in the UV for the BLAST IDs with a
{\sl GALEX} counterpart.

The (AB) UV magnitudes are corrected for extinction $A_{\lambda}$
due to dust in our Galaxy, and converted into observed flux
densities $S_{\nu_{\rm obs}}$. Rest-frame UV luminosities are
calculated as:
\begin{equation}\label{eq:UV_lum}
L^{\rm rf}_{\rm UV} = 4\pi~S_{\nu_{\rm obs}}~D_{\rm
L}^2(z)~\nu_{\rm obs},
\end{equation}
where $D_{\rm L}(z)$ is the luminosity distance.

The extinction coefficients used in the analysis are estimated
following the prescription of \citet{Wyder2007}, and the color
excesses E($B-V$) as measured from DIRBE/{\sl IRAS} dust maps
\citep{Schlegel1998} are listed in Table~\ref{tab:catalog2}.

\subsection{UV Star-Formation Rates}\label{subsec:SFR_UV}
\begin{figure}[htbp]
\includegraphics[scale=0.33,angle=270]{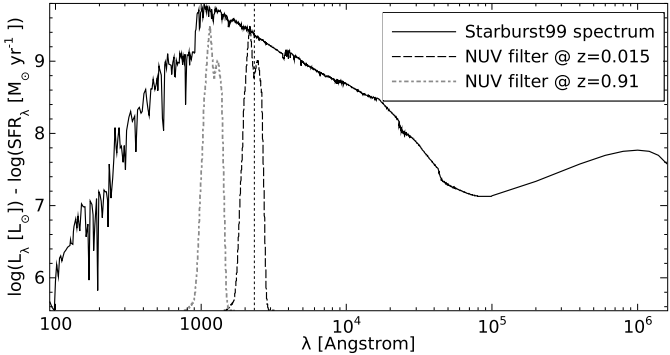}
\caption{Synthetic spectrum computed with {\it starburst99}
\citep{Leitherer1999}, under the assumptions of solar metallicity
and Salpeter \citeyearpar{Salpeter1955} IMF from 0.1 to 100
M$_{\sun}$. Following Equations~\ref{eq:SFR_NUV} and
\ref{eq:K_NUV}, the $K$-correction factor for the NUV, $K_{\rm
NUV}(z)$, is computed by averaging the synthetic spectrum over the
broad {\sl GALEX} filter profile, also shown (in arbitrary units),
blueshifted for reference in the rest frame of the nearest and
farthest object in our UV subsample. The same can be done for the
FUV filter (not shown here).} \label{fig:spectrum}
\end{figure}

Star-formation rates in the UV are estimated following the
approach of \citet[][and references therein]{Iglesias-Paramo2006}.
These are related to rest-frame luminosities in the FUV and NUV by
using a synthetic spectrum obtained with {\it
starburst99}\,\footnote{Under the same assumptions of
\citet{Iglesias-Paramo2006}: continuous star formation, recent
star-formation time scale $\sim10^8$\,yr, solar metallicity and
Salpeter \citeyearpar{Salpeter1955} IMF from 0.1 to 100
M$_{\sun}$.} \citep[{\it sb99};][]{Leitherer1999} for a
star-forming galaxy. In the wavelength range 1000--3000\,\AA, the
shape of the spectrum (shown in Figure~\ref{fig:spectrum}) is very
weakly dependent on the underlying stellar populations
\citep[e.g.][]{Kennicutt98}, and has a $\lambda^{-2}$ slope.

NUV star-formation rates are estimated using the equation:
\begin{equation}\label{eq:SFR_NUV}
{\rm log~SFR}_{\rm NUV}\left[\frac{{\rm M}_{\sun}}{\rm yr}\right]
= {\rm log~}L^{\rm rf}_{\rm NUV}[{\rm L}_{\sun}] - K_{\rm NUV}(z),
\end{equation}
where $L^{\rm rf}_{\rm NUV}$ is the rest-frame luminosity
calculated from the observed near-UV magnitude using
Equation~\ref{eq:UV_lum}. $K_{\rm NUV}(z)$ is a redshift-dependent
numerical factor which incorporates the $K$-correction, and is
derived from {\it sb99}, integrating over the {\sl GALEX} filter
profile $f_{\rm NUV}$:
\begin{equation}\label{eq:K_NUV}
K_{\rm NUV}(z)=\frac{\int ({\rm log~}L_{\lambda}^{{\it sb99}}[{\rm
L}_{\sun}] -{\rm log~SFR}_{\lambda}^{{\it sb99}}[\frac{{\rm
M}_{\sun}}{\rm yr}])~f_{\rm NUV}~ d\lambda^{\rm rf}}{\int f_{\rm
NUV}~d\lambda^{\rm rf}}.
\end{equation}
SFR$_{\rm FUV}$ and $K_{\rm FUV}(z)$ are obtained in a totally
analogous way. The values of $K_{\rm FUV}(z=0)$ and $K_{\rm
NUV}(z=0)$ are the same as those used by
\citet{Iglesias-Paramo2006} at $z=0$. The photometric errors
described in Section \ref{subsec:GALEX} are propagated in the
estimate of the uncertainties on the UV SFRs.

A redshift limitation arises when the observed NUV and FUV sample
the rest-frame Lyman continuum. This happens at $z \sim 0.36$ in
the FUV, and $z \sim 0.91$ in the NUV. Hereafter we exclude
sources beyond these redshift limits, as their inferred SFRs would
be unreliable. In order to have a more uniform and sufficiently
large sample, in what follows we only consider the NUV subset,
which counts 89 sources (see
Figure~\ref{fig:redshift_distr_NUV_FUV} for their redshift
distribution). As anticipated, the UV luminosities/SFRs are not
corrected for intrinsic dust extinction, and are combined in the
next section with FIR luminosities to build an estimator of total
SFR that is independent of extinction models.

\section{Total SFR}\label{sec:tot_SFR}
We now have two separate estimators for the star-formation rates
in our galaxy sample, SFR$_{\rm dust}$ and SFR$_{\rm NUV}$. Each
of these is expected to have different biases and short-comings.
One can clearly do better at estimating the SFR by combining the
two estimators in some way. The best way to do this is not obvious
though, since it depends on how each of the estimators was
calibrated, on the assumptions that went into them, on the range
of galaxy SEDs being studied, and on how these relate to local
galaxies that were used for calibration, including radiative
transfer effects and other complications. Because of this, we
choose to follow a prescription to estimate the total SFR in a
galaxy which has already been used by several authors
\citep{Bell2003,Hirashita2003,Iglesias-Paramo2006,Buat2007}, so
that we can at least compare our results to those of several
related studies.

In order to estimate the total SFR (SFR$_{\rm tot}$) in our
sample, we combine the contribution from the obscured star
formation with the unobscured star formation:
\begin{equation}\label{eq:tot_SFR}
{\rm SFR}_{\rm tot} = {\rm SFR}_{\rm NUV} + (1-\eta) \times {\rm
SFR}_{\rm dust}.
\end{equation}
A correction factor $(1-\eta)$ is applied to the dust contribution
to account for the IR emission from older stellar populations.
Following \citet{Bell2003} and \citet{Iglesias-Paramo2006}, we use
different values of $\eta$ depending on whether the object in
question is more likely to be a starburst ($\eta \sim 0.09$ for
$L_{\rm FIR} \geq 10^{11}$\,L$_{\sun}$) or a normal star-forming
galaxy ($\eta \sim 0.32$ for $L_{\rm FIR} \leq
10^{11}$\,L$_{\sun}$). As anticipated in Section
\ref{subsec:SFR_FIR}, this method can account for both the
contrasting effects that come into play when we try to estimate
the total SFR budget for an inhomogeneous sample of objects.
Namely, $\eta$ parametrizes the contribution to dust heating from
older stellar populations as a function of the integrated FIR
luminosity, whereas the contribution from the UV luminosity
guarantees that all the UV photons that manage to escape the
galaxy, due to the reduced optical depth of the dust, are actually
taken into account.

We briefly recall here that the main selection effects of our
sample are, on the one hand, that the rest-frame $L_{\rm FIR}$
increases steadily with redshift (see
Figure~\ref{fig:L_FIR_MIPS_vs_BLAST} and D09), and on the other
hand that the UV luminosity estimates are not reliable beyond
$z\sim 0.9$. Moreover, we stress the importance of the blending
effects reported in Section \ref{subsec:deboost}, which may lead
to misidentifications, particularly in BGS-Deep (sources in gray).

\begin{figure}[htbp]
\epsscale{1.1} \plotone{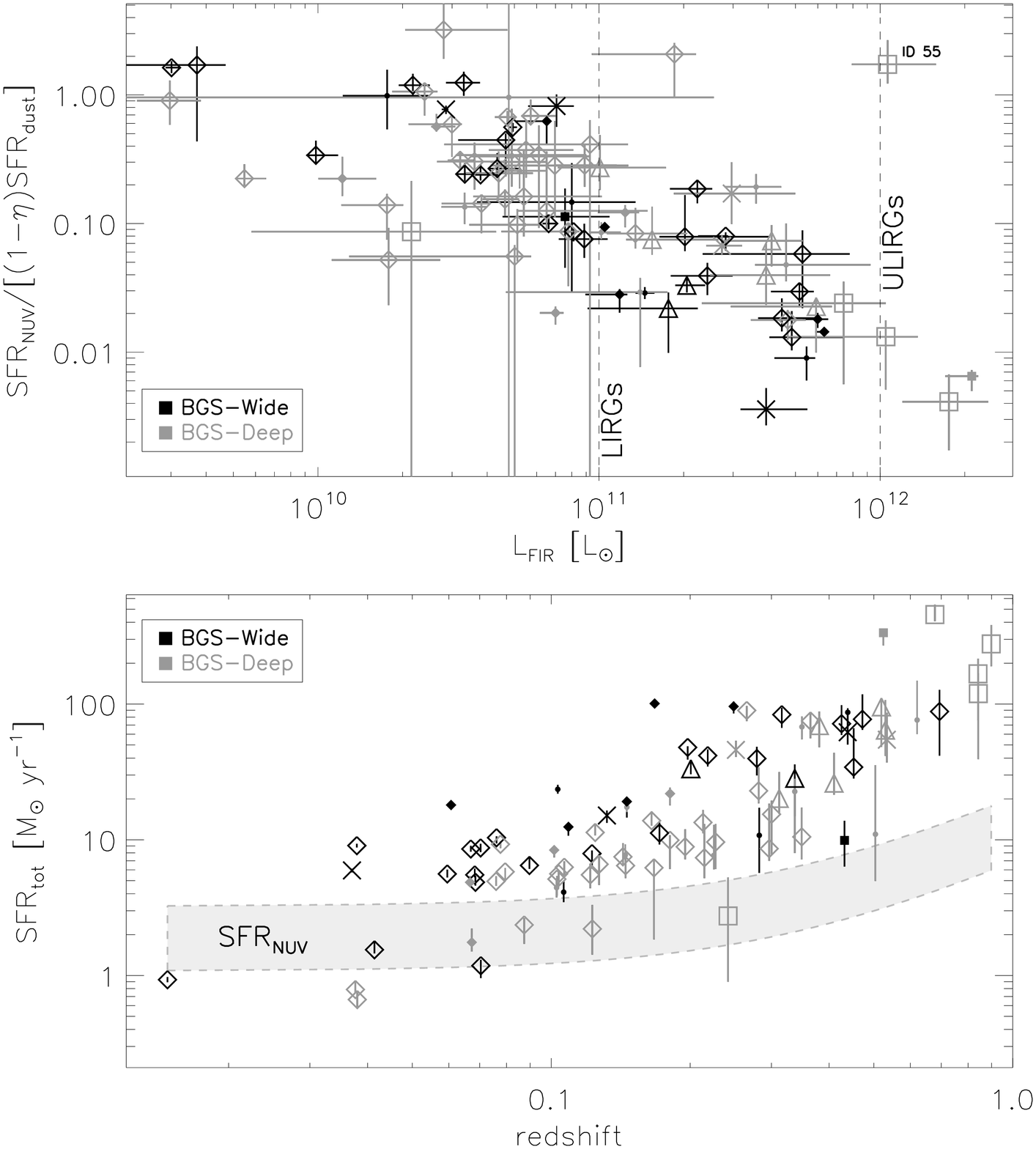} \caption{Top panel: ratio
of SFR estimated from the NUV only to SFR estimated from dust
only, as a function of the FIR luminosity. Note that SFR$_{\rm
dust}$ is corrected by a factor $(1-\eta)$ to account for the IR
emission from old stellar populations (see text). Bottom panel:
total SFR (SFR$_{\rm tot}$, see Equation~\ref{eq:tot_SFR}) as a
function of redshift. The gray shaded area shows the 1$\sigma$
confidence interval of a power-law fit to SFR$_{\rm NUV}\propto
z^{1.6}$. Symbols are as in Figure~\ref{fig:LFIR_vs_HaREW}. Filled
squares indicate that the source is a quasar (see Section
\ref{sec:AGN}).}\label{fig:total_SFR}
\end{figure}

The results of this analysis are shown in
Figure~\ref{fig:total_SFR}. In the top panel we plot the ratio of
SFR$_{\rm NUV}$ to $(1-\eta)$SFR$_{\rm dust}$ as a function of the
FIR luminosity. With the exception of a few outliers\footnote{In
particular, ID\#55 could be a misidentification because there is a
secondary counterpart, see E09.}, there is a clear trend, namely
the NUV contribution is more important at low-$L_{\rm FIR}$
(low-$z$), whereas star formation is mainly obscured at $L_{\rm
FIR} \gtrsim 10^{11}$\,L$_{\sun}$, $z \gtrsim 0.5$. The same
effect is evident in the bottom panel, where we plot SFR$_{\rm
tot}$ as a function of redshift. The gray shaded area shows the
1$\sigma$ confidence interval of a power-law fit SFR$_{\rm
NUV}\propto z^{1.6}$. Most sources with SFR$_{\rm tot}$ larger
than a few M$_{\sun}$\,yr$^{-1}$ have negligible contribution from
the UV. This is consistent with what \citet{Takeuchi2009} find in
the local Universe for a FIR-selected sample: at SFR$_{\rm tot}>
20$\,M$_{\sun}$\,yr$^{-1}$, the fraction of directly visible SFR
(SFR$_{\rm NUV}$) decreases. A very similar trend is also observed
at higher redshifts by \citet{Buat2008}, with a
24\,$\mu$m-selected sample at $0\leq z\leq0.7$ that closely
resembles our sample at those redshifts, in terms of dynamic
ranges and FIR-to-UV ratios.

Such a behavior in the individual BLAST IDs can be related to the
greater evolution of the total FIR luminosity density with respect
to the optical-UV one, as reported for instance by
\citet{Pascale2009}. On the other hand, we stress that at $L_{\rm
FIR} \lesssim 10^{11}$\,L$_{\sun}$, $z \lesssim 0.25$, FIR-only
observations would lead to underestimates of the total SFR of at
least a factor of 2.

By comparing our sample in Figure~\ref{fig:total_SFR} with the
{\sl IRAS}/FIR-selected local sample of
\citet{Iglesias-Paramo2006}, we notice that the overlap is quite
modest and limited to $L_{\rm FIR} \lesssim 10^{10}$\,L$_{\sun}$,
$z \lesssim 0.1$ sources. We point out that this conclusion should
not be diminished by considerations on the extent of the local
volume sampled by the BLAST survey.

At the very high luminosity end, only two objects (one of which is
flagged as quasar, see Section \ref{sec:AGN}) with $z \leq 0.91$
have a UV counterpart. We have thus investigated the 30 galaxies
with $L_{\rm FIR} \geq 10^{12}$\,L$_{\sun}$ in the full set of
BLAST IDs, finding that 16 are flagged as quasars, most of which
are optically bright. At $z>1$, the optical $U$ and $g$ bands
probe the rest-frame UV, and we calculate that these objects would
virtually populate the top-right corner of the upper panel of
Figure~\ref{fig:total_SFR}. However, the UV emission from quasars
is strongly contaminated by the active nucleus, and cannot be
directly associated with recent star formation. Of the remaining
14 ULIRGs with no AGN signatures, only 4 have optical magnitudes,
and would occupy the bottom-right corner, indicating severe dust
attenuation. We can therefore argue that, even if our subset of
objects lacks the abundance of most luminous IR galaxies detected
in the SHADES survey \citep[see][]{Coppin2008,Serjeant2008},
SCUBA-like sources will likely lie in the bottom-right corner and
beyond, following the same trend of increasing dust attenuation at
higher FIR luminosities. This is a first hint that our analysis
begins to detect SCUBA galaxies, which are known to overlap
considerably with the fainter BLAST galaxy population, following
joint studies of LABOCA 870\,$\mu$m and BLAST data
\citep{Dunlop2010,Chapin2010}. We will discuss this in more detail
in Section \ref{sec:masses}.

The 24\,$\mu$m-selected sample described by \citet{LeFloch2005}
most resembles our $z\leq0.9$ sample in terms of $L_{\rm
FIR}$--$z$ parameter space, although our objects are in general
more massive, as we will see in Section \ref{sec:masses}. This, in
combination with Figure~\ref{fig:LFIR_vs_HaREW}, points to the
conclusion that the BLAST counterparts detected in this survey at
$z\lesssim1$ are mostly run-of-the-mill star-forming galaxies.
Finally, given the steep number counts at the BLAST wavelengths
\citep{Patanchon2009} and the smaller beam sizes of {\sl
Herschel}, we expect SPIRE to detect roughly a factor of 10 more
sources than BLAST, probing fainter fluxes and therefore higher
redshifts. Figure~\ref{fig:total_SFR} suggest that SPIRE will
likely fill the $10^{11} \lesssim L_{\rm FIR} \lesssim 2 \times
10^{12}$\,L$_{\sun}$ region \citep[see e.g.][]{Chapin2010}, but
probably will not be dominated by SCUBA-like sources.

\section{AGN fraction and quasars}\label{sec:AGN}
In this section we describe the AGN and quasar content of our
sample, and we investigate whether the submm emission that we see
with BLAST is mainly due to the host galaxy or to the active
nucleus.

AGN are identified using spectroscopic and photometric methods,
and the information is listed in Table~\ref{tab:catalog1}. Of the
82 sources in our sample with optical spectra, 56 have a
measurement of the line ratio [NII]/H$\alpha$; 14 of these have
[NII]/H$\alpha \gtrsim 0.6$ and we flag them as AGN \citep[][and
references therein]{Kauffmann2003,Miller2003}. Broad emission
lines, such as CIII] 190.9 and CIV 154.9, which appear in the
accessible waveband at $z>1$, are used to identify 5 additional
sources as quasars. A search on NED yields that 10 more sources in
our sample are classified as AGN by other authors.

Active galaxies can also be identified using a number of
photometric empirical methods. Quasars occupy a distinct region in
the IRAC color space by virtue of their strong, red continua in
the MIR \citep{Lacy2004}. IRAC fluxes are available for 205
sources and we use the 3 color-color cut prescriptions of
\citet{Hatziminaoglou2005}, \citet{Stern2005} and
\citet{Marsden2009}. Optical magnitudes and postage stamp images
are also available for 114 sources, along with radio fluxes for
107 sources from D09. A source is considered a quasar when it is
compact\footnote{By ``compact'' we mean objects unresolved in the
optical and MIR, with linear sizes $\lesssim 3$ kpc at $z\gtrsim
1$.} and satisfies the 3 aforementioned color-color cut
prescriptions. If only 2 color-color cuts prescriptions are
satisfied, we also require the source to be either radio-loud
($L_{\rm 1.4GHz} \gtrsim 10^{39}$ W), optically bright ($L_{\it
U/g} \gtrsim 10^{11}$\,L$_{\sun}$), or one of the 10 NED AGN.

Using these empirical methods, we find 24 quasars plus 10
additional sources showing weaker yet significant quasar activity,
when the above conditions are near the threshold. The 5 quasars
identified spectroscopically are all contained in this photometric
list. Of the 14 spectroscopically identified AGN, 10 are
definitely not compact, but rather spiral in shape (see next
Section on morphology), and mostly radio-quiet. We believe that
these objects are Seyfert galaxies
\citep[e.g.][]{CidFernandes2010}.

In conclusion, we have assessed that about 15\% of the galaxies in
our sample show strong indication of having an active nucleus and
an additional 6\% have weaker yet significant evidence.
\citet{Chapin2010} found a comparable proportion\footnote{Only
sources with a redshift estimate.} of sources with excess radio
and/or MIR that can be interpreted as an AGN signature. Several
recent observations find close association of AGN activity and
young star formation \citep{Silverman2009}, consistent with a
scenario in which the FIR/submm emission is mainly due to star
formation ongoing in the host galaxy, rather than to emission from
a dusty torus obscuring the inner regions of the active nucleus
\citep{Wiebe2009,Coppin2010,Muzzin2010,Hatziminaoglou2010,Shao2010,Elbaz2010}.
In addition, our AGN selection criteria, which use optical and MIR
data, tend to favor type-1 AGN, i.e. unobscured Seyfert galaxies
and quasars. This is definitely the case for the IRAC color-color
selection methods, as reported by \citet{Hatziminaoglou2005} and
\citet{Stern2005}, but it is also corroborated by the fact that
most of the quasars we have identified are optically bright. We
aim to address this issue in greater detail in a future paper.

\section{Morphology}\label{sec:morph}
We have assigned a broad morphological classification to 137
(60\%) of the BLAST IDs presented in this paper, based upon visual
inspection of UV, optical and MIR postage stamp images (see
Section \ref{subsec:IRAC_optical}) centered at [$\alpha_{\rm
BLAST}$, $\delta_{\rm BLAST}$]. A selection of cut-outs is shown
in Figure~\ref{fig:postage_stamps}.

In addition to the visual examination of the multiwavelength
images, we corroborated our choice with ancillary information
(when available), such as: (a) location on the color-magnitude
diagram, typically ($U-r$) vs $M_r$; (b) spectral features; (c) UV
detection; (d) FIR luminosity. Our findings are listed in the
``morphology'' column in Table~\ref{tab:catalog1} and summarized
in Table~\ref{tab:morphology}.

\begin{deluxetable}{llcr}
\tablewidth{0pt} \small \tablecaption{Broad morphological
classification of BLAST IDs\label{tab:morphology}}\tablehead{
\colhead{type} & \colhead{sub-type} & \colhead{number} &
\colhead{frequency}} \startdata
spiral              & & 69 & 50\% \\
 & Seyfert            &  8 &  6\% \\
compact             & & 52 & 38\% \\
 & quasar             & 31 & 23\% \\
 & blue compact       &  5 &  4\% \\
 & red compact        &  3 &  2\% \\
elliptical          & &  8 &  6\% \\
interacting system  & &  7 &  5\% \\
irregular           & &  1 &  $<$1\%\\
\enddata
\tablecomments{Morphological classification available for 137 out
of 227 BLAST IDs (60\%), based upon visual inspection of UV,
optical and mid-IR (MIR) postage stamp images (see Section
\ref{subsec:IRAC_optical}). By ``compact'' here we mean objects
unresolved in the optical and MIR, with linear sizes $\lesssim 3$
kpc. By ``interacting system'' we mean a visually obvious physical
association of two or more objects.}
\end{deluxetable}

At low redshift we find predominantly spirals, whereas most of the
BLAST sources identified at high redshift are compact and show AGN
signatures. This is probably a selection bias, as the fraction of
submm sources identified at other wavelengths is known to
gradually decreases with $z$ (see D09), and the most distant
sources are often identified only thanks to their extreme radio
and/or optical emission, due to the AGN. In fact, the study by
\citet{Dunlop2010} shows that a deep survey at 250\,$\mu$m not
only reveals low-$z$ spirals, but also extreme dust-enshrouded
starburst galaxies at $z\sim 2$. The latter tend to be missed in
our selection, because they are typically extremely faint in the
optical/UV, unless they also host an AGN.

We point out here that this broad morphological scheme should not
be regarded as meaningful on a source-by-source basis, but rather
be considered as guidance for interpreting the other results of
this paper. For this purpose, we encoded the morphological
information in Figures~\ref{fig:LFIR_vs_HaREW},
\ref{fig:total_SFR}, \ref{fig:Mass_vs_redshift},
\ref{fig:Mass_vs_L_FIR} and \ref{fig:SSFRs_vs_mass}.

\section{Stellar masses}\label{sec:masses}
Stellar masses ($M_{\star}$) are computed by \citet{Dye2010a} for
a subset of 92 sources in our sample with counterparts in a
minimum of 5 bands, from the optical to NIR. The distribution has
median of $10^{10.9}$\,M$_{\sun}$, and inter-quartile range of
$10^{10.6}$--$10^{11.2}$\,M$_{\sun}$.

\begin{figure}[htbp]
\epsscale{1.1} \plotone{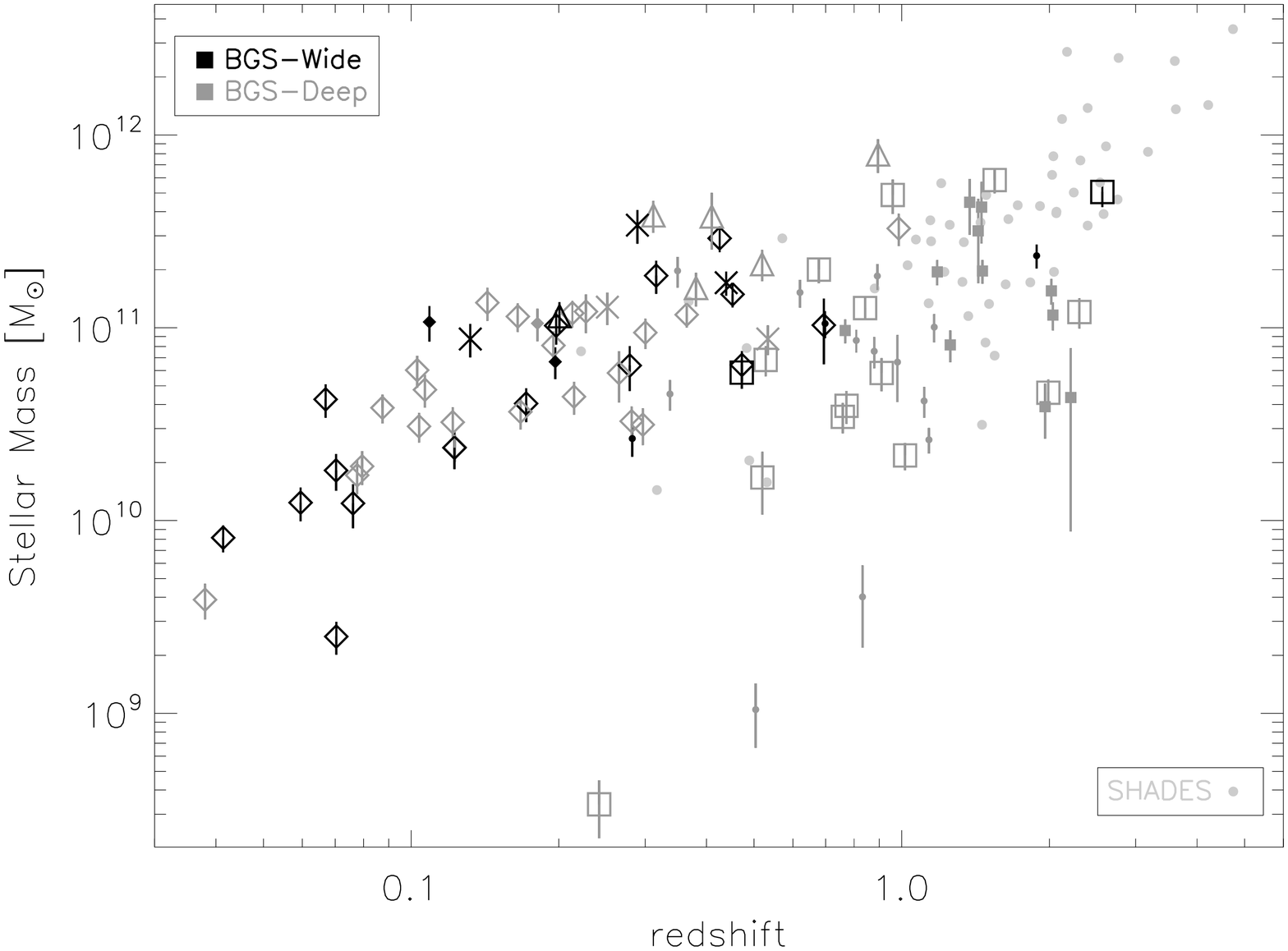} \caption{Stellar
mass as a function of redshift for the whole subset of 92 sources
described in Section \ref{sec:masses}. Symbols are as in
Figure~\ref{fig:LFIR_vs_HaREW}. Filled squares indicate that the
source is a quasar. We overplot SHADES sources \citep{Dye2008a} as
light gray filled circles.}\label{fig:Mass_vs_redshift}
\end{figure}

These stellar masses are plotted vs. redshift in
Figure~\ref{fig:Mass_vs_redshift}; we also show for comparison the
stellar masses of SCUBA sources in SHADES, computed by
\citet{Dye2008a} using a methodology and photometry almost
identical to ours.

Except for 3 outliers (that may well be misidentifications as they
all lie in BGS-Deep), the monotonic trend of increasing stellar
masses is the result of multiple selection effects; sources at a
given redshift are not detected with arbitrarily low, or
arbitrarily high stellar masses. As we discuss later in this
section, there is an approximately constant relation between
$L_{\rm FIR}$ and stellar masses in our sample. Low-luminosity
sources (with low stellar masses) are excluded at a given redshift
because of sensitivity. On the other hand, sources with $L_{\rm
FIR}$ (and stellar masses) above a certain threshold are excluded
from our sample despite the well-documented strongly evolving FIR
luminosity function \citep[E09;][]{Dye2010b,Eales2010}; our
present study simply does not go deep enough to start detecting
the bulk of high-$z$ (and higher volume density) $k$-corrected
sources. In particular, sources with $M_{\star}\gtrsim
10^{12}$\,M$_{\sun}$, which are present in the SHADES sample, are
absent from ours. Indeed, these very massive sources are not
detected among 24\,$\mu$m-selected samples, down to a flux density
level of $\sim$20\,$\mu$Jy \citep[GOODS survey, see
e.g.][]{LeFloch2005,Perez-Gonzalez2005,Caputi2006,Elbaz2007,Santini2009}\footnote{All
the authors cited above adopt a Salpeter
\citeyearpar{Salpeter1955} IMF. \citet{Caputi2006},
\citet{Santini2009}, \citet{Dye2008a,Dye2010a} estimate the
stellar masses by means of an optical--to--NIR SED fit of each
galaxy at the determined redshift. \citet{LeFloch2005} and
\citet{Perez-Gonzalez2005} simply convert, respectively, $V$- and
$K$-band luminosities into stellar masses. Finally,
\citet{Elbaz2007} compute stellar masses by modeling the stellar
populations of each galaxy using stellar absorption-line
indices.}. The 24\,$\mu$m catalog used by D09 to find counterparts
to the BLAST sources goes down to the same depth, therefore we are
only left with the radio catalogs. It is indeed possible that our
analysis is missing very massive galaxies that, though having a
radio ID, do not have an estimate of stellar mass because
measurements are not available in a minimum of 5 optical/NIR
bands. An accurate account of the selection effects at work for
$M_{\star}\gtrsim 10^{12}$\,M$_{\sun}$, which is beyond the scope
of this work, would not invalidate the results of the rest of this
paper.

Our subsample is composed of relatively massive objects, with a
significant fraction of sources (45\%) with stellar masses greater
than $10^{11}$\,M$_{\sun}$. This fraction soars to 84\% in the
SHADES survey, whereas the majority of sources detected at
24\,$\mu$m in deep surveys of the CDFS (down to a flux density
level of $\sim$20\,$\mu$Jy) have
$M_{\star}\leq10^{11}$\,M$_{\sun}$
\citep[e.g.][]{LeFloch2005,Perez-Gonzalez2005,Caputi2006,Elbaz2007,Santini2009}.
However, a direct comparison of the detection rates of massive
galaxies among these surveys is very difficult because of the
dissimilar comoving volumes probed; in fact, BLAST samples a
volume roughly 14 (57) times larger than SHADES
(GOODS)\footnote{Based on the following redshift depth and sky
area covered by, respectively, the GOODS survey, the SHADES survey
and the present BLAST study: $\sim140$\,arcmin$^2$ out to
$z\sim3$; $\sim320$\,arcmin$^2$ out to $z\sim5$; and
$\sim4.15$\,deg$^2$ out to $z\sim2$.}. Furthermore, it would be
necessary to quantify the numerous selection effects and the
different shape of the stellar mass function at the wavelengths in
question.

Nevertheless, BLAST observes a significant number of large,
massive and actively star-forming galaxies (typically spirals, see
Section \ref{sec:morph}), which qualitatively appear to link the
24\,$\mu$m and SCUBA populations at $0<z<2$. With the deep
24\,$\mu$m GOODS survey, other authors seem to be already
detecting this linking population (in particular
\citealt{Caputi2006} and \citealt{Elbaz2007}), but their most
massive sources at $0<z<1$ all have long ($\geq4$\,Gyr)
star-formation time scales (defined as the ratio of already
assembled stellar mass over the recent SFR, see later in this
section), indicating prolonged star formation histories. On the
contrary, about 60\% of our galaxies in the same $M_{\star}$--$z$
range have star-formation time scales shorter than 4\,Gyr,
consistent with the findings that submm-selected $M_{\star}\gtrsim
10^{11}$\,M$_{\sun}$ systems at $z\geq0.5$ form their stellar mass
predominantly at late and at early times, but less so when the
galaxies are middle-aged \citep{Dye2010a,Dye2008a}. These figures
indicate that the moderately massive population detected at
$0<z<1$ by BLAST is more actively forming stars than the equally
massive 24\,$\mu$m-selected galaxies in the same redshift range.
One might wonder whether this observation arises just as a
consequence of a selection effect in the shallower BLAST sample;
although our data do not allow us to investigate the stellar
masses of fainter BLAST galaxies, a thorough examination of the
$M_{\star}$ distribution at $0<z<1$ in the GOODS survey
\citep[e.g. Fig. 7 of][]{Caputi2006} does not suggest that the
exclusion of the fainter 24\,$\mu$m sources (below e.g.
83\,$\mu$Jy, the 80\% completeness limit in the CDFS) would
dramatically alter the proportions of galaxies with stellar mass
above and below $10^{11}$\,M$_{\sun}$. It is certainly possible
that a cut at a brighter 24\,$\mu$m flux density would bias high
the detection rate of massive galaxies; however, the massive BLAST
galaxies at $z\leq1$ have a median SFR of
$\sim70$\,M$_{\sun}$\,yr$^{-1}$ that equals the maximum SFR among
the likewise massive and aged galaxies in GOODS. This would still
be true if the 24\,$\mu$m sample were shallower.

\begin{figure}[htbp]
\epsscale{1.1} \plotone{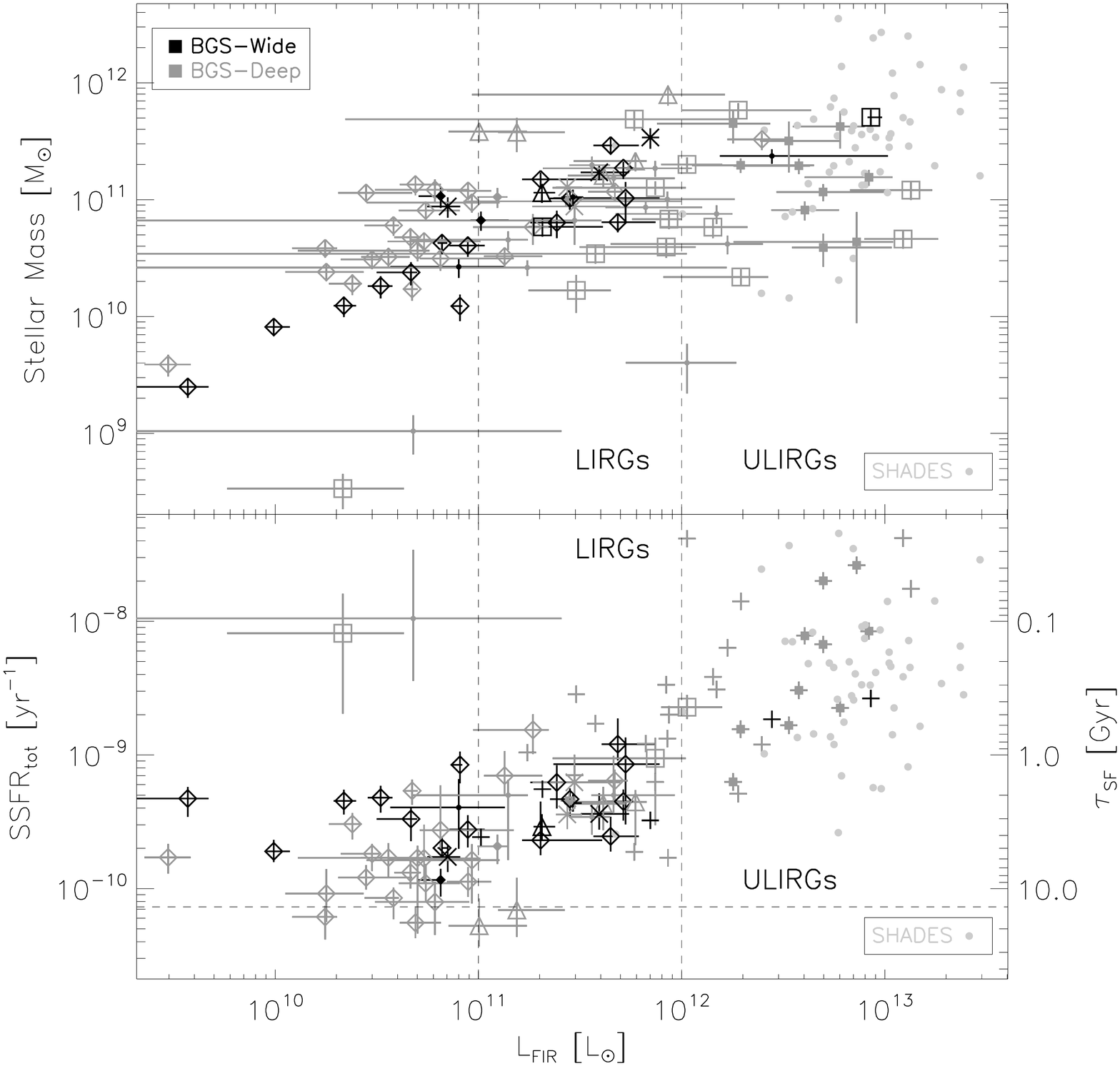} \caption{Top
panel: stellar mass as a function of FIR luminosity for the whole
subset of 92 sources described in Section \ref{sec:masses}. Bottom
panel: specific total SFR (SSFR$_{\rm tot}$) as a function of FIR
luminosity for the subset of 55 sources at $z\leq 0.9$ that have
an estimate of SFR$_{\rm tot}$. Symbols are as in
Figure~\ref{fig:LFIR_vs_HaREW}. For the remaining 37 sources, we
assume SFR$_{\rm tot}=(1-\eta)$\,SFR$_{\rm dust}$ as they all have
$L_{\rm FIR} \gtrsim 10^{11}$\,L$_{\sun}$; these are shown as
crosses without error bars. The right-hand ordinate shows the
corresponding star-formation time scales, defined as $\tau_{\rm
SF}={\rm SSFR}^{-1}$. Filled squares indicate that the source is a
quasar. The horizontal dashed line shows the inverse of the age of
the Universe. We overplot in both panels SHADES sources
\citep{Dye2008a} as light gray filled
circles.}\label{fig:Mass_vs_L_FIR}
\end{figure}

\begin{figure}[htbp]
\epsscale{1.1} \plotone{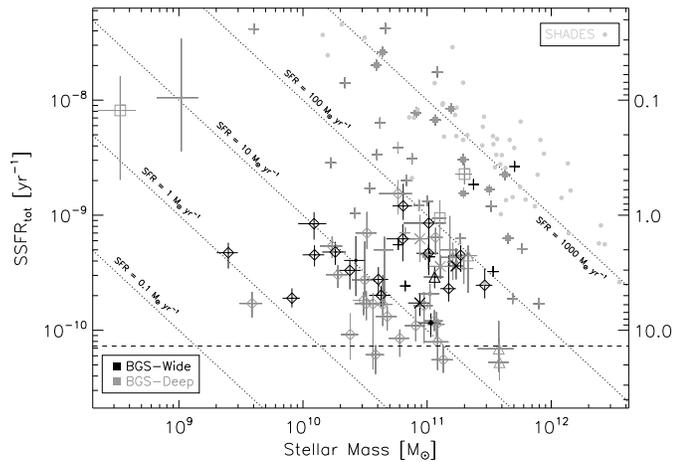} \caption{Specific total
SFR (SSFR$_{\rm tot}$) as a function of stellar mass for the
subset of 55 sources at $z\leq 0.9$ that have an estimate of
SFR$_{\rm tot}$. Symbols are as in Figure~\ref{fig:LFIR_vs_HaREW}.
For the remaining 37 sources, we assume SFR$_{\rm
tot}=(1-\eta)$\,SFR$_{\rm dust}$ as they all have $L_{\rm FIR}
\gtrsim 10^{11}$\,L$_{\sun}$; these are shown as crosses without
error bars. The right-hand ordinate shows the corresponding
star-formation time scales, defined as $\tau_{\rm SF}={\rm
SSFR}^{-1}$. Dotted isolines correspond to constant SFRs, under
the assumption that $M_{\star}$ is the galaxy's total stellar
mass. The horizontal dashed line shows the inverse of the age of
the Universe. We overplot SHADES sources \citep{Dye2008a} as light
gray filled circles.}\label{fig:SSFRs_vs_mass}
\end{figure}

Moreover, Figure~\ref{fig:Mass_vs_redshift} exhibits, in the range
$1< z < 2$, a substantial overlap between BLAST and SCUBA sources.
Therefore, assuming that the BGS is a representative field, our
data suggest that the BLAST galaxies seem to connect the
24\,$\mu$m and SCUBA populations, in terms of both stellar mass
and star-formation activity. Figures~\ref{fig:Mass_vs_L_FIR} and
\ref{fig:SSFRs_vs_mass} further corroborate this conclusion. It is
worth reminding the reader that the $M_{\star}$ estimates are
based on the optical/NIR fluxes of BLAST IDs and do not employ any
BLAST-specific photometric data.

Figure~\ref{fig:Mass_vs_L_FIR} plots stellar masses (top panel)
and SSFR$_{\rm tot}$ (bottom panel) vs. $L_{\rm FIR}$ for the
subset of 55 sources at $z\leq 0.9$ that have an estimate of both
these quantities. There are 37 additional sources in our catalog
with $L_{\rm FIR} \gtrsim 10^{11}$\,L$_{\sun}$ and stellar mass
estimates, but no reliable SFR$_{\rm NUV}$. These are included in
Figure~\ref{fig:Mass_vs_L_FIR}, because in this case SFR$_{\rm
tot}\simeq (1-\eta)$\,SFR$_{\rm dust}$ (see Section
\ref{sec:tot_SFR}). SHADES sources are also shown in this figure.
\citet{Dye2010c} estimates their FIR luminosities using a
two-component SED fit from \citet{Dunne2001} that has cold/hot
ratio of 186, with $T_{\rm hot}=44$\,K and $T_{\rm cold}=20$\,K.
SFRs are estimated using Equation~\ref{eq:SFR_FIR} and corrected
by ($1-\eta$). Finally, star-formation time scales, defined as
$\tau_{\rm SF}={\rm SSFR}^{-1}$, are shown as secondary $y$-axis.

BLAST IDs selected in BGS-Wide show a positive correlation between
their stellar masses and $L_{\rm FIR}$, but there is no strong
evidence for a correlation between SSFR$_{\rm tot}$ and FIR
luminosities. Although BLAST IDs selected in BGS-Deep appear to
have different trends, one should be cautious as the they are, in
general, less reliable than the IDs in BGS-Wide. However, BGS-Deep
sources can be used to study bulk properties under appropriate
caveats. The emerging picture appears to confirm
Figure~\ref{fig:Mass_vs_redshift}, in which there is a
non-negligible overlap between the BLAST and SCUBA populations in
the range $1< z < 2$. In particular, the high luminosity tail of
the BLAST sample appears to encroach on the SHADES sources in
terms of both $L_{\rm FIR}$ and $M_{\star}$, bridging the gap with
the lower-redshift Universe populated by 24\,$\mu$m sources and by
run-of-the-mill star-forming BLAST galaxies, with $\tau_{\rm SF}$
spanning the interval 1--10\,Gyr. A considerable overlap between
fainter BLAST sources and 870\,$\mu$m-selected galaxies has
already been established by \citet{Dunlop2010} and
\citet{Chapin2010}, but it is important to have confirmed an
additional, less direct, connection with our shallower BLAST
sample, by means of a comparable analysis to that of SHADES.

We have investigated if a temporal connection between the two
populations is allowed by the data, in a scenario where the BLAST
sources are SCUBA sources fading at the end of their late
star-formation burst \citep{Borys2005,Dye2008a}. However,
\citet{Dye2010a} seem to rule out this possibility, because the
higher-$z$, more massive BLAST IDs are observed during a
star-formation burst lasting too briefly in redshift to allow this
connection. This disconnection is consistent with the phenomenon
of downsizing observed in optically-selected samples of galaxies
\citep[e.g.][]{Heavens2004}.

The approximately flat trend between SSFR$_{\rm tot}$ with FIR
luminosity of Figure~\ref{fig:Mass_vs_L_FIR} evidenced by the
BLAST IDs selected in BGS-Wide is consistent with
\citet{Serjeant2008}. The inclusion of BGS-Deep sources at high
FIR luminosities seems to suggest a different, mild trend of
increasing SSFR$_{\rm tot}$, also reported by \citet{Santini2009}
and \citet{Rodighiero2010}. The data available to us do not
manifest enough evidence to support either scenario. Larger
samples now accessible with {\sl Herschel} will shed more light on
the evolution of the specific star-formation rate.

In Figure~\ref{fig:SSFRs_vs_mass}, we plot SSFR$_{\rm tot}$ vs.
stellar mass, for BLAST and SHADES sources. The dotted isolines
correspond to constant SFRs, under the assumption that $M_{\star}$
is the galaxy's total stellar mass. We do not find any clear
correlation between specific total star-formation rate and stellar
mass, which is not surprising as we are sampling a population of
young, active, star-forming galaxies \citep[see
also][]{Santini2009}. Expectedly, the bulk of SHADES sources
occupies a well-defined region of the plane, around the isoline of
SFR $=1000$\,M$_{\sun}$\,yr$^{-1}$, whereas practically all the
BLAST counterparts at $z\leq0.9$ lie below the isoline of SFR
$=100$\,M$_{\sun}$\,yr$^{-1}$. The gap is again filled by the
BLAST IDs at higher redshift.

We can compare our results in Figure~\ref{fig:SSFRs_vs_mass} with
\citet{Buat2008}, who derived mean relationships between observed
SSFR and stellar mass at $z=0$ and $z=0.7$, and confronted these
with models based on a progressive infall of gas into the galactic
disk, starting at high $z$. Both their data and models exhibit a
flat distribution of SSFR for galaxies with masses between
$10^{10}$ and $10^{11}$\,M$_{\sun}$. Our $z\leq0.9$ subset of star
forming galaxies shares a similar behavior, as well as the dynamic
ranges. On the other hand, we can also compare the high-$z$ tail
of the BLAST IDs with the $z>0.85$ sample of
\citet{Rodighiero2010}: although the scatter is quite large in
both subsets, we observe the same negative trend of SSFR with
$M_{\star}$, again consistent with downsizing.

The in-depth analysis of the bright BLAST counterparts reveals a
population with an intrinsic dichotomy in terms of star-formation
rate, stellar mass and morphology. The bulk of BLAST counterparts
at $z\lesssim1$ appear to be run-of-the-mill star-forming spiral
galaxies, with intermediate stellar masses (median
$M_{\star}\sim7\times10^{10}$\,M$_{\sun}$) and approximately
constant specific star-formation rates ($\tau_{\rm SF}$ in the
range 1--10\,Gyr); in addition, they form stars more actively than
the equally massive and aged 24\,$\mu$m sources. On the other
hand, the high-$z$ BLAST counterparts significantly overlap with
the SCUBA population, and the observed trends of SSFR, albeit
inconclusive, suggest stronger evolution and downsizing. In
conclusion, our study suggests that the BLAST galaxies may act as
linking population between the star-forming 24\,$\mu$m sources and
the more extreme SCUBA starbursts.

\section{Concluding remarks}\label{sec:summary}
We have carried out a panchromatic study of individual bright
BLAST galaxies identified at other wavelengths, extending the
analysis of previous BLAST papers. Our basic results are as
follows.

\begin{enumerate}\addtolength{\itemsep}{-0.5\baselineskip}
    \item The flux densities of BLAST sources are boosted due to a combination of Eddington bias, source confusion and
    blending. We have developed a Monte Carlo method to quantify
    these biases, both in confusion-limited maps and in maps dominated by instrumental
    noise. The boosting effects are more pronounced in the
    confusion-limited regime, and become more important as the
    wavelength increases. In addition, flux densities are heavily
    correlated among the BLAST bands, again more prominently in
    BGS-Deep. We account for all these effects coherently while
    calculating the FIR luminosities of BLAST galaxies. We have
    also shown how crucial the BLAST/SPIRE photometry is to estimate without bias the FIR luminosity of a galaxy, especially at high redshift.
    \item We have measured that star formation is predominantly obscured at $L_{\rm FIR} \gtrsim
10^{11}$\,L$_{\sun}$, $z \gtrsim 0.5$. On the other hand,
unobscured star formation is important at $L_{\rm FIR} \lesssim
10^{11}$\,L$_{\sun}$, $z \lesssim 0.25$ and FIR-only evaluations
of SFR would lead to underestimates up to a factor of 2. This is
probably a direct consequence of the well documented stronger
evolution of the FIR luminosity density with respect to the
optical-UV one.
    \item We have compared, in terms of $L_{\rm FIR}$--$z$ parameter space, the BLAST counterparts to the {\sl IRAS}/FIR-selected sample
    of local galaxies, to the 24\,$\mu$m-selected sample observed
    by {\sl Spitzer}, and to the SCUBA 850\,$\mu$m-selected
    sample. The overlap with the local {\sl IRAS} sample is minimal and
    this conclusion should not be belittled by the extent of local volume surveyed by
    BLAST. Similarly, our sample lacks the abundance of most luminous IR
galaxies detected in the SHADES survey, but the high-$L_{\rm
FIR}$, high-$z$ tail of the BLAST counterparts seems to overlap
with the SCUBA population. The 24\,$\mu$m-selected sample
resembles the most the bulk of BLAST IDs in terms of $L_{\rm FIR}$
and redshift distribution.
    \item We have assessed that 15\% of the galaxies in our sample show strong indication of an active nucleus and an
additional 6\% have weaker yet significant evidence. In
particular, these are predominantly type-1 AGN, i.e. unobscured
Seyfert galaxies and quasars. The AGN fraction and the SFRs
inferred for these objects are comparable to recent observations
at similar wavelengths and point to a scenario in which the
submillimeter emission detected by BLAST is mainly due to star
formation ongoing in the host galaxy, rather than to emission from
a dusty torus obscuring the inner regions of the active nucleus.

\item We have computed stellar masses for a subset of 92 BLAST
counterparts. These appear to be relatively massive objects, with
median mass of $10^{10.9}$\,M$_{\sun}$, and inter-quartile range
of $10^{10.6}$--$10^{11.2}$\,M$_{\sun}$. In particular, a
significant fraction of them fill the region of
$M_{\star}\sim10^{11}$\,M$_{\sun}$ at $z\lesssim 1$ that is
practically vacant in the SCUBA surveys, and sparsely populated by
24\,$\mu$m-selected samples. Although the dissimilar volumes
sampled by these surveys discourage a direct comparison of the
detection rates of massive galaxies, our study suggest that the
BLAST counterparts seem to link the 24\,$\mu$m and SCUBA
populations, in terms of both stellar mass and star-formation
activity.

\item We have highlighted a dichotomy in the BLAST population in terms of star-formation rate,
stellar mass and morphology. The bulk of BLAST counterparts at
$z\lesssim1$ are run-of-the-mill star-forming galaxies, typically
spiral in shape, with intermediate stellar masses and nearly
constant specific star-formation rates. On the other hand, the
higher redshift BLAST counterparts significantly overlap with the
SCUBA population, and the observed trends of SSFR, albeit
inconclusive, suggest stronger evolution. Other BLAST studies have
already described the significant overlap existing between fainter
BLAST sources and 870\,$\mu$m-selected galaxies, but here we have
established an additional link with a shallower BLAST sample, via
an analysis equivalent to that of SHADES.

\item We rule out a temporal connection between the BLAST and SCUBA
populations, in a scenario where BLAST sources would correspond to
SCUBA galaxies whose burst of star formation is ceasing. This
disconnection is consistent with the downsizing observed in
optical samples.
\end{enumerate}

The findings described in this paper represent a taste of what
should be possible with a significantly larger sample of sources.
The increased sensitivity and resolution of the {\sl Herschel}
Space Observatory, which recently started operation, will soon
provide vastly increased numbers of sources. This will enable
significantly reduced uncertainties and therefore much improved
constraints on models of galaxy evolution and formation.
Nevertheless, the BLAST data have provided a very valuable
benchmark for the {\sl Herschel} data and the various analyses
that will emerge for some time to come. Furthermore, the results
in this paper probably will not immediately become obsolete, as
even the much more sensitive SPIRE surveys will have to face the
lack of deeper ancillary data, especially in the optical/NIR and
in the radio. Identifying the precise location of the submm
sources will require either deep and very wide-area VLA data, or a
combination of MIPS 24\,$\mu$m and PACS, or ultimately ALMA.
Finally, in order to study the rest-frame optical/NIR of the $z>2$
submm galaxies in much more detail than BLAST or SCUBA, future
studies will really require instruments like {\sl WFC3} or {\sl
JWST}.

\acknowledgments{We acknowledge the support of NASA through grant
numbers NAG5-12785, NAG5-13301, and NNGO-6GI11G, the NSF Office of
Polar Programs, the Canadian Space Agency, the Natural Sciences
and Engineering Research Council (NSERC) of Canada, and the UK
Science and Technology Facilities Council (STFC). This paper
relies on observations made with the AAOmega spectrograph on the
Anglo-Australian Telescope, and we thank the staff of the
telescope and especially those involved in the development of the
spectrograph. We are grateful to Heath Jones for his help with the
observations and Rob Sharp for his help with the {\it 2dfdr}
data-reduction pipeline. This work makes use of the {\it Runz}
redshift-fitting code developed by Will Sutherland, Will Saunders,
Russell Cannon and Scott Croom, and we are grateful to Scott Croom
for making this available to us. This research also made use of
the NASA/IPAC Extragalactic Database (NED), operated by the Jet
Propulsion Laboratory, under contract with NASA. Finally, this
work is based in part on observations made with the {\sl Spitzer}
Space Telescope and the Galaxy Evolution Explorer {\sl GALEX},
which are operated by the Jet Propulsion Laboratory, California
Institute of Technology under a contract with NASA.

We acknowledge Stefanie Walch, Giorgio Savini, Locke Spencer and
Karina Caputi for helpful discussions and comments. We thank David
Shupe and Jason Surace for providing the depths of the SWIRE MIPS
and IRAC maps of the CDFS. Finally, we thank the anonymous referee
for his/her insightful comments and suggestions.}

{\it Facilities:} \facility{BLAST}, \facility{AAT (AAOmega)},
\facility{{\sl GALEX}}, \facility{{\sl Spitzer} (MIPS, IRAC)}

\appendix

\section{Postage stamps}
Postage stamp images for a selection of low redshift BLAST IDs.
The images are all $2\arcmin\times 2\arcmin$ in size. Every row
shows a BLAST source, imaged at three different bands:
\textit{left}, {\sl GALEX} NUV filter (centered at 2315\,\AA);
\textit{center}, RGB combination of the $U\,g\,r$ filters from the
SWIRE optical survey; \textit{right}, 3.6\,$\mu$m IRAC band. The
complete set of full-color cut-outs can be found at
http://blastexperiment.info/results\_images/moncelsi/

\begin{figure}
\figurenum{A1} \epsscale{0.95} \plotone{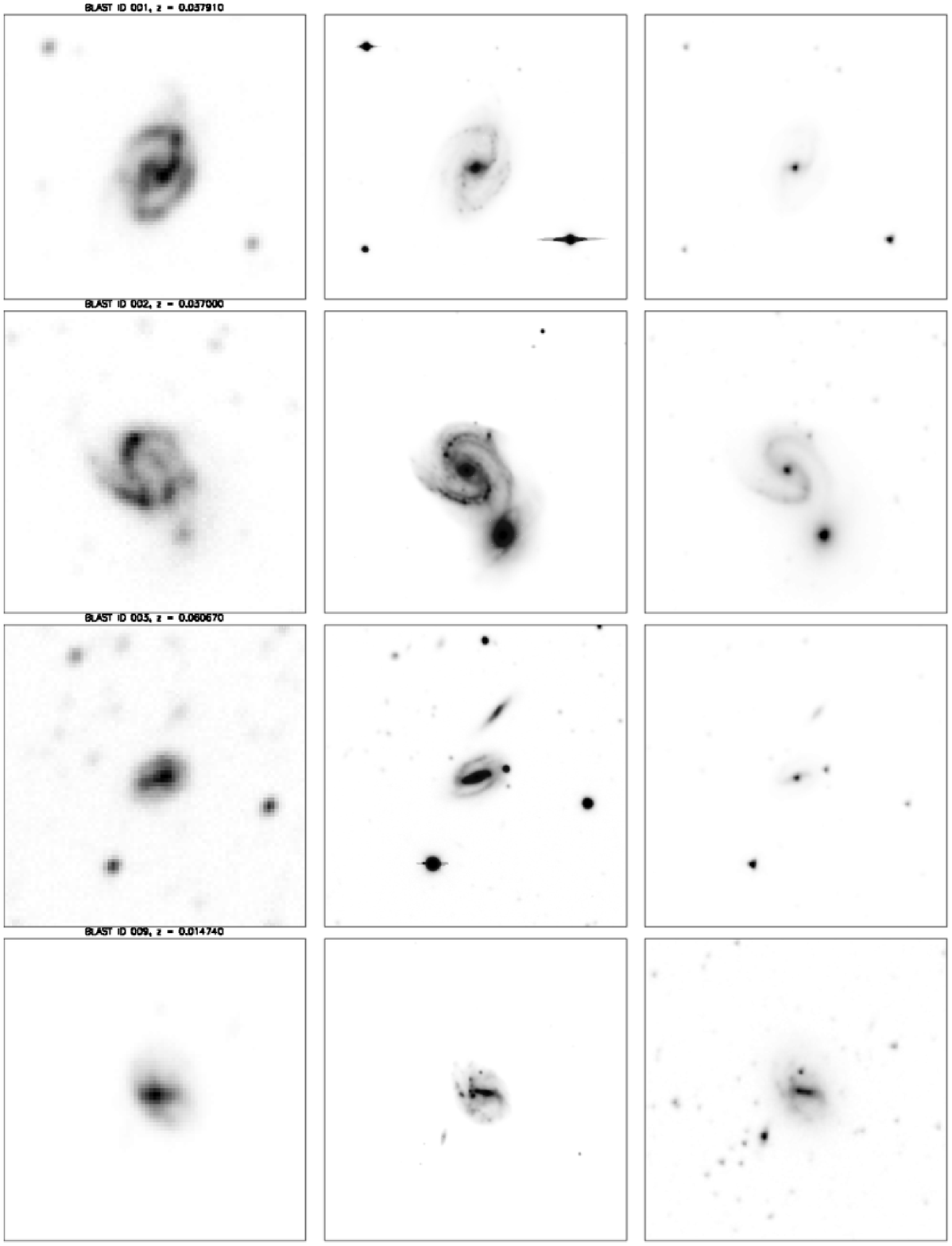}
\label{fig:postage_stamps}\caption{}
\end{figure}

\clearpage
\begin{figure}
\epsscale{0.95} \plotone{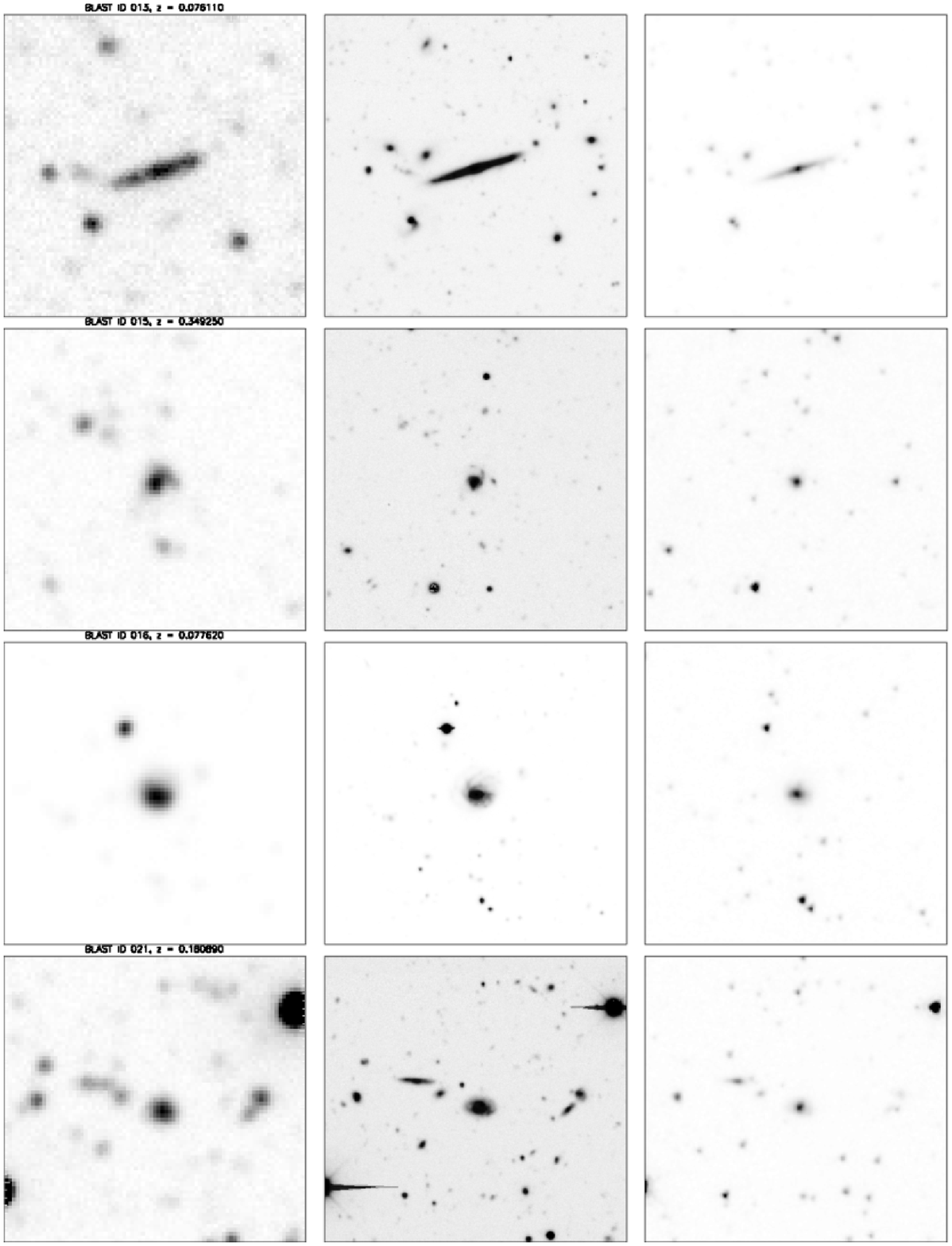}
\end{figure}
\clearpage

\clearpage
\begin{figure}
\epsscale{0.95} \plotone{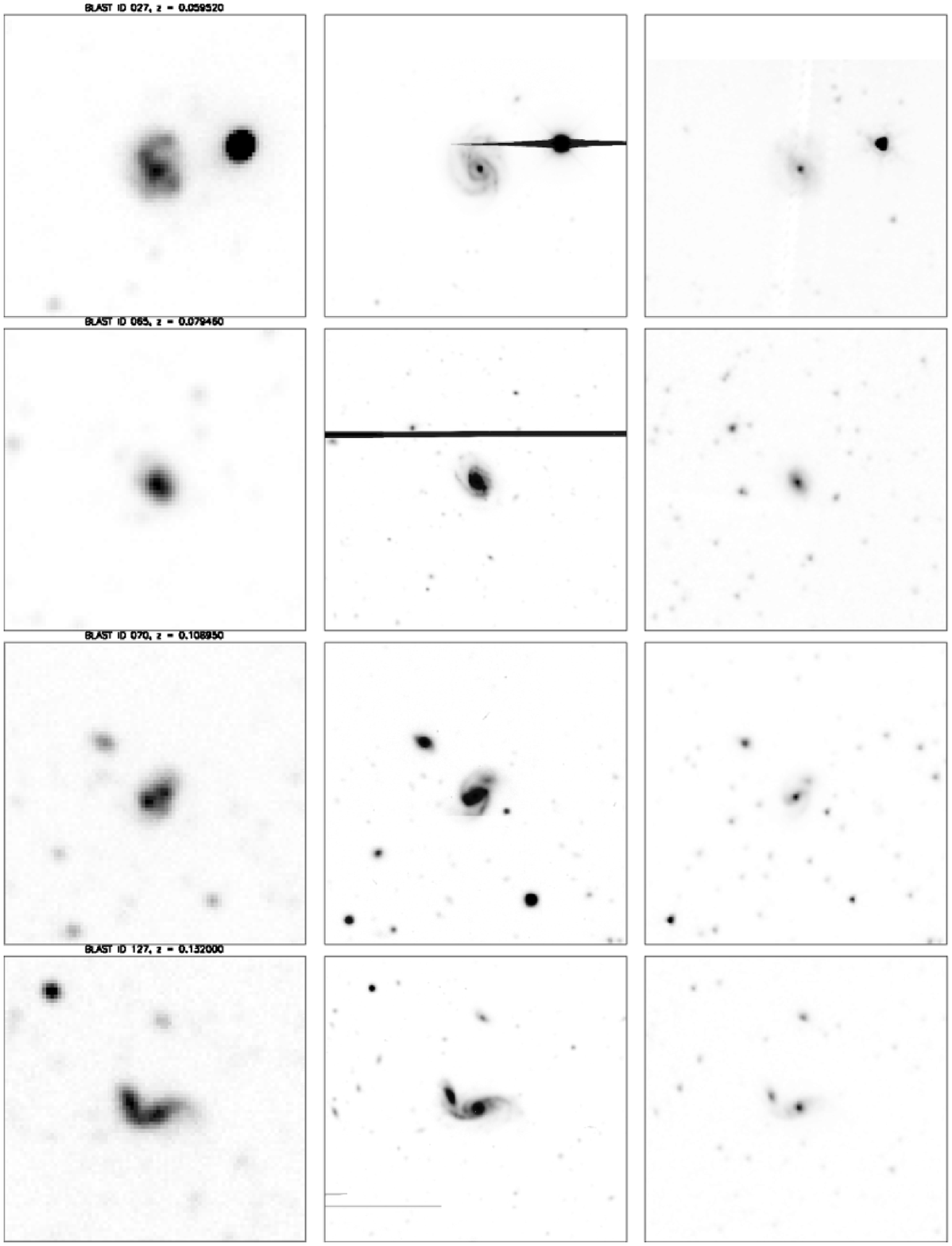}
\end{figure}
\clearpage

\clearpage
\begin{figure}
\epsscale{0.95} \plotone{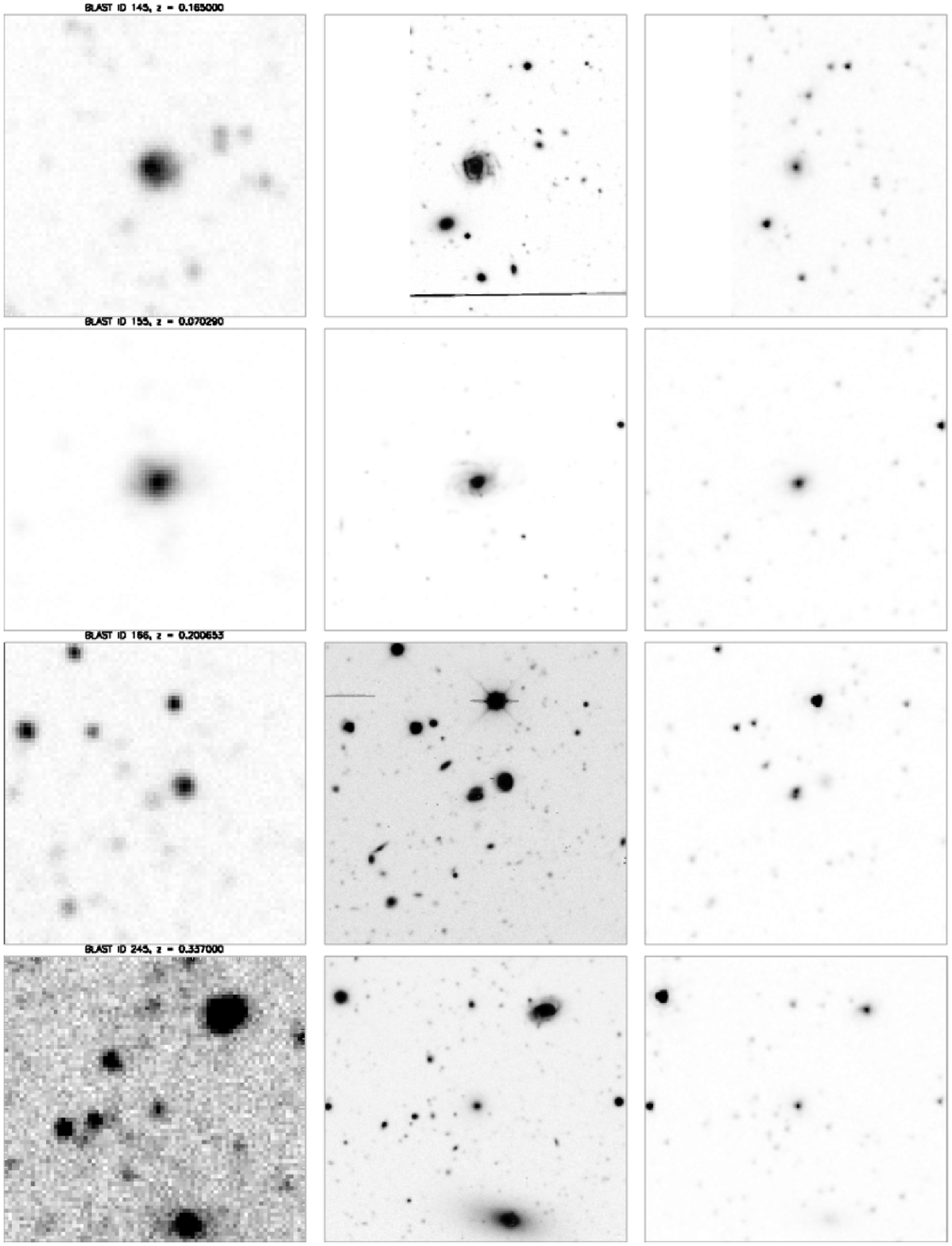}
\end{figure}
\clearpage

\begin{figure}
\epsscale{0.95} \plotone{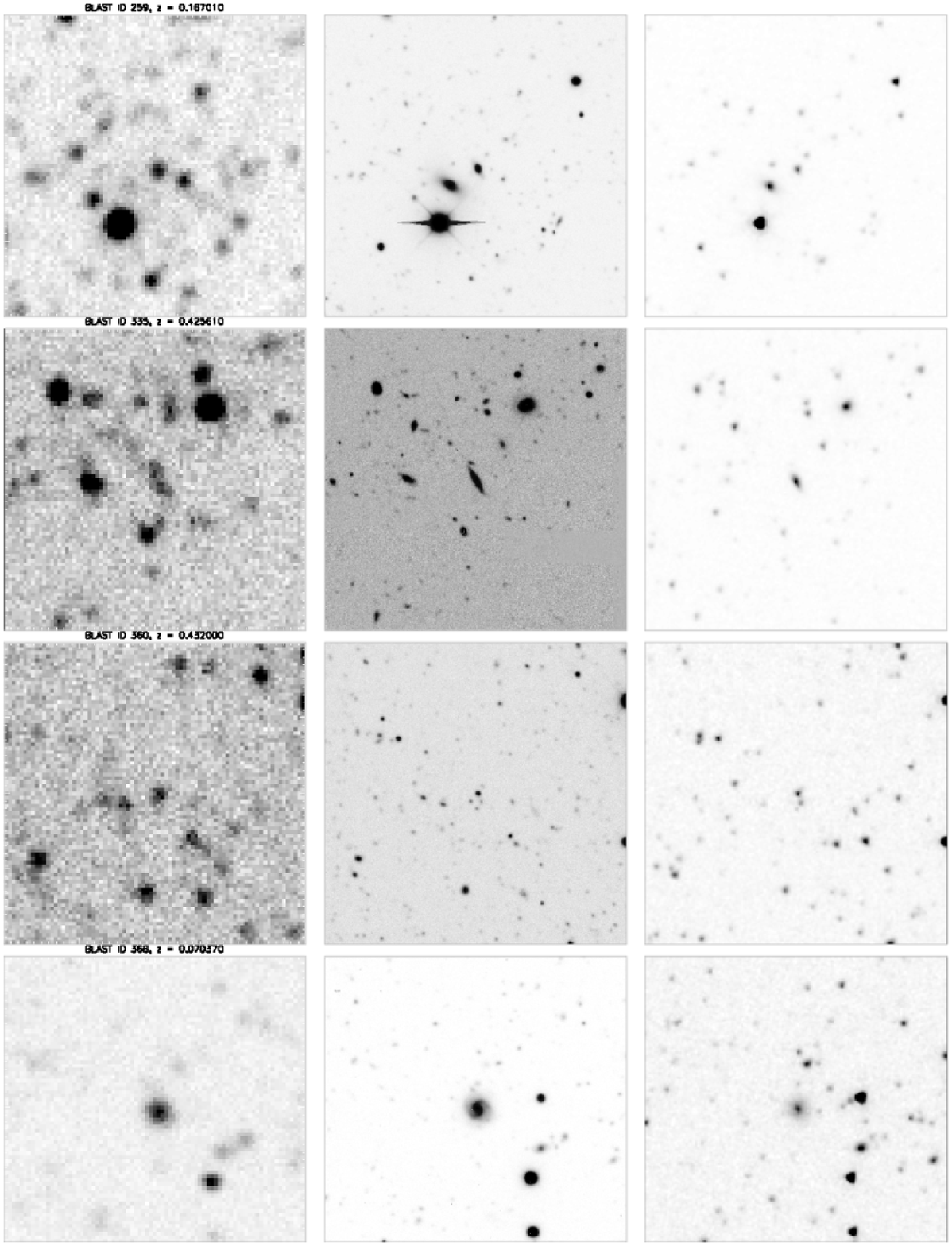}
\end{figure}
\clearpage

\section{Data tables}
\renewcommand{\thefootnote}{\alph{footnote}}
\begin{deluxetable}{rlllclccccccc}
\tablenum{B1} \tabletypesize{\scriptsize}\tablewidth{0pt} \rotate
\tablecolumns{13} \setlength{\tabcolsep}{0.05in}
\tablecaption{Primary counterparts to $\geq5\sigma$ BLAST sources:
redshift, spectral and morphological
information\label{tab:catalog1}} \tablehead{
  \colhead{ID} &
  \colhead{BLAST name} &
  \colhead{$\alpha_{\rm BLAST}$} &
  \colhead{$\delta_{\rm BLAST}$} &
  \colhead{deep} &
  \colhead{$z$} &
  \colhead{flag spec-$z$} &
  \colhead{provenance} &
  \colhead{H$\alpha$ EW$_{\rm rf}$} &
  \colhead{[NII]/H$\alpha$} &
  \colhead{AGN flag} &
  \colhead{Q flag} &
  \colhead{morphology} \\}
\startdata
  1 & BLAST J032921$-$280803 & 52.33792 & $-$28.13348 & 0 & 0.03791 & 1 & AAO & 11.4 $\pm$ 1.0 & 0.64 & AGN? & ... & S\\
  2 & BLAST J032956$-$284631 & 52.48567 & $-$28.77572 & 0 & 0.037 & 1 & AAO & 4.6 $\pm$ 5.2 & 0.56 & ... & ... & IS\\
  3 & BLAST J032741$-$282325 & 51.921225 & $-$28.38895 & 0 & 0.06067 & 1 & AAO & 15.8 $\pm$ 3.0 & 0.74 & AGN & ... & S\\
  4 & BLAST J033235$-$275530 & 53.146165 & $-$27.92571 & 1 & 0.03764 & 1 & AAO & 7.1 $\pm$ 2.0 & 0.60 & ... & ... & S\\
  5 & BLAST J033131$-$272842 & 52.880575 & $-$27.479735 & 1 & 0.06668 & 1 & AAO & 13.4 $\pm$ 0.9 & 1.33 & AGN & ... & S\\
  6 & BLAST J033229$-$274415 & 53.12448 & $-$27.740165 & 1 & 0.07593 & 1 & AAO & 38.4 $\pm$ 1.6 & 0.43 & ... & ... & S\\
  7 & BLAST J033250$-$273420 & 53.20818 & $-$27.57581 & 1 & 0.25126 & 1 & AAO & 31.3 $\pm$ 3.6 & 0.44 & ... & ... & IS?\\
  8 & BLAST J033548$-$274920 & 53.954945 & $-$27.821905 & 0 & 0.16752 & 1 & AAO & 13.1 $\pm$ 3.3 & 1.18 & AGN & ... & ...\\
\enddata
\tablecomments{Reading from the left, the columns are: the BLAST
identification number; the full IAU name of the BLAST source; the
position of the counterpart (the arithmetic mean between the two
sets of coordinates if both the radio and 24\,$\mu$m counterparts
are present); flag indicating whether the source is located within
BGS-Deep; the redshift; flag indicating whether the redshift is
spectroscopic or photometric; the provenance of the redshift (see
Section \ref{subsec:redshifts} for details); the H$\alpha$
rest-frame Equivalent Width (EW$_{\rm rf}$) from AAOmega spectra,
in \AA, with uncertainty; the ratio of the flux in the [NII] 658.3
line to the flux in the H$\alpha$ line, from AAOmega spectra;
column assessing the presence of an AGN in the host galaxy, based
on line ratios ([NII]/H$\alpha >$0.6,
\citealt{Kauffmann2003,Miller2003}), or of a quasar, based solely
on the broadness of the lines (we also indicate with ``NED''
objects flagged as AGN in NED); column assessing whether the
objects is a quasar (Q), based solely on optical and mid-IR (IRAC)
colors (see Section \ref{sec:AGN} for details); morphological
classification: S=spiral, IS = interacting system, E = elliptical,
C = compact, RC = red compact, BC = blue compact (see Section
\ref{sec:morph} for details).}
\tablenotetext{a}{\citet{Colless2003}}
\tablenotetext{b}{\citet{Ratcliffe1998}}
\tablenotetext{c}{\citet{Ravikumar2007}}
\end{deluxetable}

\renewcommand{\thefootnote}{\arabic{footnote}}

\begin{deluxetable}{rlllccccccc}
\tablenum{B2} \tabletypesize{\scriptsize}\rotate \tablewidth{0pt}
\tablecolumns{11} \setlength{\tabcolsep}{0.05in}
\tablecaption{Primary counterparts to $\geq5\sigma$ BLAST sources:
UV and FIR properties\label{tab:catalog2}} \tablehead{
  \colhead{ID} &
  \colhead{GALEX name} &
  \colhead{$\alpha_{\rm GALEX}$} &
  \colhead{$\delta_{\rm GALEX}$} &
  \colhead{FUV} &
  \colhead{NUV} &
  \colhead{E$(B-V)$} &
  \colhead{SFR$_{\rm FUV}$} &
  \colhead{SFR$_{\rm NUV}$} &
  \colhead{$L_{\rm FIR}$} &
  \colhead{$M_{\star}$} \\}
\startdata
1&GALEX J032920.6$-$280800&52.336213&$-$28.133591&17.23 $\pm$ 0.12&16.795 $\pm$0.067&0.0071&2.01 $\pm$ 0.22&3.25 $\pm$ 0.2&4.94$_{-0.16}^{+0.17}$&...\\
2&GALEX J032956.4$-$284633&52.485265&$-$28.775954&17.78 $\pm$ 0.12&17.051 $\pm$0.067&0.01456&1.22 $\pm$ 0.14&2.59 $\pm$ 0.16&2.85$_{-0.10}^{+0.06}$&...\\
3&GALEX J032740.9$-$282320&51.920727&$-$28.389056&19.09 $\pm$ 0.12&18.637 $\pm$0.067&0.00862&0.94 $\pm$ 0.1&1.55 $\pm$ 0.1&10.5$_{-0.4}^{+0.2}$&...\\
4&GALEX J033235.0$-$275532&53.14599&$-$27.925756&20.49 $\pm$ 0.12&20.176 $\pm$ 0.068&0.00864&0.1 $\pm$ 0.01&0.14 $\pm$ 0.01&0.55$_{-0.04}^{+0.10}$&...\\
5&GALEX J033131.3$-$272846&52.880468&$-$27.479551&19.23 $\pm$ 0.12&18.716 $\pm$0.067&0.00995&1.01 $\pm$ 0.11&1.76 $\pm$ 0.11&2.64$_{-0.11}^{+0.44}$&...\\
6&GALEX J033229.8$-$274423&53.124378&$-$27.73994&19.97 $\pm$ 0.12&19.435 $\pm$ 0.067&0.0092&0.66 $\pm$ 0.07&1.17 $\pm$ 0.07&3.21$_{-0.31}^{+0.37}$&...\\
7&GALEX J033249.8$-$273433&53.207886&$-$27.575957&22.08 $\pm$ 0.13&21.065 $\pm$0.068&0.00827&1.16 $\pm$ 0.14&2.89 $\pm$ 0.18&27.4$_{-3.3}^{+4.9}$&12.791 $\pm$2.442\\
8&GALEX J033549.0$-$274919&53.95423&$-$27.822048&21.95 $\pm$ 0.13&20.976 $\pm$ 0.073&0.01343&0.56 $\pm$ 0.07&1.43 $\pm$ 0.1&63.2$_{-4.4}^{+1.2}$&...\\
\enddata
\tablecomments{Reading from the left, the columns are: the BLAST
identification number; the full IAU name of the {\sl GALEX}
counterpart to the BLAST source; the position of the {\sl GALEX}
counterpart; the flux in the FUV filter, in magnitudes, with
uncertainty; the flux in the NUV filter, in magnitudes, with
uncertainty; Galactic extinction correction as from
\citet{Schlegel1998}, in magnitudes; star-formation rate as
estimated from the FUV flux (see Equation~\ref{eq:SFR_NUV}), in
M$_{\sun}$\,yr$^{-1}$, with uncertainty (note that we listed
SFR$_{\rm FUV}$ for all FUV sources, even if only those with $z<
0.36$ are to be considered reliable, see Section
\ref{subsec:SFR_UV}); star-formation rate as estimated from the
NUV flux (see Equation~\ref{eq:SFR_NUV}), in
M$_{\sun}$\,yr$^{-1}$, with uncertainty (note that we listed
SFR$_{\rm NUV}$ for all NUV sources, even if only those with $z<
0.91$ are to be considered reliable, see Section
\ref{subsec:SFR_UV}); rest-frame bolometric FIR luminosity of the
BLAST ID, in $10^{10}$\,L$_{\sun}$, with upper and lower
uncertainties (note that we quote the mode, and 68\% c.l. of the
distribution obtained, see Section \ref{subsec:FIR_lum} for
details); stellar mass of the BLAST ID, in $10^{10}$\,M$_{\sun}$
(see Section \ref{sec:masses} for details), with uncertainty.}
\end{deluxetable}

\bibliography{ms}

\begin{thebibliography}{116}
\expandafter\ifx\csname natexlab\endcsname\relax\def\natexlab#1{#1}\fi

\bibitem[{{Balogh} {et~al.}(2004){Balogh}, {Eke}, {Miller}, {Lewis}, {Bower},
  {Couch}, {Nichol}, {Bland-Hawthorn}, {Baldry}, {Baugh}, {Bridges}, {Cannon},
  {Cole}, {Colless}, {Collins}, {Cross}, {Dalton}, {de Propris}, {Driver},
  {Efstathiou}, {Ellis}, {Frenk}, {Glazebrook}, {Gomez}, {Gray}, {Hawkins},
  {Jackson}, {Lahav}, {Lumsden}, {Maddox}, {Madgwick}, {Norberg}, {Peacock},
  {Percival}, {Peterson}, {Sutherland}, \& {Taylor}}]{Balogh2004}
{Balogh}, M., {Eke}, V., {Miller}, C., {Lewis}, I., {Bower}, R., {Couch}, W.,
  {Nichol}, R., {Bland-Hawthorn}, J., {Baldry}, I.~K., {Baugh}, C., {Bridges},
  T., {Cannon}, R., {Cole}, S., {Colless}, M., {Collins}, C., {Cross}, N.,
  {Dalton}, G., {de Propris}, R., {Driver}, S.~P., {Efstathiou}, G., {Ellis},
  R.~S., {Frenk}, C.~S., {Glazebrook}, K., {Gomez}, P., {Gray}, A., {Hawkins},
  E., {Jackson}, C., {Lahav}, O., {Lumsden}, S., {Maddox}, S., {Madgwick}, D.,
  {Norberg}, P., {Peacock}, J.~A., {Percival}, W., {Peterson}, B.~A.,
  {Sutherland}, W., \& {Taylor}, K. 2004, \mnras, 348, 1355

\bibitem[{{Barger} {et~al.}(1998){Barger}, {Cowie}, {Sanders}, {Fulton},
  {Taniguchi}, {Sato}, {Kawara}, \& {Okuda}}]{Barger1998}
{Barger}, A.~J., {Cowie}, L.~L., {Sanders}, D.~B., {Fulton}, E., {Taniguchi},
  Y., {Sato}, Y., {Kawara}, K., \& {Okuda}, H. 1998, \nat, 394, 248

\bibitem[{{Bell}(2003)}]{Bell2003}
{Bell}, E.~F. 2003, \apj, 586, 794

\bibitem[{{B{\'e}thermin} {et~al.}(2010){B{\'e}thermin}, {Dole}, {Beelen}, \&
  {Aussel}}]{Bethermin2010}
{B{\'e}thermin}, M., {Dole}, H., {Beelen}, A., \& {Aussel}, H. 2010, \aap, 512,
  A78+

\bibitem[{{Blain} {et~al.}(2003){Blain}, {Barnard}, \& {Chapman}}]{Blain2003}
{Blain}, A.~W., {Barnard}, V.~E., \& {Chapman}, S.~C. 2003, \mnras, 338, 733

\bibitem[{{Blain} {et~al.}(1999){Blain}, {Kneib}, {Ivison}, \&
  {Smail}}]{Blain1999}
{Blain}, A.~W., {Kneib}, J., {Ivison}, R.~J., \& {Smail}, I. 1999, \apjl, 512,
  L87

\bibitem[{{Borys} {et~al.}(2005){Borys}, {Smail}, {Chapman}, {Blain},
  {Alexander}, \& {Ivison}}]{Borys2005}
{Borys}, C., {Smail}, I., {Chapman}, S.~C., {Blain}, A.~W., {Alexander}, D.~M.,
  \& {Ivison}, R.~J. 2005, \apj, 635, 853

\bibitem[{{Buat} {et~al.}(2008){Buat}, {Boissier}, {Burgarella}, {Takeuchi},
  {Le Floc'h}, {Marcillac}, {Huang}, {Nagashima}, \& {Enoki}}]{Buat2008}
{Buat}, V., {Boissier}, S., {Burgarella}, D., {Takeuchi}, T.~T., {Le Floc'h},
  E., {Marcillac}, D., {Huang}, J., {Nagashima}, M., \& {Enoki}, M. 2008, \aap,
  483, 107

\bibitem[{{Buat} {et~al.}(2010){Buat}, {Giovannoli}, {Burgarella}, {Altieri},
  {Amblard}, {Arumugam}, {Aussel}, {Babbedge}, {Blain}, {Bock}, {Boselli},
  {Castro-Rodriguez}, {Cava}, {Chanial}, {Clements}, {Conley}, {Conversi},
  {Cooray}, {Dowell}, {Dwek}, {Eales}, {Elbaz}, {Fox}, {Franceschini}, {Gear},
  {Glenn}, {Griffin}, {Halpern}, {Hatziminaoglou}, {Heinis}, {Ibar}, {Isaak},
  {Ivison}, {Lagache}, {Levenson}, {Lonsdale}, {Lu}, {Madden}, {Maffei},
  {Magdis}, {Mainetti}, {Marchetti}, {Morrison}, {Nguyen}, {O'Halloran},
  {Oliver}, {Omont}, {Owen}, {Page}, {Pannella}, {Panuzzo}, {Papageorgiou},
  {Pearson}, {Perez-Fournon}, {Pohlen}, {Rigopoulou}, {Rizzo}, {Roseboom},
  {Rowan-Robinson}, {Sanchez Portal}, {Schulz}, {Seymour}, {Shupe}, {Smith},
  {Stevens}, {Strazzullo}, {Symeonidis}, {Trichas}, {Tugwell}, {Vaccari},
  {Valiante}, {Valtchanov}, {Vigroux}, {Wang}, {Ward}, {Wright}, {Xu}, \&
  {Zemcov}}]{Buat2010}
{Buat}, V., {Giovannoli}, E., {Burgarella}, D., {Altieri}, B., {Amblard}, A.,
  {Arumugam}, V., {Aussel}, H., {Babbedge}, T., {Blain}, A., {Bock}, J.,
  {Boselli}, A., {Castro-Rodriguez}, N., {Cava}, A., {Chanial}, P., {Clements},
  D.~L., {Conley}, A., {Conversi}, L., {Cooray}, A., {Dowell}, C.~D., {Dwek},
  E., {Eales}, S., {Elbaz}, D., {Fox}, M., {Franceschini}, A., {Gear}, W.,
  {Glenn}, J., {Griffin}, M., {Halpern}, M., {Hatziminaoglou}, E., {Heinis},
  S., {Ibar}, E., {Isaak}, K., {Ivison}, R.~J., {Lagache}, G., {Levenson}, L.,
  {Lonsdale}, C.~J., {Lu}, N., {Madden}, S., {Maffei}, B., {Magdis}, G.,
  {Mainetti}, G., {Marchetti}, L., {Morrison}, G.~E., {Nguyen}, H.~T.,
  {O'Halloran}, B., {Oliver}, S.~J., {Omont}, A., {Owen}, F.~N., {Page}, M.~J.,
  {Pannella}, M., {Panuzzo}, P., {Papageorgiou}, A., {Pearson}, C.~P.,
  {Perez-Fournon}, I., {Pohlen}, M., {Rigopoulou}, D., {Rizzo}, D., {Roseboom},
  I.~G., {Rowan-Robinson}, M., {Sanchez Portal}, M., {Schulz}, B., {Seymour},
  N., {Shupe}, D.~L., {Smith}, A.~J., {Stevens}, J.~A., {Strazzullo}, V.,
  {Symeonidis}, M., {Trichas}, M., {Tugwell}, K.~E., {Vaccari}, M., {Valiante},
  E., {Valtchanov}, I., {Vigroux}, L., {Wang}, L., {Ward}, R., {Wright}, G.,
  {Xu}, C.~K., \& {Zemcov}, M. 2010, ArXiv e-prints

\bibitem[{{Buat} {et~al.}(2007){Buat}, {Takeuchi}, {Iglesias-P{\'a}ramo}, {Xu},
  {Burgarella}, {Boselli}, {Barlow}, {Bianchi}, {Donas}, {Forster}, {Friedman},
  {Heckman}, {Lee}, {Madore}, {Martin}, {Milliard}, {Morissey}, {Neff}, {Rich},
  {Schiminovich}, {Seibert}, {Small}, {Szalay}, {Welsh}, {Wyder}, \&
  {Yi}}]{Buat2007}
{Buat}, V., {Takeuchi}, T.~T., {Iglesias-P{\'a}ramo}, J., {Xu}, C.~K.,
  {Burgarella}, D., {Boselli}, A., {Barlow}, T., {Bianchi}, L., {Donas}, J.,
  {Forster}, K., {Friedman}, P.~G., {Heckman}, T.~M., {Lee}, Y., {Madore},
  B.~F., {Martin}, D.~C., {Milliard}, B., {Morissey}, P., {Neff}, S., {Rich},
  M., {Schiminovich}, D., {Seibert}, M., {Small}, T., {Szalay}, A.~S., {Welsh},
  B., {Wyder}, T., \& {Yi}, S.~K. 2007, \apjs, 173, 404

\bibitem[{{Caputi} {et~al.}(2006){Caputi}, {Dole}, {Lagache}, {McLure},
  {Puget}, {Rieke}, {Dunlop}, {Le Floc'h}, {Papovich}, \&
  {P{\'e}rez-Gonz{\'a}lez}}]{Caputi2006}
{Caputi}, K.~I., {Dole}, H., {Lagache}, G., {McLure}, R.~J., {Puget}, J.,
  {Rieke}, G.~H., {Dunlop}, J.~S., {Le Floc'h}, E., {Papovich}, C., \&
  {P{\'e}rez-Gonz{\'a}lez}, P.~G. 2006, \apj, 637, 727

\bibitem[{{Chapin} {et~al.}(2010){Chapin}, {Chapman}, {Coppin}, {Devlin},
  {Dunlop}, {Greve}, {Halpern}, {Hasselfied}, {Hughes}, {Ivison}, {Marsden},
  {Moncelsi}, {Netterfield}, {Pascale}, {Scott}, {Smail}, {Viero}, {Walter},
  {Weiss}, \& {van der Werf}}]{Chapin2010}
{Chapin}, E.~L., {Chapman}, S.~C., {Coppin}, K.~E., {Devlin}, M.~J., {Dunlop},
  J.~S., {Greve}, T.~R., {Halpern}, M., {Hasselfied}, M.~F., {Hughes}, D.~H.,
  {Ivison}, R.~J., {Marsden}, G., {Moncelsi}, L., {Netterfield}, C.~B.,
  {Pascale}, E., {Scott}, D., {Smail}, I., {Viero}, M., {Walter}, F., {Weiss},
  A., \& {van der Werf}, P. 2010, ArXiv e-prints

\bibitem[{{Chapman} {et~al.}(2005){Chapman}, {Blain}, {Smail}, \&
  {Ivison}}]{Chapman2005}
{Chapman}, S.~C., {Blain}, A.~W., {Smail}, I., \& {Ivison}, R.~J. 2005, \apj,
  622, 772

\bibitem[{{Chary} \& {Elbaz}(2001)}]{Chary2001}
{Chary}, R. \& {Elbaz}, D. 2001, \apj, 556, 562

\bibitem[{{Cid Fernandes} {et~al.}(2010){Cid Fernandes}, {Stasi{\'n}ska},
  {Schlickmann}, {Mateus}, {Vale Asari}, {Schoenell}, \&
  {Sodr{\'e}}}]{CidFernandes2010}
{Cid Fernandes}, R., {Stasi{\'n}ska}, G., {Schlickmann}, M.~S., {Mateus}, A.,
  {Vale Asari}, N., {Schoenell}, W., \& {Sodr{\'e}}, L. 2010, \mnras, 403, 1036

\bibitem[{{Colless} {et~al.}(2003){Colless}, {Peterson}, {Jackson}, {Peacock},
  {Cole}, {Norberg}, {Baldry}, {Baugh}, {Bland-Hawthorn}, {Bridges}, {Cannon},
  {Collins}, {Couch}, {Cross}, {Dalton}, {De Propris}, {Driver}, {Efstathiou},
  {Ellis}, {Frenk}, {Glazebrook}, {Lahav}, {Lewis}, {Lumsden}, {Maddox},
  {Madgwick}, {Sutherland}, \& {Taylor}}]{Colless2003}
{Colless}, M., {Peterson}, B.~A., {Jackson}, C., {Peacock}, J.~A., {Cole}, S.,
  {Norberg}, P., {Baldry}, I.~K., {Baugh}, C.~M., {Bland-Hawthorn}, J.,
  {Bridges}, T., {Cannon}, R., {Collins}, C., {Couch}, W., {Cross}, N.,
  {Dalton}, G., {De Propris}, R., {Driver}, S.~P., {Efstathiou}, G., {Ellis},
  R.~S., {Frenk}, C.~S., {Glazebrook}, K., {Lahav}, O., {Lewis}, I., {Lumsden},
  S., {Maddox}, S., {Madgwick}, D., {Sutherland}, W., \& {Taylor}, K. 2003,
  ArXiv Astrophysics e-prints

\bibitem[{{Coppin} {et~al.}(2006){Coppin}, {Chapin}, {Mortier}, {Scott},
  {Borys}, {Dunlop}, {Halpern}, {Hughes}, {Pope}, {Scott}, {Serjeant}, {Wagg},
  {Alexander}, {Almaini}, {Aretxaga}, {Babbedge}, {Best}, {Blain}, {Chapman},
  {Clements}, {Crawford}, {Dunne}, {Eales}, {Edge}, {Farrah}, {Gazta{\~n}aga},
  {Gear}, {Granato}, {Greve}, {Fox}, {Ivison}, {Jarvis}, {Jenness}, {Lacey},
  {Lepage}, {Mann}, {Marsden}, {Martinez-Sansigre}, {Oliver}, {Page},
  {Peacock}, {Pearson}, {Percival}, {Priddey}, {Rawlings}, {Rowan-Robinson},
  {Savage}, {Seigar}, {Sekiguchi}, {Silva}, {Simpson}, {Smail}, {Stevens},
  {Takagi}, {Vaccari}, {van Kampen}, \& {Willott}}]{Coppin2006}
{Coppin}, K., {Chapin}, E.~L., {Mortier}, A.~M.~J., {Scott}, S.~E., {Borys},
  C., {Dunlop}, J.~S., {Halpern}, M., {Hughes}, D.~H., {Pope}, A., {Scott}, D.,
  {Serjeant}, S., {Wagg}, J., {Alexander}, D.~M., {Almaini}, O., {Aretxaga},
  I., {Babbedge}, T., {Best}, P.~N., {Blain}, A., {Chapman}, S., {Clements},
  D.~L., {Crawford}, M., {Dunne}, L., {Eales}, S.~A., {Edge}, A.~C., {Farrah},
  D., {Gazta{\~n}aga}, E., {Gear}, W.~K., {Granato}, G.~L., {Greve}, T.~R.,
  {Fox}, M., {Ivison}, R.~J., {Jarvis}, M.~J., {Jenness}, T., {Lacey}, C.,
  {Lepage}, K., {Mann}, R.~G., {Marsden}, G., {Martinez-Sansigre}, A.,
  {Oliver}, S., {Page}, M.~J., {Peacock}, J.~A., {Pearson}, C.~P., {Percival},
  W.~J., {Priddey}, R.~S., {Rawlings}, S., {Rowan-Robinson}, M., {Savage},
  R.~S., {Seigar}, M., {Sekiguchi}, K., {Silva}, L., {Simpson}, C., {Smail},
  I., {Stevens}, J.~A., {Takagi}, T., {Vaccari}, M., {van Kampen}, E., \&
  {Willott}, C.~J. 2006, \mnras, 372, 1621

\bibitem[{{Coppin} {et~al.}(2005){Coppin}, {Halpern}, {Scott}, {Borys}, \&
  {Chapman}}]{Coppin2005}
{Coppin}, K., {Halpern}, M., {Scott}, D., {Borys}, C., \& {Chapman}, S. 2005,
  \mnras, 357, 1022

\bibitem[{{Coppin} {et~al.}(2008){Coppin}, {Halpern}, {Scott}, {Borys},
  {Dunlop}, {Dunne}, {Ivison}, {Wagg}, {Aretxaga}, {Battistelli}, {Benson},
  {Blain}, {Chapman}, {Clements}, {Dye}, {Farrah}, {Hughes}, {Jenness}, {van
  Kampen}, {Lacey}, {Mortier}, {Pope}, {Priddey}, {Serjeant}, {Smail},
  {Stevens}, \& {Vaccari}}]{Coppin2008}
{Coppin}, K., {Halpern}, M., {Scott}, D., {Borys}, C., {Dunlop}, J., {Dunne},
  L., {Ivison}, R., {Wagg}, J., {Aretxaga}, I., {Battistelli}, E., {Benson},
  A., {Blain}, A., {Chapman}, S., {Clements}, D., {Dye}, S., {Farrah}, D.,
  {Hughes}, D., {Jenness}, T., {van Kampen}, E., {Lacey}, C., {Mortier}, A.,
  {Pope}, A., {Priddey}, R., {Serjeant}, S., {Smail}, I., {Stevens}, J., \&
  {Vaccari}, M. 2008, \mnras, 384, 1597

\bibitem[{{Coppin} {et~al.}(2010){Coppin}, {Pope}, {Men{\'e}ndez-Delmestre},
  {Alexander}, {Dunlop}, {Egami}, {Gabor}, {Ibar}, {Ivison}, {Austermann},
  {Blain}, {Chapman}, {Clements}, {Dunne}, {Dye}, {Farrah}, {Hughes},
  {Mortier}, {Page}, {Rowan-Robinson}, {Scott}, {Simpson}, {Smail}, {Swinbank},
  {Vaccari}, \& {Yun}}]{Coppin2010}
{Coppin}, K., {Pope}, A., {Men{\'e}ndez-Delmestre}, K., {Alexander}, D.~M.,
  {Dunlop}, J.~S., {Egami}, E., {Gabor}, J., {Ibar}, E., {Ivison}, R.~J.,
  {Austermann}, J.~E., {Blain}, A.~W., {Chapman}, S.~C., {Clements}, D.~L.,
  {Dunne}, L., {Dye}, S., {Farrah}, D., {Hughes}, D.~H., {Mortier}, A.~M.~J.,
  {Page}, M.~J., {Rowan-Robinson}, M., {Scott}, D., {Simpson}, C., {Smail}, I.,
  {Swinbank}, A.~M., {Vaccari}, M., \& {Yun}, M.~S. 2010, \apj, 713, 503

\bibitem[{{Dale} \& {Helou}(2002)}]{DaleHelou2002}
{Dale}, D.~A. \& {Helou}, G. 2002, \apj, 576, 159

\bibitem[{{Della Valle} {et~al.}(2006){Della Valle}, {Mazzei}, {Bettoni},
  {Aussel}, \& {de Zotti}}]{DellaValle2006}
{Della Valle}, A., {Mazzei}, P., {Bettoni}, D., {Aussel}, H., \& {de Zotti}, G.
  2006, \aap, 454, 453

\bibitem[{{Devlin} {et~al.}(2009){Devlin}, {Ade}, {Aretxaga}, {Bock}, {Chapin},
  {Griffin}, {Gundersen}, {Halpern}, {Hargrave}, {Hughes}, {Klein}, {Marsden},
  {Martin}, {Mauskopf}, {Moncelsi}, {Netterfield}, {Ngo}, {Olmi}, {Pascale},
  {Patanchon}, {Rex}, {Scott}, {Semisch}, {Thomas}, {Truch}, {Tucker},
  {Tucker}, {Viero}, \& {Wiebe}}]{Devlin2009}
{Devlin}, M.~J., {Ade}, P.~A.~R., {Aretxaga}, I., {Bock}, J.~J., {Chapin},
  E.~L., {Griffin}, M., {Gundersen}, J.~O., {Halpern}, M., {Hargrave}, P.~C.,
  {Hughes}, D.~H., {Klein}, J., {Marsden}, G., {Martin}, P.~G., {Mauskopf}, P.,
  {Moncelsi}, L., {Netterfield}, C.~B., {Ngo}, H., {Olmi}, L., {Pascale}, E.,
  {Patanchon}, G., {Rex}, M., {Scott}, D., {Semisch}, C., {Thomas}, N.,
  {Truch}, M.~D.~P., {Tucker}, C., {Tucker}, G.~S., {Viero}, M.~P., \& {Wiebe},
  D.~V. 2009, \nat, 458, 737

\bibitem[{{Devlin} {et~al.}(2004){Devlin}, {Ade}, {Aretxaga}, {Bock}, {Chung},
  {Chapin}, {Dicker}, {Griffin}, {Gundersen}, {Halpern}, {Hargrave}, {Hughes},
  {Klein}, {Marsden}, {Martin}, {Mauskopf}, {Netterfield}, {Olmi}, {Pascale},
  {Rex}, {Scott}, {Semisch}, {Truch}, {Tucker}, {Tucker}, {Turner}, \&
  {Weibe}}]{devlin04}
{Devlin}, M.~J., {Ade}, P.~A.~R., {Aretxaga}, I., {Bock}, J.~J., {Chung}, J.,
  {Chapin}, E., {Dicker}, S.~R., {Griffin}, M., {Gundersen}, J., {Halpern}, M.,
  {Hargrave}, P., {Hughes}, D., {Klein}, J., {Marsden}, G., {Martin}, P.,
  {Mauskopf}, P.~D., {Netterfield}, B., {Olmi}, L., {Pascale}, E., {Rex}, M.,
  {Scott}, D., {Semisch}, C., {Truch}, M., {Tucker}, C., {Tucker}, G.,
  {Turner}, A.~D., \& {Weibe}, D. 2004, in Presented at the Society of
  Photo-Optical Instrumentation Engineers (SPIE) Conference, Vol. 5498,
  Millimeter and Submillimeter Detectors for Astronomy II. Edited by Jonas
  Zmuidzinas, Wayne S. Holland and Stafford Withington Proceedings of the SPIE,
  Volume 5498, pp. 42-54 (2004)., ed. J.~Zmuidzinas, W.~S. Holland, \&
  S.~Withington, 42--54

\bibitem[{{Dickinson} \& {FIDEL team}(2007)}]{Dickinson2007}
{Dickinson}, M. \& {FIDEL team}. 2007, in Bulletin of the American Astronomical
  Society, Vol.~38, Bulletin of the American Astronomical Society, 822--+

\bibitem[{{Dickinson} {et~al.}(2003){Dickinson}, {Giavalisco}, \& {The Goods
  Team}}]{Dickinson2003}
{Dickinson}, M., {Giavalisco}, M., \& {The Goods Team}. 2003, in The Mass of
  Galaxies at Low and High Redshift, ed. {R.~Bender \& A.~Renzini}, 324--+

\bibitem[{{Dole} {et~al.}(2006){Dole}, {Lagache}, {Puget}, {Caputi},
  {Fern{\'a}ndez-Conde}, {Le Floc'h}, {Papovich}, {P{\'e}rez-Gonz{\'a}lez},
  {Rieke}, \& {Blaylock}}]{Dole2006}
{Dole}, H., {Lagache}, G., {Puget}, J.-L., {Caputi}, K.~I.,
  {Fern{\'a}ndez-Conde}, N., {Le Floc'h}, E., {Papovich}, C.,
  {P{\'e}rez-Gonz{\'a}lez}, P.~G., {Rieke}, G.~H., \& {Blaylock}, M. 2006,
  \aap, 451, 417

\bibitem[{{Dunlop} {et~al.}(2010){Dunlop}, {Ade}, {Bock}, {Chapin},
  {Cirasuolo}, {Coppin}, {Devlin}, {Griffin}, {Greve}, {Gundersen}, {Halpern},
  {Hargrave}, {Hughes}, {Ivison}, {Klein}, {Kovacs}, {Marsden}, {Mauskopf},
  {Netterfield}, {Olmi}, {Pascale}, {Patanchon}, {Rex}, {Scott}, {Semisch},
  {Smail}, {Targett}, {Thomas}, {Truch}, {Tucker}, {Tucker}, {Viero}, {Walter},
  {Wardlow}, {Weiss}, \& {Wiebe}}]{Dunlop2010}
{Dunlop}, J.~S., {Ade}, P.~A.~R., {Bock}, J.~J., {Chapin}, E.~L., {Cirasuolo},
  M., {Coppin}, K.~E.~K., {Devlin}, M.~J., {Griffin}, M., {Greve}, T.~R.,
  {Gundersen}, J.~O., {Halpern}, M., {Hargrave}, P.~C., {Hughes}, D.~H.,
  {Ivison}, R.~J., {Klein}, J., {Kovacs}, A., {Marsden}, G., {Mauskopf}, P.,
  {Netterfield}, C.~B., {Olmi}, L., {Pascale}, E., {Patanchon}, G., {Rex}, M.,
  {Scott}, D., {Semisch}, C., {Smail}, I., {Targett}, T.~A., {Thomas}, N.,
  {Truch}, M.~D.~P., {Tucker}, C., {Tucker}, G.~S., {Viero}, M.~P., {Walter},
  F., {Wardlow}, J.~L., {Weiss}, A., \& {Wiebe}, D.~V. 2010, \mnras, 1354

\bibitem[{{Dunne} {et~al.}(2000){Dunne}, {Eales}, {Edmunds}, {Ivison},
  {Alexander}, \& {Clements}}]{Dunne2000}
{Dunne}, L., {Eales}, S., {Edmunds}, M., {Ivison}, R., {Alexander}, P., \&
  {Clements}, D.~L. 2000, \mnras, 315, 115

\bibitem[{{Dunne} \& {Eales}(2001)}]{Dunne2001}
{Dunne}, L. \& {Eales}, S.~A. 2001, \mnras, 327, 697

\bibitem[{{Dwek} {et~al.}(1998){Dwek}, {Arendt}, {Hauser}, {Fixsen}, {Kelsall},
  {Leisawitz}, {Pei}, {Wright}, {Mather}, {Moseley}, {Odegard}, {Shafer},
  {Silverberg}, \& {Weiland}}]{Dwek1998}
{Dwek}, E., {Arendt}, R.~G., {Hauser}, M.~G., {Fixsen}, D., {Kelsall}, T.,
  {Leisawitz}, D., {Pei}, Y.~C., {Wright}, E.~L., {Mather}, J.~C., {Moseley},
  S.~H., {Odegard}, N., {Shafer}, R., {Silverberg}, R.~F., \& {Weiland}, J.~L.
  1998, \apj, 508, 106

\bibitem[{{Dye}(2008)}]{Dye2008b}
{Dye}, S. 2008, \mnras, 389, 1293

\bibitem[{Dye(2010)}]{Dye2010c}
Dye, S. 2010, private communication

\bibitem[{{Dye} {et~al.}(2009){Dye}, {Ade}, {Bock}, {Chapin}, {Devlin},
  {Dunlop}, {Eales}, {Griffin}, {Gundersen}, {Halpern}, {Hargrave}, {Hughes},
  {Klein}, {Magnelli}, {Marsden}, {Mauskopf}, {Moncelsi}, {Netterfield},
  {Olmi}, {Pascale}, {Patanchon}, {Rex}, {Scott}, {Semisch}, {Targett},
  {Thomas}, {Truch}, {Tucker}, {Tucker}, {Viero}, \& {Wiebe}}]{Dye2009}
{Dye}, S., {Ade}, P.~A.~R., {Bock}, J.~J., {Chapin}, E.~L., {Devlin}, M.~J.,
  {Dunlop}, J.~S., {Eales}, S.~A., {Griffin}, M., {Gundersen}, J.~O.,
  {Halpern}, M., {Hargrave}, P.~C., {Hughes}, D.~H., {Klein}, J., {Magnelli},
  B., {Marsden}, G., {Mauskopf}, P., {Moncelsi}, L., {Netterfield}, C.~B.,
  {Olmi}, L., {Pascale}, E., {Patanchon}, G., {Rex}, M., {Scott}, D.,
  {Semisch}, C., {Targett}, T., {Thomas}, N., {Truch}, M.~D.~P., {Tucker}, C.,
  {Tucker}, G.~S., {Viero}, M.~P., \& {Wiebe}, D.~V. 2009, \apj, 703, 285

\bibitem[{{Dye} {et~al.}(2010{\natexlab{a}}){Dye}, {Dunne}, {Eales}, {Smith},
  {Amblard}, {Auld}, {Baes}, {Baldry}, {Bamford}, {Blain}, {Bonfield},
  {Bremer}, {Burgarella}, {Buttiglione}, {Cameron}, {Cava}, {Clements},
  {Cooray}, {Croom}, {Dariush}, {de Zotti}, {Driver}, {Dunlop}, {Frayer},
  {Fritz}, {Gardner}, {Gomez}, {Gonzalez-Nuevo}, {Herranz}, {Hill}, {Hopkins},
  {Ibar}, {Ivison}, {Jarvis}, {Jones}, {Kelvin}, {Lagache}, {Leeuw}, {Liske},
  {Lopez-Caniego}, {Loveday}, {Maddox}, {Micha{\l}owski}, {Negrello},
  {Norberg}, {Page}, {Parkinson}, {Pascale}, {Peacock}, {Pohlen}, {Popescu},
  {Prescott}, {Rigopoulou}, {Robotham}, {Rigby}, {Rodighiero}, {Samui},
  {Scott}, {Serjeant}, {Sharp}, {Sibthorpe}, {Temi}, {Thompson}, {Tuffs},
  {Valtchanov}, {van der Werf}, {van Kampen}, \& {Verma}}]{Dye2010b}
{Dye}, S., {Dunne}, L., {Eales}, S., {Smith}, D.~J.~B., {Amblard}, A., {Auld},
  R., {Baes}, M., {Baldry}, I.~K., {Bamford}, S., {Blain}, A.~W., {Bonfield},
  D.~G., {Bremer}, M., {Burgarella}, D., {Buttiglione}, S., {Cameron}, E.,
  {Cava}, A., {Clements}, D.~L., {Cooray}, A., {Croom}, S., {Dariush}, A., {de
  Zotti}, G., {Driver}, S., {Dunlop}, J.~S., {Frayer}, D., {Fritz}, J.,
  {Gardner}, J.~P., {Gomez}, H.~L., {Gonzalez-Nuevo}, J., {Herranz}, D.,
  {Hill}, D., {Hopkins}, A., {Ibar}, E., {Ivison}, R.~J., {Jarvis}, M.~J.,
  {Jones}, D.~H., {Kelvin}, L., {Lagache}, G., {Leeuw}, L., {Liske}, J.,
  {Lopez-Caniego}, M., {Loveday}, J., {Maddox}, S., {Micha{\l}owski}, M.~J.,
  {Negrello}, M., {Norberg}, P., {Page}, M.~J., {Parkinson}, H., {Pascale}, E.,
  {Peacock}, J.~A., {Pohlen}, M., {Popescu}, C., {Prescott}, M., {Rigopoulou},
  D., {Robotham}, A., {Rigby}, E., {Rodighiero}, G., {Samui}, S., {Scott}, D.,
  {Serjeant}, S., {Sharp}, R., {Sibthorpe}, B., {Temi}, P., {Thompson}, M.~A.,
  {Tuffs}, R., {Valtchanov}, I., {van der Werf}, P.~P., {van Kampen}, E., \&
  {Verma}, A. 2010{\natexlab{a}}, \aap, 518, L10+

\bibitem[{{Dye} {et~al.}(2010{\natexlab{b}}){Dye}, {Eales}, {Moncelsi}, \&
  {Pascale}}]{Dye2010a}
{Dye}, S., {Eales}, S., {Moncelsi}, L., \& {Pascale}, E. 2010{\natexlab{b}},
  \mnras, L104+

\bibitem[{{Dye} {et~al.}(2008){Dye}, {Eales}, {Aretxaga}, {Serjeant}, {Dunlop},
  {Babbedge}, {Chapman}, {Cirasuolo}, {Clements}, {Coppin}, {Dunne}, {Egami},
  {Farrah}, {Ivison}, {van Kampen}, {Pope}, {Priddey}, {Rieke}, {Schael},
  {Scott}, {Simpson}, {Takagi}, {Takata}, \& {Vaccari}}]{Dye2008a}
{Dye}, S., {Eales}, S.~A., {Aretxaga}, I., {Serjeant}, S., {Dunlop}, J.~S.,
  {Babbedge}, T.~S.~R., {Chapman}, S.~C., {Cirasuolo}, M., {Clements}, D.~L.,
  {Coppin}, K.~E.~K., {Dunne}, L., {Egami}, E., {Farrah}, D., {Ivison}, R.~J.,
  {van Kampen}, E., {Pope}, A., {Priddey}, R., {Rieke}, G.~H., {Schael}, A.~M.,
  {Scott}, D., {Simpson}, C., {Takagi}, T., {Takata}, T., \& {Vaccari}, M.
  2008, \mnras, 386, 1107

\bibitem[{{Dye} {et~al.}(2007){Dye}, {Eales}, {Ashby}, {Huang}, {Egami},
  {Brodwin}, {Lilly}, \& {Webb}}]{Dye2007}
{Dye}, S., {Eales}, S.~A., {Ashby}, M.~L.~N., {Huang}, J., {Egami}, E.,
  {Brodwin}, M., {Lilly}, S., \& {Webb}, T. 2007, \mnras, 375, 725

\bibitem[{{Eales} {et~al.}(2009){Eales}, {Chapin}, {Devlin}, {Dye}, {Halpern},
  {Hughes}, {Marsden}, {Mauskopf}, {Moncelsi}, {Netterfield}, {Pascale},
  {Patanchon}, {Raymond}, {Rex}, {Scott}, {Semisch}, {Siana}, {Truch}, \&
  {Viero}}]{Eales2009}
{Eales}, S., {Chapin}, E.~L., {Devlin}, M.~J., {Dye}, S., {Halpern}, M.,
  {Hughes}, D.~H., {Marsden}, G., {Mauskopf}, P., {Moncelsi}, L.,
  {Netterfield}, C.~B., {Pascale}, E., {Patanchon}, G., {Raymond}, G., {Rex},
  M., {Scott}, D., {Semisch}, C., {Siana}, B., {Truch}, M.~D.~P., \& {Viero},
  M.~P. 2009, \apj, 707, 1779

\bibitem[{{Eales} {et~al.}(2010){Eales}, {Raymond}, {Roseboom}, {Altieri},
  {Amblard}, {Arumugam}, {Auld}, {Aussel}, {Babbedge}, {Blain}, {Bock},
  {Boselli}, {Brisbin}, {Buat}, {Burgarella}, {Castro-Rodr{\'{\i}}guez},
  {Cava}, {Chanial}, {Clements}, {Conley}, {Conversi}, {Cooray}, {Dowell},
  {Dwek}, {Dye}, {Elbaz}, {Farrah}, {Fox}, {Franceschini}, {Gear}, {Glenn},
  {Gonz{\'a}lez Solares}, {Griffin}, {Harwit}, {Hatziminaoglou}, {Huang},
  {Ibar}, {Isaak}, {Ivison}, {Lagache}, {Levenson}, {Lonsdale}, {Lu}, {Madden},
  {Maffei}, {Mainetti}, {Marchetti}, {Morrison}, {Mortier}, {Nguyen},
  {O'Halloran}, {Oliver}, {Omont}, {Owen}, {Page}, {Pannella}, {Panuzzo},
  {Papageorgiou}, {Pearson}, {P{\'e}rez-Fournon}, {Pohlen}, {Rawlings},
  {Rigopoulou}, {Rizzo}, {Rowan-Robinson}, {S{\'a}nchez Portal}, {Schulz},
  {Scott}, {Seymour}, {Shupe}, {Smith}, {Stevens}, {Strazzullo}, {Symeonidis},
  {Trichas}, {Tugwell}, {Vaccari}, {Valtchanov}, {Vigroux}, {Wang}, {Ward},
  {Wright}, {Xu}, \& {Zemcov}}]{Eales2010}
{Eales}, S.~A., {Raymond}, G., {Roseboom}, I.~G., {Altieri}, B., {Amblard}, A.,
  {Arumugam}, V., {Auld}, R., {Aussel}, H., {Babbedge}, T., {Blain}, A.,
  {Bock}, J., {Boselli}, A., {Brisbin}, D., {Buat}, V., {Burgarella}, D.,
  {Castro-Rodr{\'{\i}}guez}, N., {Cava}, A., {Chanial}, P., {Clements}, D.~L.,
  {Conley}, A., {Conversi}, L., {Cooray}, A., {Dowell}, C.~D., {Dwek}, E.,
  {Dye}, S., {Elbaz}, D., {Farrah}, D., {Fox}, M., {Franceschini}, A., {Gear},
  W., {Glenn}, J., {Gonz{\'a}lez Solares}, E.~A., {Griffin}, M., {Harwit}, M.,
  {Hatziminaoglou}, E., {Huang}, J., {Ibar}, E., {Isaak}, K., {Ivison}, R.~J.,
  {Lagache}, G., {Levenson}, L., {Lonsdale}, C.~J., {Lu}, N., {Madden}, S.,
  {Maffei}, B., {Mainetti}, G., {Marchetti}, L., {Morrison}, G.~E., {Mortier},
  A.~M.~J., {Nguyen}, H.~T., {O'Halloran}, B., {Oliver}, S.~J., {Omont}, A.,
  {Owen}, F.~N., {Page}, M.~J., {Pannella}, M., {Panuzzo}, P., {Papageorgiou},
  A., {Pearson}, C.~P., {P{\'e}rez-Fournon}, I., {Pohlen}, M., {Rawlings},
  J.~I., {Rigopoulou}, D., {Rizzo}, D., {Rowan-Robinson}, M., {S{\'a}nchez
  Portal}, M., {Schulz}, B., {Scott}, D., {Seymour}, N., {Shupe}, D.~L.,
  {Smith}, A.~J., {Stevens}, J.~A., {Strazzullo}, V., {Symeonidis}, M.,
  {Trichas}, M., {Tugwell}, K.~E., {Vaccari}, M., {Valtchanov}, I., {Vigroux},
  L., {Wang}, L., {Ward}, R., {Wright}, G., {Xu}, C.~K., \& {Zemcov}, M. 2010,
  \aap, 518, L23+

\bibitem[{{Elbaz} {et~al.}(2007){Elbaz}, {Daddi}, {Le Borgne}, {Dickinson},
  {Alexander}, {Chary}, {Starck}, {Brandt}, {Kitzbichler}, {MacDonald},
  {Nonino}, {Popesso}, {Stern}, \& {Vanzella}}]{Elbaz2007}
{Elbaz}, D., {Daddi}, E., {Le Borgne}, D., {Dickinson}, M., {Alexander}, D.~M.,
  {Chary}, R., {Starck}, J., {Brandt}, W.~N., {Kitzbichler}, M., {MacDonald},
  E., {Nonino}, M., {Popesso}, P., {Stern}, D., \& {Vanzella}, E. 2007, \aap,
  468, 33

\bibitem[{{Elbaz} {et~al.}(2010){Elbaz}, {Hwang}, {Magnelli}, {Daddi},
  {Aussel}, {Altieri}, {Amblard}, {Andreani}, {Arumugam}, {Auld}, {Babbedge},
  {Berta}, {Blain}, {Bock}, {Bongiovanni}, {Boselli}, {Buat}, {Burgarella},
  {Castro-Rodriguez}, {Cava}, {Cepa}, {Chanial}, {Chary}, {Cimatti},
  {Clements}, {Conley}, {Conversi}, {Cooray}, {Dickinson}, {Dominguez},
  {Dowell}, {Dunlop}, {Dwek}, {Eales}, {Farrah}, {F{\"o}rster Schreiber},
  {Fox}, {Franceschini}, {Gear}, {Genzel}, {Glenn}, {Griffin}, {Gruppioni},
  {Halpern}, {Hatziminaoglou}, {Ibar}, {Isaak}, {Ivison}, {Lagache}, {Le
  Borgne}, {Le Floc'h}, {Levenson}, {Lu}, {Lutz}, {Madden}, {Maffei}, {Magdis},
  {Mainetti}, {Maiolino}, {Marchetti}, {Mortier}, {Nguyen}, {Nordon},
  {O'Halloran}, {Okumura}, {Oliver}, {Omont}, {Page}, {Panuzzo},
  {Papageorgiou}, {Pearson}, {Perez Fournon}, {P{\'e}rez Garc{\'{\i}}a},
  {Poglitsch}, {Pohlen}, {Popesso}, {Pozzi}, {Rawlings}, {Rigopoulou},
  {Riguccini}, {Rizzo}, {Rodighiero}, {Roseboom}, {Rowan-Robinson},
  {Saintonge}, {Sanchez Portal}, {Santini}, {Sauvage}, {Schulz}, {Scott},
  {Seymour}, {Shao}, {Shupe}, {Smith}, {Stevens}, {Sturm}, {Symeonidis},
  {Tacconi}, {Trichas}, {Tugwell}, {Vaccari}, {Valtchanov}, {Vieira},
  {Vigroux}, {Wang}, {Ward}, {Wright}, {Xu}, \& {Zemcov}}]{Elbaz2010}
{Elbaz}, D., {Hwang}, H.~S., {Magnelli}, B., {Daddi}, E., {Aussel}, H.,
  {Altieri}, B., {Amblard}, A., {Andreani}, P., {Arumugam}, V., {Auld}, R.,
  {Babbedge}, T., {Berta}, S., {Blain}, A., {Bock}, J., {Bongiovanni}, A.,
  {Boselli}, A., {Buat}, V., {Burgarella}, D., {Castro-Rodriguez}, N., {Cava},
  A., {Cepa}, J., {Chanial}, P., {Chary}, R., {Cimatti}, A., {Clements}, D.~L.,
  {Conley}, A., {Conversi}, L., {Cooray}, A., {Dickinson}, M., {Dominguez}, H.,
  {Dowell}, C.~D., {Dunlop}, J.~S., {Dwek}, E., {Eales}, S., {Farrah}, D.,
  {F{\"o}rster Schreiber}, N., {Fox}, M., {Franceschini}, A., {Gear}, W.,
  {Genzel}, R., {Glenn}, J., {Griffin}, M., {Gruppioni}, C., {Halpern}, M.,
  {Hatziminaoglou}, E., {Ibar}, E., {Isaak}, K., {Ivison}, R.~J., {Lagache},
  G., {Le Borgne}, D., {Le Floc'h}, E., {Levenson}, L., {Lu}, N., {Lutz}, D.,
  {Madden}, S., {Maffei}, B., {Magdis}, G., {Mainetti}, G., {Maiolino}, R.,
  {Marchetti}, L., {Mortier}, A.~M.~J., {Nguyen}, H.~T., {Nordon}, R.,
  {O'Halloran}, B., {Okumura}, K., {Oliver}, S.~J., {Omont}, A., {Page}, M.~J.,
  {Panuzzo}, P., {Papageorgiou}, A., {Pearson}, C.~P., {Perez Fournon}, I.,
  {P{\'e}rez Garc{\'{\i}}a}, A.~M., {Poglitsch}, A., {Pohlen}, M., {Popesso},
  P., {Pozzi}, F., {Rawlings}, J.~I., {Rigopoulou}, D., {Riguccini}, L.,
  {Rizzo}, D., {Rodighiero}, G., {Roseboom}, I.~G., {Rowan-Robinson}, M.,
  {Saintonge}, A., {Sanchez Portal}, M., {Santini}, P., {Sauvage}, M.,
  {Schulz}, B., {Scott}, D., {Seymour}, N., {Shao}, L., {Shupe}, D.~L.,
  {Smith}, A.~J., {Stevens}, J.~A., {Sturm}, E., {Symeonidis}, M., {Tacconi},
  L., {Trichas}, M., {Tugwell}, K.~E., {Vaccari}, M., {Valtchanov}, I.,
  {Vieira}, J., {Vigroux}, L., {Wang}, L., {Ward}, R., {Wright}, G., {Xu},
  C.~K., \& {Zemcov}, M. 2010, \aap, 518, L29+

\bibitem[{{Fazio} {et~al.}(2004){Fazio}, {Hora}, {Allen}, {Ashby}, {Barmby},
  {Deutsch}, {Huang}, {Kleiner}, {Marengo}, {Megeath}, {Melnick}, {Pahre},
  {Patten}, {Polizotti}, {Smith}, {Taylor}, {Wang}, {Willner}, {Hoffmann},
  {Pipher}, {Forrest}, {McMurty}, {McCreight}, {McKelvey}, {McMurray}, {Koch},
  {Moseley}, {Arendt}, {Mentzell}, {Marx}, {Losch}, {Mayman}, {Eichhorn},
  {Krebs}, {Jhabvala}, {Gezari}, {Fixsen}, {Flores}, {Shakoorzadeh}, {Jungo},
  {Hakun}, {Workman}, {Karpati}, {Kichak}, {Whitley}, {Mann}, {Tollestrup},
  {Eisenhardt}, {Stern}, {Gorjian}, {Bhattacharya}, {Carey}, {Nelson},
  {Glaccum}, {Lacy}, {Lowrance}, {Laine}, {Reach}, {Stauffer}, {Surace},
  {Wilson}, {Wright}, {Hoffman}, {Domingo}, \& {Cohen}}]{Fazio2004}
{Fazio}, G.~G., {Hora}, J.~L., {Allen}, L.~E., {Ashby}, M.~L.~N., {Barmby}, P.,
  {Deutsch}, L.~K., {Huang}, J., {Kleiner}, S., {Marengo}, M., {Megeath},
  S.~T., {Melnick}, G.~J., {Pahre}, M.~A., {Patten}, B.~M., {Polizotti}, J.,
  {Smith}, H.~A., {Taylor}, R.~S., {Wang}, Z., {Willner}, S.~P., {Hoffmann},
  W.~F., {Pipher}, J.~L., {Forrest}, W.~J., {McMurty}, C.~W., {McCreight},
  C.~R., {McKelvey}, M.~E., {McMurray}, R.~E., {Koch}, D.~G., {Moseley}, S.~H.,
  {Arendt}, R.~G., {Mentzell}, J.~E., {Marx}, C.~T., {Losch}, P., {Mayman}, P.,
  {Eichhorn}, W., {Krebs}, D., {Jhabvala}, M., {Gezari}, D.~Y., {Fixsen},
  D.~J., {Flores}, J., {Shakoorzadeh}, K., {Jungo}, R., {Hakun}, C., {Workman},
  L., {Karpati}, G., {Kichak}, R., {Whitley}, R., {Mann}, S., {Tollestrup},
  E.~V., {Eisenhardt}, P., {Stern}, D., {Gorjian}, V., {Bhattacharya}, B.,
  {Carey}, S., {Nelson}, B.~O., {Glaccum}, W.~J., {Lacy}, M., {Lowrance},
  P.~J., {Laine}, S., {Reach}, W.~T., {Stauffer}, J.~A., {Surace}, J.~A.,
  {Wilson}, G., {Wright}, E.~L., {Hoffman}, A., {Domingo}, G., \& {Cohen}, M.
  2004, \apjs, 154, 10

\bibitem[{{Fixsen} {et~al.}(1998){Fixsen}, {Dwek}, {Mather}, {Bennett}, \&
  {Shafer}}]{Fixsen98}
{Fixsen}, D.~J., {Dwek}, E., {Mather}, J.~C., {Bennett}, C.~L., \& {Shafer},
  R.~A. 1998, \apj, 508, 123

\bibitem[{{Frayer} {et~al.}(2009){Frayer}, {Sanders}, {Surace}, {Aussel},
  {Salvato}, {Le Floc'h}, {Huynh}, {Scoville}, {Afonso-Luis}, {Bhattacharya},
  {Capak}, {Fadda}, {Fu}, {Helou}, {Ilbert}, {Kartaltepe}, {Koekemoer}, {Lee},
  {Murphy}, {Sargent}, {Schinnerer}, {Sheth}, {Shopbell}, {Shupe}, \&
  {Yan}}]{Frayer2009}
{Frayer}, D.~T., {Sanders}, D.~B., {Surace}, J.~A., {Aussel}, H., {Salvato},
  M., {Le Floc'h}, E., {Huynh}, M.~T., {Scoville}, N.~Z., {Afonso-Luis}, A.,
  {Bhattacharya}, B., {Capak}, P., {Fadda}, D., {Fu}, H., {Helou}, G.,
  {Ilbert}, O., {Kartaltepe}, J.~S., {Koekemoer}, A.~M., {Lee}, N., {Murphy},
  E., {Sargent}, M.~T., {Schinnerer}, E., {Sheth}, K., {Shopbell}, P.~L.,
  {Shupe}, D.~L., \& {Yan}, L. 2009, \aj, 138, 1261

\bibitem[{{Gawiser} {et~al.}(2006){Gawiser}, {van Dokkum}, {Herrera}, {Maza},
  {Castander}, {Infante}, {Lira}, {Quadri}, {Toner}, {Treister}, {Urry},
  {Altmann}, {Assef}, {Christlein}, {Coppi}, {Dur{\'a}n}, {Franx}, {Galaz},
  {Huerta}, {Liu}, {L{\'o}pez}, {M{\'e}ndez}, {Moore}, {Rubio}, {Ruiz}, {Toft},
  \& {Yi}}]{Gawiser2006}
{Gawiser}, E., {van Dokkum}, P.~G., {Herrera}, D., {Maza}, J., {Castander},
  F.~J., {Infante}, L., {Lira}, P., {Quadri}, R., {Toner}, R., {Treister}, E.,
  {Urry}, C.~M., {Altmann}, M., {Assef}, R., {Christlein}, D., {Coppi}, P.~S.,
  {Dur{\'a}n}, M.~F., {Franx}, M., {Galaz}, G., {Huerta}, L., {Liu}, C.,
  {L{\'o}pez}, S., {M{\'e}ndez}, R., {Moore}, D.~C., {Rubio}, M., {Ruiz},
  M.~T., {Toft}, S., \& {Yi}, S.~K. 2006, \apjs, 162, 1

\bibitem[{{Gispert} {et~al.}(2000){Gispert}, {Lagache}, \&
  {Puget}}]{Gispert2000}
{Gispert}, R., {Lagache}, G., \& {Puget}, J.~L. 2000, \aap, 360, 1

\bibitem[{{Griffin} {et~al.}(2010){Griffin}, {Abergel}, {Abreu}, {Ade},
  {Andr{\'e}}, {Augueres}, {Babbedge}, {Bae}, {Baillie}, {Baluteau}, {Barlow},
  {Bendo}, {Benielli}, {Bock}, {Bonhomme}, {Brisbin}, {Brockley-Blatt},
  {Caldwell}, {Cara}, {Castro-Rodriguez}, {Cerulli}, {Chanial}, {Chen},
  {Clark}, {Clements}, {Clerc}, {Coker}, {Communal}, {Conversi}, {Cox},
  {Crumb}, {Cunningham}, {Daly}, {Davis}, {de Antoni}, {Delderfield}, {Devin},
  {di Giorgio}, {Didschuns}, {Dohlen}, {Donati}, {Dowell}, {Dowell}, {Duband},
  {Dumaye}, {Emery}, {Ferlet}, {Ferrand}, {Fontignie}, {Fox}, {Franceschini},
  {Frerking}, {Fulton}, {Garcia}, {Gastaud}, {Gear}, {Glenn}, {Goizel},
  {Griffin}, {Grundy}, {Guest}, {Guillemet}, {Hargrave}, {Harwit}, {Hastings},
  {Hatziminaoglou}, {Herman}, {Hinde}, {Hristov}, {Huang}, {Imhof}, {Isaak},
  {Israelsson}, {Ivison}, {Jennings}, {Kiernan}, {King}, {Lange}, {Latter},
  {Laurent}, {Laurent}, {Leeks}, {Lellouch}, {Levenson}, {Li}, {Li},
  {Lilienthal}, {Lim}, {Liu}, {Lu}, {Madden}, {Mainetti}, {Marliani}, {McKay},
  {Mercier}, {Molinari}, {Morris}, {Moseley}, {Mulder}, {Mur}, {Naylor},
  {Nguyen}, {O'Halloran}, {Oliver}, {Olofsson}, {Olofsson}, {Orfei}, {Page},
  {Pain}, {Panuzzo}, {Papageorgiou}, {Parks}, {Parr-Burman}, {Pearce},
  {Pearson}, {P{\'e}rez-Fournon}, {Pinsard}, {Pisano}, {Podosek}, {Pohlen},
  {Polehampton}, {Pouliquen}, {Rigopoulou}, {Rizzo}, {Roseboom}, {Roussel},
  {Rowan-Robinson}, {Rownd}, {Saraceno}, {Sauvage}, {Savage}, {Savini},
  {Sawyer}, {Scharmberg}, {Schmitt}, {Schneider}, {Schulz}, {Schwartz},
  {Shafer}, {Shupe}, {Sibthorpe}, {Sidher}, {Smith}, {Smith}, {Smith},
  {Spencer}, {Stobie}, {Sudiwala}, {Sukhatme}, {Surace}, {Stevens}, {Swinyard},
  {Trichas}, {Tourette}, {Triou}, {Tseng}, {Tucker}, {Turner}, {Vaccari},
  {Valtchanov}, {Vigroux}, {Virique}, {Voellmer}, {Walker}, {Ward}, {Waskett},
  {Weilert}, {Wesson}, {White}, {Whitehouse}, {Wilson}, {Winter}, {Woodcraft},
  {Wright}, {Xu}, {Zavagno}, {Zemcov}, {Zhang}, \& {Zonca}}]{Griffin2010}
{Griffin}, M.~J., {Abergel}, A., {Abreu}, A., {Ade}, P.~A.~R., {Andr{\'e}}, P.,
  {Augueres}, J., {Babbedge}, T., {Bae}, Y., {Baillie}, T., {Baluteau}, J.,
  {Barlow}, M.~J., {Bendo}, G., {Benielli}, D., {Bock}, J.~J., {Bonhomme}, P.,
  {Brisbin}, D., {Brockley-Blatt}, C., {Caldwell}, M., {Cara}, C.,
  {Castro-Rodriguez}, N., {Cerulli}, R., {Chanial}, P., {Chen}, S., {Clark},
  E., {Clements}, D.~L., {Clerc}, L., {Coker}, J., {Communal}, D., {Conversi},
  L., {Cox}, P., {Crumb}, D., {Cunningham}, C., {Daly}, F., {Davis}, G.~R., {de
  Antoni}, P., {Delderfield}, J., {Devin}, N., {di Giorgio}, A., {Didschuns},
  I., {Dohlen}, K., {Donati}, M., {Dowell}, A., {Dowell}, C.~D., {Duband}, L.,
  {Dumaye}, L., {Emery}, R.~J., {Ferlet}, M., {Ferrand}, D., {Fontignie}, J.,
  {Fox}, M., {Franceschini}, A., {Frerking}, M., {Fulton}, T., {Garcia}, J.,
  {Gastaud}, R., {Gear}, W.~K., {Glenn}, J., {Goizel}, A., {Griffin}, D.~K.,
  {Grundy}, T., {Guest}, S., {Guillemet}, L., {Hargrave}, P.~C., {Harwit}, M.,
  {Hastings}, P., {Hatziminaoglou}, E., {Herman}, M., {Hinde}, B., {Hristov},
  V., {Huang}, M., {Imhof}, P., {Isaak}, K.~J., {Israelsson}, U., {Ivison},
  R.~J., {Jennings}, D., {Kiernan}, B., {King}, K.~J., {Lange}, A.~E.,
  {Latter}, W., {Laurent}, G., {Laurent}, P., {Leeks}, S.~J., {Lellouch}, E.,
  {Levenson}, L., {Li}, B., {Li}, J., {Lilienthal}, J., {Lim}, T., {Liu},
  S.~J., {Lu}, N., {Madden}, S., {Mainetti}, G., {Marliani}, P., {McKay}, D.,
  {Mercier}, K., {Molinari}, S., {Morris}, H., {Moseley}, H., {Mulder}, J.,
  {Mur}, M., {Naylor}, D.~A., {Nguyen}, H., {O'Halloran}, B., {Oliver}, S.,
  {Olofsson}, G., {Olofsson}, H., {Orfei}, R., {Page}, M.~J., {Pain}, I.,
  {Panuzzo}, P., {Papageorgiou}, A., {Parks}, G., {Parr-Burman}, P., {Pearce},
  A., {Pearson}, C., {P{\'e}rez-Fournon}, I., {Pinsard}, F., {Pisano}, G.,
  {Podosek}, J., {Pohlen}, M., {Polehampton}, E.~T., {Pouliquen}, D.,
  {Rigopoulou}, D., {Rizzo}, D., {Roseboom}, I.~G., {Roussel}, H.,
  {Rowan-Robinson}, M., {Rownd}, B., {Saraceno}, P., {Sauvage}, M., {Savage},
  R., {Savini}, G., {Sawyer}, E., {Scharmberg}, C., {Schmitt}, D., {Schneider},
  N., {Schulz}, B., {Schwartz}, A., {Shafer}, R., {Shupe}, D.~L., {Sibthorpe},
  B., {Sidher}, S., {Smith}, A., {Smith}, A.~J., {Smith}, D., {Spencer}, L.,
  {Stobie}, B., {Sudiwala}, R., {Sukhatme}, K., {Surace}, C., {Stevens}, J.~A.,
  {Swinyard}, B.~M., {Trichas}, M., {Tourette}, T., {Triou}, H., {Tseng}, S.,
  {Tucker}, C., {Turner}, A., {Vaccari}, M., {Valtchanov}, I., {Vigroux}, L.,
  {Virique}, E., {Voellmer}, G., {Walker}, H., {Ward}, R., {Waskett}, T.,
  {Weilert}, M., {Wesson}, R., {White}, G.~J., {Whitehouse}, N., {Wilson},
  C.~D., {Winter}, B., {Woodcraft}, A.~L., {Wright}, G.~S., {Xu}, C.~K.,
  {Zavagno}, A., {Zemcov}, M., {Zhang}, L., \& {Zonca}, E. 2010, \aap, 518, L3+

\bibitem[{{Hatziminaoglou} {et~al.}(2010){Hatziminaoglou}, {Omont}, {Stevens},
  {Amblard}, {Arumugam}, {Auld}, {Aussel}, {Babbedge}, {Blain}, {Bock},
  {Boselli}, {Buat}, {Burgarella}, {Castro-Rodr{\'{\i}}guez}, {Cava},
  {Chanial}, {Clements}, {Conley}, {Conversi}, {Cooray}, {Dowell}, {Dwek},
  {Dye}, {Eales}, {Elbaz}, {Farrah}, {Fox}, {Franceschini}, {Gear}, {Glenn},
  {Gonz{\'a}lez Solares}, {Griffin}, {Halpern}, {Ibar}, {Isaak}, {Ivison},
  {Lagache}, {Levenson}, {Lu}, {Madden}, {Maffei}, {Mainetti}, {Marchetti},
  {Mortier}, {Nguyen}, {O'Halloran}, {Oliver}, {Page}, {Panuzzo},
  {Papageorgiou}, {Pearson}, {P{\'e}rez-Fournon}, {Pohlen}, {Rawlings},
  {Rigopoulou}, {Rizzo}, {Roseboom}, {Rowan-Robinson}, {Sanchez Portal},
  {Schulz}, {Scott}, {Seymour}, {Shupe}, {Smith}, {Symeonidis}, {Trichas},
  {Tugwell}, {Vaccari}, {Valtchanov}, {Vigroux}, {Wang}, {Ward}, {Wright},
  {Xu}, \& {Zemcov}}]{Hatziminaoglou2010}
{Hatziminaoglou}, E., {Omont}, A., {Stevens}, J.~A., {Amblard}, A., {Arumugam},
  V., {Auld}, R., {Aussel}, H., {Babbedge}, T., {Blain}, A., {Bock}, J.,
  {Boselli}, A., {Buat}, V., {Burgarella}, D., {Castro-Rodr{\'{\i}}guez}, N.,
  {Cava}, A., {Chanial}, P., {Clements}, D.~L., {Conley}, A., {Conversi}, L.,
  {Cooray}, A., {Dowell}, C.~D., {Dwek}, E., {Dye}, S., {Eales}, S., {Elbaz},
  D., {Farrah}, D., {Fox}, M., {Franceschini}, A., {Gear}, W., {Glenn}, J.,
  {Gonz{\'a}lez Solares}, E.~A., {Griffin}, M., {Halpern}, M., {Ibar}, E.,
  {Isaak}, K., {Ivison}, R.~J., {Lagache}, G., {Levenson}, L., {Lu}, N.,
  {Madden}, S., {Maffei}, B., {Mainetti}, G., {Marchetti}, L., {Mortier},
  A.~M.~J., {Nguyen}, H.~T., {O'Halloran}, B., {Oliver}, S.~J., {Page}, M.~J.,
  {Panuzzo}, P., {Papageorgiou}, A., {Pearson}, C.~P., {P{\'e}rez-Fournon}, I.,
  {Pohlen}, M., {Rawlings}, J.~I., {Rigopoulou}, D., {Rizzo}, D., {Roseboom},
  I.~G., {Rowan-Robinson}, M., {Sanchez Portal}, M., {Schulz}, B., {Scott}, D.,
  {Seymour}, N., {Shupe}, D.~L., {Smith}, A.~J., {Symeonidis}, M., {Trichas},
  M., {Tugwell}, K.~E., {Vaccari}, M., {Valtchanov}, I., {Vigroux}, L., {Wang},
  L., {Ward}, R., {Wright}, G., {Xu}, C.~K., \& {Zemcov}, M. 2010, \aap, 518,
  L33+

\bibitem[{{Hatziminaoglou} {et~al.}(2005){Hatziminaoglou}, {P{\'e}rez-Fournon},
  {Polletta}, {Afonso-Luis}, {Hern{\'a}n-Caballero}, {Montenegro-Montes},
  {Lonsdale}, {Xu}, {Franceschini}, {Rowan-Robinson}, {Babbedge}, {Smith},
  {Surace}, {Shupe}, {Fang}, {Farrah}, {Oliver}, {Gonz{\'a}lez-Solares}, \&
  {Serjeant}}]{Hatziminaoglou2005}
{Hatziminaoglou}, E., {P{\'e}rez-Fournon}, I., {Polletta}, M., {Afonso-Luis},
  A., {Hern{\'a}n-Caballero}, A., {Montenegro-Montes}, F.~M., {Lonsdale}, C.,
  {Xu}, C.~K., {Franceschini}, A., {Rowan-Robinson}, M., {Babbedge}, T.,
  {Smith}, H.~E., {Surace}, J., {Shupe}, D., {Fang}, F., {Farrah}, D.,
  {Oliver}, S., {Gonz{\'a}lez-Solares}, E.~A., \& {Serjeant}, S. 2005, \aj,
  129, 1198

\bibitem[{{Hauser} \& {Dwek}(2001)}]{Hauser2001}
{Hauser}, M.~G. \& {Dwek}, E. 2001, \araa, 39, 249

\bibitem[{{Heavens} {et~al.}(2004){Heavens}, {Panter}, {Jimenez}, \&
  {Dunlop}}]{Heavens2004}
{Heavens}, A., {Panter}, B., {Jimenez}, R., \& {Dunlop}, J. 2004, \nat, 428,
  625

\bibitem[{{Hildebrand}(1983)}]{Hildebrand1983}
{Hildebrand}, R.~H. 1983, \qjras, 24, 267

\bibitem[{{Hinshaw} {et~al.}(2009){Hinshaw}, {Weiland}, {Hill}, {Odegard},
  {Larson}, {Bennett}, {Dunkley}, {Gold}, {Greason}, {Jarosik}, {Komatsu},
  {Nolta}, {Page}, {Spergel}, {Wollack}, {Halpern}, {Kogut}, {Limon}, {Meyer},
  {Tucker}, \& {Wright}}]{Hinshaw2009}
{Hinshaw}, G., {Weiland}, J.~L., {Hill}, R.~S., {Odegard}, N., {Larson}, D.,
  {Bennett}, C.~L., {Dunkley}, J., {Gold}, B., {Greason}, M.~R., {Jarosik}, N.,
  {Komatsu}, E., {Nolta}, M.~R., {Page}, L., {Spergel}, D.~N., {Wollack}, E.,
  {Halpern}, M., {Kogut}, A., {Limon}, M., {Meyer}, S.~S., {Tucker}, G.~S., \&
  {Wright}, E.~L. 2009, \apjs, 180, 225

\bibitem[{{Hirashita} {et~al.}(2003){Hirashita}, {Buat}, \&
  {Inoue}}]{Hirashita2003}
{Hirashita}, H., {Buat}, V., \& {Inoue}, A.~K. 2003, \aap, 410, 83

\bibitem[{{Hopkins} {et~al.}(2003){Hopkins}, {Miller}, {Nichol}, {Connolly},
  {Bernardi}, {G{\'o}mez}, {Goto}, {Tremonti}, {Brinkmann}, {Ivezi{\'c}}, \&
  {Lamb}}]{Hopkins2003}
{Hopkins}, A.~M., {Miller}, C.~J., {Nichol}, R.~C., {Connolly}, A.~J.,
  {Bernardi}, M., {G{\'o}mez}, P.~L., {Goto}, T., {Tremonti}, C.~A.,
  {Brinkmann}, J., {Ivezi{\'c}}, {\v Z}., \& {Lamb}, D.~Q. 2003, \apj, 599, 971

\bibitem[{{Hughes} {et~al.}(1998){Hughes}, {Serjeant}, {Dunlop},
  {Rowan-Robinson}, {Blain}, {Mann}, {Ivison}, {Peacock}, {Efstathiou}, {Gear},
  {Oliver}, {Lawrence}, {Longair}, {Goldschmidt}, \& {Jenness}}]{Hughes1998}
{Hughes}, D.~H., {Serjeant}, S., {Dunlop}, J., {Rowan-Robinson}, M., {Blain},
  A., {Mann}, R.~G., {Ivison}, R., {Peacock}, J., {Efstathiou}, A., {Gear}, W.,
  {Oliver}, S., {Lawrence}, A., {Longair}, M., {Goldschmidt}, P., \& {Jenness},
  T. 1998, \nat, 394, 241

\bibitem[{{Iglesias-P{\'a}ramo} {et~al.}(2004){Iglesias-P{\'a}ramo}, {Buat},
  {Donas}, {Boselli}, \& {Milliard}}]{Iglesias-Paramo2004}
{Iglesias-P{\'a}ramo}, J., {Buat}, V., {Donas}, J., {Boselli}, A., \&
  {Milliard}, B. 2004, \aap, 419, 109

\bibitem[{{Iglesias-P{\'a}ramo} {et~al.}(2006){Iglesias-P{\'a}ramo}, {Buat},
  {Takeuchi}, {Xu}, {Boissier}, {Boselli}, {Burgarella}, {Madore}, {Gil de
  Paz}, {Bianchi}, {Barlow}, {Byun}, {Donas}, {Forster}, {Friedman}, {Heckman},
  {Jelinski}, {Lee}, {Malina}, {Martin}, {Milliard}, {Morrissey}, {Neff},
  {Rich}, {Schiminovich}, {Seibert}, {Siegmund}, {Small}, {Szalay}, {Welsh}, \&
  {Wyder}}]{Iglesias-Paramo2006}
{Iglesias-P{\'a}ramo}, J., {Buat}, V., {Takeuchi}, T.~T., {Xu}, K., {Boissier},
  S., {Boselli}, A., {Burgarella}, D., {Madore}, B.~F., {Gil de Paz}, A.,
  {Bianchi}, L., {Barlow}, T.~A., {Byun}, Y.-I., {Donas}, J., {Forster}, K.,
  {Friedman}, P.~G., {Heckman}, T.~M., {Jelinski}, P.~N., {Lee}, Y.-W.,
  {Malina}, R.~F., {Martin}, D.~C., {Milliard}, B., {Morrissey}, P.~F., {Neff},
  S.~G., {Rich}, R.~M., {Schiminovich}, D., {Seibert}, M., {Siegmund},
  O.~H.~W., {Small}, T., {Szalay}, A.~S., {Welsh}, B.~Y., \& {Wyder}, T.~K.
  2006, \apjs, 164, 38

\bibitem[{{Inoue}(2002)}]{Inoue2002}
{Inoue}, A.~K. 2002, \apjl, 570, L97

\bibitem[{{Ivison} {et~al.}(2010){Ivison}, {Alexander}, {Biggs}, {Brandt},
  {Chapin}, {Coppin}, {Devlin}, {Dickinson}, {Dunlop}, {Dye}, {Eales},
  {Frayer}, {Halpern}, {Hughes}, {Ibar}, {Kov{\'a}cs}, {Marsden}, {Moncelsi},
  {Netterfield}, {Pascale}, {Patanchon}, {Rafferty}, {Rex}, {Schinnerer},
  {Scott}, {Semisch}, {Smail}, {Swinbank}, {Truch}, {Tucker}, {Viero},
  {Walter}, {Wei{\ss}}, {Wiebe}, \& {Xue}}]{Ivison2010}
{Ivison}, R.~J., {Alexander}, D.~M., {Biggs}, A.~D., {Brandt}, W.~N., {Chapin},
  E.~L., {Coppin}, K.~E.~K., {Devlin}, M.~J., {Dickinson}, M., {Dunlop}, J.,
  {Dye}, S., {Eales}, S.~A., {Frayer}, D.~T., {Halpern}, M., {Hughes}, D.~H.,
  {Ibar}, E., {Kov{\'a}cs}, A., {Marsden}, G., {Moncelsi}, L., {Netterfield},
  C.~B., {Pascale}, E., {Patanchon}, G., {Rafferty}, D.~A., {Rex}, M.,
  {Schinnerer}, E., {Scott}, D., {Semisch}, C., {Smail}, I., {Swinbank}, A.~M.,
  {Truch}, M.~D.~P., {Tucker}, G.~S., {Viero}, M.~P., {Walter}, F., {Wei{\ss}},
  A., {Wiebe}, D.~V., \& {Xue}, Y.~Q. 2010, \mnras, 402, 245

\bibitem[{{Kauffmann} {et~al.}(2003){Kauffmann}, {Heckman}, {Tremonti},
  {Brinchmann}, {Charlot}, {White}, {Ridgway}, {Brinkmann}, {Fukugita}, {Hall},
  {Ivezi{\'c}}, {Richards}, \& {Schneider}}]{Kauffmann2003}
{Kauffmann}, G., {Heckman}, T.~M., {Tremonti}, C., {Brinchmann}, J., {Charlot},
  S., {White}, S.~D.~M., {Ridgway}, S.~E., {Brinkmann}, J., {Fukugita}, M.,
  {Hall}, P.~B., {Ivezi{\'c}}, {\v Z}., {Richards}, G.~T., \& {Schneider},
  D.~P. 2003, \mnras, 346, 1055

\bibitem[{{Kennicutt}(1998)}]{Kennicutt98}
{Kennicutt}, Jr., R.~C. 1998, \araa, 36, 189

\bibitem[{{Kriek} {et~al.}(2008){Kriek}, {van Dokkum}, {Franx}, {Illingworth},
  {Marchesini}, {Quadri}, {Rudnick}, {Taylor}, {F{\"o}rster Schreiber},
  {Gawiser}, {Labb{\'e}}, {Lira}, \& {Wuyts}}]{Kriek2008}
{Kriek}, M., {van Dokkum}, P.~G., {Franx}, M., {Illingworth}, G.~D.,
  {Marchesini}, D., {Quadri}, R., {Rudnick}, G., {Taylor}, E.~N., {F{\"o}rster
  Schreiber}, N.~M., {Gawiser}, E., {Labb{\'e}}, I., {Lira}, P., \& {Wuyts}, S.
  2008, \apj, 677, 219

\bibitem[{{Lacy} {et~al.}(2004){Lacy}, {Storrie-Lombardi}, {Sajina},
  {Appleton}, {Armus}, {Chapman}, {Choi}, {Fadda}, {Fang}, {Frayer},
  {Heinrichsen}, {Helou}, {Im}, {Marleau}, {Masci}, {Shupe}, {Soifer},
  {Surace}, {Teplitz}, {Wilson}, \& {Yan}}]{Lacy2004}
{Lacy}, M., {Storrie-Lombardi}, L.~J., {Sajina}, A., {Appleton}, P.~N.,
  {Armus}, L., {Chapman}, S.~C., {Choi}, P.~I., {Fadda}, D., {Fang}, F.,
  {Frayer}, D.~T., {Heinrichsen}, I., {Helou}, G., {Im}, M., {Marleau}, F.~R.,
  {Masci}, F., {Shupe}, D.~L., {Soifer}, B.~T., {Surace}, J., {Teplitz}, H.~I.,
  {Wilson}, G., \& {Yan}, L. 2004, \apjs, 154, 166

\bibitem[{{Le Floc'h} {et~al.}(2005){Le Floc'h}, {Papovich}, {Dole}, {Bell},
  {Lagache}, {Rieke}, {Egami}, {P{\'e}rez-Gonz{\'a}lez}, {Alonso-Herrero},
  {Rieke}, {Blaylock}, {Engelbracht}, {Gordon}, {Hines}, {Misselt}, {Morrison},
  \& {Mould}}]{LeFloch2005}
{Le Floc'h}, E., {Papovich}, C., {Dole}, H., {Bell}, E.~F., {Lagache}, G.,
  {Rieke}, G.~H., {Egami}, E., {P{\'e}rez-Gonz{\'a}lez}, P.~G.,
  {Alonso-Herrero}, A., {Rieke}, M.~J., {Blaylock}, M., {Engelbracht}, C.~W.,
  {Gordon}, K.~D., {Hines}, D.~C., {Misselt}, K.~A., {Morrison}, J.~E., \&
  {Mould}, J. 2005, \apj, 632, 169

\bibitem[{{Leitherer} {et~al.}(1999){Leitherer}, {Schaerer}, {Goldader},
  {Gonz{\'a}lez Delgado}, {Robert}, {Kune}, {de Mello}, {Devost}, \&
  {Heckman}}]{Leitherer1999}
{Leitherer}, C., {Schaerer}, D., {Goldader}, J.~D., {Gonz{\'a}lez Delgado},
  R.~M., {Robert}, C., {Kune}, D.~F., {de Mello}, D.~F., {Devost}, D., \&
  {Heckman}, T.~M. 1999, \apjs, 123, 3

\bibitem[{{Lonsdale} {et~al.}(2004){Lonsdale}, {Polletta}, {Surace}, {Shupe},
  {Fang}, {Xu}, {Smith}, {Siana}, {Rowan-Robinson}, {Babbedge}, {Oliver},
  {Pozzi}, {Davoodi}, {Owen}, {Padgett}, {Frayer}, {Jarrett}, {Masci},
  {O'Linger}, {Conrow}, {Farrah}, {Morrison}, {Gautier}, {Franceschini},
  {Berta}, {Perez-Fournon}, {Hatziminaoglou}, {Afonso-Luis}, {Dole}, {Stacey},
  {Serjeant}, {Pierre}, {Griffin}, \& {Puetter}}]{Lonsdale2004}
{Lonsdale}, C., {Polletta}, M.~d.~C., {Surace}, J., {Shupe}, D., {Fang}, F.,
  {Xu}, C.~K., {Smith}, H.~E., {Siana}, B., {Rowan-Robinson}, M., {Babbedge},
  T., {Oliver}, S., {Pozzi}, F., {Davoodi}, P., {Owen}, F., {Padgett}, D.,
  {Frayer}, D., {Jarrett}, T., {Masci}, F., {O'Linger}, J., {Conrow}, T.,
  {Farrah}, D., {Morrison}, G., {Gautier}, N., {Franceschini}, A., {Berta}, S.,
  {Perez-Fournon}, I., {Hatziminaoglou}, E., {Afonso-Luis}, A., {Dole}, H.,
  {Stacey}, G., {Serjeant}, S., {Pierre}, M., {Griffin}, M., \& {Puetter}, R.
  2004, \apjs, 154, 54

\bibitem[{{Magnelli} {et~al.}(2009){Magnelli}, {Elbaz}, {Chary}, {Dickinson},
  {Le Borgne}, {Frayer}, \& {Willmer}}]{Magnelli2009}
{Magnelli}, B., {Elbaz}, D., {Chary}, R.~R., {Dickinson}, M., {Le Borgne}, D.,
  {Frayer}, D.~T., \& {Willmer}, C.~N.~A. 2009, \aap, 496, 57

\bibitem[{{Marsden} {et~al.}(2009){Marsden}, {Ade}, {Bock}, {Chapin}, {Devlin},
  {Dicker}, {Griffin}, {Gundersen}, {Halpern}, {Hargrave}, {Hughes}, {Klein},
  {Mauskopf}, {Magnelli}, {Moncelsi}, {Netterfield}, {Ngo}, {Olmi}, {Pascale},
  {Patanchon}, {Rex}, {Scott}, {Semisch}, {Thomas}, {Truch}, {Tucker},
  {Tucker}, {Viero}, \& {Wiebe}}]{Marsden2009}
{Marsden}, G., {Ade}, P.~A.~R., {Bock}, J.~J., {Chapin}, E.~L., {Devlin},
  M.~J., {Dicker}, S.~R., {Griffin}, M., {Gundersen}, J.~O., {Halpern}, M.,
  {Hargrave}, P.~C., {Hughes}, D.~H., {Klein}, J., {Mauskopf}, P., {Magnelli},
  B., {Moncelsi}, L., {Netterfield}, C.~B., {Ngo}, H., {Olmi}, L., {Pascale},
  E., {Patanchon}, G., {Rex}, M., {Scott}, D., {Semisch}, C., {Thomas}, N.,
  {Truch}, M.~D.~P., {Tucker}, C., {Tucker}, G.~S., {Viero}, M.~P., \& {Wiebe},
  D.~V. 2009, \apj, 707, 1729

\bibitem[{{Martin} {et~al.}(2005){Martin}, {Fanson}, {Schiminovich},
  {Morrissey}, {Friedman}, {Barlow}, {Conrow}, {Grange}, {Jelinsky},
  {Milliard}, {Siegmund}, {Bianchi}, {Byun}, {Donas}, {Forster}, {Heckman},
  {Lee}, {Madore}, {Malina}, {Neff}, {Rich}, {Small}, {Surber}, {Szalay},
  {Welsh}, \& {Wyder}}]{Martin2005}
{Martin}, D.~C., {Fanson}, J., {Schiminovich}, D., {Morrissey}, P., {Friedman},
  P.~G., {Barlow}, T.~A., {Conrow}, T., {Grange}, R., {Jelinsky}, P.~N.,
  {Milliard}, B., {Siegmund}, O.~H.~W., {Bianchi}, L., {Byun}, Y., {Donas}, J.,
  {Forster}, K., {Heckman}, T.~M., {Lee}, Y., {Madore}, B.~F., {Malina}, R.~F.,
  {Neff}, S.~G., {Rich}, R.~M., {Small}, T., {Surber}, F., {Szalay}, A.~S.,
  {Welsh}, B., \& {Wyder}, T.~K. 2005, \apjl, 619, L1

\bibitem[{{Mazzei} {et~al.}(2007){Mazzei}, {Della Valle}, \&
  {Bettoni}}]{Mazzei2007}
{Mazzei}, P., {Della Valle}, A., \& {Bettoni}, D. 2007, \aap, 462, 21

\bibitem[{{Miller} {et~al.}(2003){Miller}, {Nichol}, {G{\'o}mez}, {Hopkins}, \&
  {Bernardi}}]{Miller2003}
{Miller}, C.~J., {Nichol}, R.~C., {G{\'o}mez}, P.~L., {Hopkins}, A.~M., \&
  {Bernardi}, M. 2003, \apj, 597, 142

\bibitem[{{Miller} {et~al.}(2008){Miller}, {Fomalont}, {Kellermann},
  {Mainieri}, {Norman}, {Padovani}, {Rosati}, \& {Tozzi}}]{Miller2008}
{Miller}, N.~A., {Fomalont}, E.~B., {Kellermann}, K.~I., {Mainieri}, V.,
  {Norman}, C., {Padovani}, P., {Rosati}, P., \& {Tozzi}, P. 2008, \apjs, 179,
  114

\bibitem[{{Morrissey} {et~al.}(2007){Morrissey}, {Conrow}, {Barlow}, {Small},
  {Seibert}, {Wyder}, {Budav{\'a}ri}, {Arnouts}, {Friedman}, {Forster},
  {Martin}, {Neff}, {Schiminovich}, {Bianchi}, {Donas}, {Heckman}, {Lee},
  {Madore}, {Milliard}, {Rich}, {Szalay}, {Welsh}, \& {Yi}}]{Morrissey2007}
{Morrissey}, P., {Conrow}, T., {Barlow}, T.~A., {Small}, T., {Seibert}, M.,
  {Wyder}, T.~K., {Budav{\'a}ri}, T., {Arnouts}, S., {Friedman}, P.~G.,
  {Forster}, K., {Martin}, D.~C., {Neff}, S.~G., {Schiminovich}, D., {Bianchi},
  L., {Donas}, J., {Heckman}, T.~M., {Lee}, Y., {Madore}, B.~F., {Milliard},
  B., {Rich}, R.~M., {Szalay}, A.~S., {Welsh}, B.~Y., \& {Yi}, S.~K. 2007,
  \apjs, 173, 682

\bibitem[{{Murphy} {et~al.}(2009){Murphy}, {Chary}, {Alexander}, {Dickinson},
  {Magnelli}, {Morrison}, {Pope}, \& {Teplitz}}]{Murphy2009}
{Murphy}, E.~J., {Chary}, R., {Alexander}, D.~M., {Dickinson}, M., {Magnelli},
  B., {Morrison}, G., {Pope}, A., \& {Teplitz}, H.~I. 2009, \apj, 698, 1380

\bibitem[{{Muzzin} {et~al.}(2010){Muzzin}, {van Dokkum}, {Kriek}, {Labbe},
  {Cury}, {Marchesini}, \& {Franx}}]{Muzzin2010}
{Muzzin}, A., {van Dokkum}, P., {Kriek}, M., {Labbe}, I., {Cury}, I.,
  {Marchesini}, D., \& {Franx}, M. 2010, ArXiv e-prints

\bibitem[{{Nordon} {et~al.}(2010){Nordon}, {Lutz}, {Shao}, {Magnelli}, {Berta},
  {Altieri}, {Andreani}, {Aussel}, {Bongiovanni}, {Cava}, {Cepa}, {Cimatti},
  {Daddi}, {Dominguez}, {Elbaz}, {F{\"o}rster Schreiber}, {Genzel}, {Grazian},
  {Magdis}, {Maiolino}, {P{\'e}rez Garc{\'{\i}}a}, {Poglitsch}, {Popesso},
  {Pozzi}, {Riguccini}, {Rodighiero}, {Saintonge}, {Sanchez-Portal}, {Santini},
  {Sturm}, {Tacconi}, {Valtchanov}, {Wetzstein}, \& {Wieprecht}}]{Nordon2010}
{Nordon}, R., {Lutz}, D., {Shao}, L., {Magnelli}, B., {Berta}, S., {Altieri},
  B., {Andreani}, P., {Aussel}, H., {Bongiovanni}, A., {Cava}, A., {Cepa}, J.,
  {Cimatti}, A., {Daddi}, E., {Dominguez}, H., {Elbaz}, D., {F{\"o}rster
  Schreiber}, N.~M., {Genzel}, R., {Grazian}, A., {Magdis}, G., {Maiolino}, R.,
  {P{\'e}rez Garc{\'{\i}}a}, A.~M., {Poglitsch}, A., {Popesso}, P., {Pozzi},
  F., {Riguccini}, L., {Rodighiero}, G., {Saintonge}, A., {Sanchez-Portal}, M.,
  {Santini}, P., {Sturm}, E., {Tacconi}, L., {Valtchanov}, I., {Wetzstein}, M.,
  \& {Wieprecht}, E. 2010, \aap, 518, L24+

\bibitem[{{Norris} {et~al.}(2006){Norris}, {Afonso}, {Appleton}, {Boyle},
  {Ciliegi}, {Croom}, {Huynh}, {Jackson}, {Koekemoer}, {Lonsdale},
  {Middelberg}, {Mobasher}, {Oliver}, {Polletta}, {Siana}, {Smail}, \&
  {Voronkov}}]{Norris2006}
{Norris}, R.~P., {Afonso}, J., {Appleton}, P.~N., {Boyle}, B.~J., {Ciliegi},
  P., {Croom}, S.~M., {Huynh}, M.~T., {Jackson}, C.~A., {Koekemoer}, A.~M.,
  {Lonsdale}, C.~J., {Middelberg}, E., {Mobasher}, B., {Oliver}, S.~J.,
  {Polletta}, M., {Siana}, B.~D., {Smail}, I., \& {Voronkov}, M.~A. 2006, \aj,
  132, 2409

\bibitem[{{Papovich} {et~al.}(2007){Papovich}, {Rudnick}, {Le Floc'h}, {van
  Dokkum}, {Rieke}, {Taylor}, {Armus}, {Gawiser}, {Huang}, {Marcillac}, \&
  {Franx}}]{Papovich2007}
{Papovich}, C., {Rudnick}, G., {Le Floc'h}, E., {van Dokkum}, P.~G., {Rieke},
  G.~H., {Taylor}, E.~N., {Armus}, L., {Gawiser}, E., {Huang}, J., {Marcillac},
  D., \& {Franx}, M. 2007, \apj, 668, 45

\bibitem[{{Pascale} {et~al.}(2008){Pascale}, {Ade}, {Bock}, {Chapin}, {Chung},
  {Devlin}, {Dicker}, {Griffin}, {Gundersen}, {Halpern}, {Hargrave}, {Hughes},
  {Klein}, {MacTavish}, {Marsden}, {Martin}, {Martin}, {Mauskopf},
  {Netterfield}, {Olmi}, {Patanchon}, {Rex}, {Scott}, {Semisch}, {Thomas},
  {Truch}, {Tucker}, {Tucker}, {Viero}, \& {Wiebe}}]{Pascale2008}
{Pascale}, E., {Ade}, P.~A.~R., {Bock}, J.~J., {Chapin}, E.~L., {Chung}, J.,
  {Devlin}, M.~J., {Dicker}, S., {Griffin}, M., {Gundersen}, J.~O., {Halpern},
  M., {Hargrave}, P.~C., {Hughes}, D.~H., {Klein}, J., {MacTavish}, C.~J.,
  {Marsden}, G., {Martin}, P.~G., {Martin}, T.~G., {Mauskopf}, P.,
  {Netterfield}, C.~B., {Olmi}, L., {Patanchon}, G., {Rex}, M., {Scott}, D.,
  {Semisch}, C., {Thomas}, N., {Truch}, M.~D.~P., {Tucker}, C., {Tucker},
  G.~S., {Viero}, M.~P., \& {Wiebe}, D.~V. 2008, \apj, 681, 400

\bibitem[{{Pascale} {et~al.}(2009){Pascale}, {Ade}, {Bock}, {Chapin}, {Devlin},
  {Dye}, {Eales}, {Griffin}, {Gundersen}, {Halpern}, {Hargrave}, {Hughes},
  {Klein}, {Marsden}, {Mauskopf}, {Moncelsi}, {Ngo}, {Netterfield}, {Olmi},
  {Patanchon}, {Rex}, {Scott}, {Semisch}, {Thomas}, {Truch}, {Tucker},
  {Tucker}, {Viero}, \& {Wiebe}}]{Pascale2009}
{Pascale}, E., {Ade}, P.~A.~R., {Bock}, J.~J., {Chapin}, E.~L., {Devlin},
  M.~J., {Dye}, S., {Eales}, S.~A., {Griffin}, M., {Gundersen}, J.~O.,
  {Halpern}, M., {Hargrave}, P.~C., {Hughes}, D.~H., {Klein}, J., {Marsden},
  G., {Mauskopf}, P., {Moncelsi}, L., {Ngo}, H., {Netterfield}, C.~B., {Olmi},
  L., {Patanchon}, G., {Rex}, M., {Scott}, D., {Semisch}, C., {Thomas}, N.,
  {Truch}, M.~D.~P., {Tucker}, C., {Tucker}, G.~S., {Viero}, M.~P., \& {Wiebe},
  D.~V. 2009, \apj, 707, 1740

\bibitem[{{Patanchon} {et~al.}(2009){Patanchon}, {Ade}, {Bock}, {Chapin},
  {Devlin}, {Dicker}, {Griffin}, {Gundersen}, {Halpern}, {Hargrave}, {Hughes},
  {Klein}, {Marsden}, {Mauskopf}, {Moncelsi}, {Netterfield}, {Olmi}, {Pascale},
  {Rex}, {Scott}, {Semisch}, {Thomas}, {Truch}, {Tucker}, {Tucker}, {Viero}, \&
  {Wiebe}}]{Patanchon2009}
{Patanchon}, G., {Ade}, P.~A.~R., {Bock}, J.~J., {Chapin}, E.~L., {Devlin},
  M.~J., {Dicker}, S.~R., {Griffin}, M., {Gundersen}, J.~O., {Halpern}, M.,
  {Hargrave}, P.~C., {Hughes}, D.~H., {Klein}, J., {Marsden}, G., {Mauskopf},
  P., {Moncelsi}, L., {Netterfield}, C.~B., {Olmi}, L., {Pascale}, E., {Rex},
  M., {Scott}, D., {Semisch}, C., {Thomas}, N., {Truch}, M.~D.~P., {Tucker},
  C., {Tucker}, G.~S., {Viero}, M.~P., \& {Wiebe}, D.~V. 2009, \apj, 707, 1750

\bibitem[{{P{\'e}rez-Gonz{\'a}lez} {et~al.}(2005){P{\'e}rez-Gonz{\'a}lez},
  {Rieke}, {Egami}, {Alonso-Herrero}, {Dole}, {Papovich}, {Blaylock}, {Jones},
  {Rieke}, {Rigby}, {Barmby}, {Fazio}, {Huang}, \&
  {Martin}}]{Perez-Gonzalez2005}
{P{\'e}rez-Gonz{\'a}lez}, P.~G., {Rieke}, G.~H., {Egami}, E., {Alonso-Herrero},
  A., {Dole}, H., {Papovich}, C., {Blaylock}, M., {Jones}, J., {Rieke}, M.,
  {Rigby}, J., {Barmby}, P., {Fazio}, G.~G., {Huang}, J., \& {Martin}, C. 2005,
  \apj, 630, 82

\bibitem[{{Pilbratt} {et~al.}(2010){Pilbratt}, {Riedinger}, {Passvogel},
  {Crone}, {Doyle}, {Gageur}, {Heras}, {Jewell}, {Metcalfe}, {Ott}, \&
  {Schmidt}}]{Pilbratt2010}
{Pilbratt}, G.~L., {Riedinger}, J.~R., {Passvogel}, T., {Crone}, G., {Doyle},
  D., {Gageur}, U., {Heras}, A.~M., {Jewell}, C., {Metcalfe}, L., {Ott}, S., \&
  {Schmidt}, M. 2010, \aap, 518, L1+

\bibitem[{{Pope} {et~al.}(2005){Pope}, {Borys}, {Scott}, {Conselice},
  {Dickinson}, \& {Mobasher}}]{Pope2005}
{Pope}, A., {Borys}, C., {Scott}, D., {Conselice}, C., {Dickinson}, M., \&
  {Mobasher}, B. 2005, \mnras, 358, 149

\bibitem[{{Pope} {et~al.}(2006){Pope}, {Scott}, {Dickinson}, {Chary},
  {Morrison}, {Borys}, {Sajina}, {Alexander}, {Daddi}, {Frayer}, {MacDonald},
  \& {Stern}}]{Pope2006}
{Pope}, A., {Scott}, D., {Dickinson}, M., {Chary}, R., {Morrison}, G., {Borys},
  C., {Sajina}, A., {Alexander}, D.~M., {Daddi}, E., {Frayer}, D., {MacDonald},
  E., \& {Stern}, D. 2006, \mnras, 370, 1185

\bibitem[{{Puget} {et~al.}(1996){Puget}, {Abergel}, {Bernard}, {Boulanger},
  {Burton}, {Desert}, \& {Hartmann}}]{puget96}
{Puget}, J.-L., {Abergel}, A., {Bernard}, J.-P., {Boulanger}, F., {Burton},
  W.~B., {Desert}, F.-X., \& {Hartmann}, D. 1996, \aap, 308, L5+

\bibitem[{{Ratcliffe} {et~al.}(1998){Ratcliffe}, {Shanks}, {Parker},
  {Broadbent}, {Watson}, {Oates}, {Collins}, \& {Fong}}]{Ratcliffe1998}
{Ratcliffe}, A., {Shanks}, T., {Parker}, Q.~A., {Broadbent}, A., {Watson},
  F.~G., {Oates}, A.~P., {Collins}, C.~A., \& {Fong}, R. 1998, \mnras, 300, 417

\bibitem[{{Ravikumar} {et~al.}(2007){Ravikumar}, {Puech}, {Flores}, {Proust},
  {Hammer}, {Lehnert}, {Rawat}, {Amram}, {Balkowski}, {Burgarella}, {Cassata},
  {Cesarsky}, {Cimatti}, {Combes}, {Daddi}, {Dannerbauer}, {di Serego
  Alighieri}, {Elbaz}, {Guiderdoni}, {Kembhavi}, {Liang}, {Pozzetti},
  {Vergani}, {Vernet}, {Wozniak}, \& {Zheng}}]{Ravikumar2007}
{Ravikumar}, C.~D., {Puech}, M., {Flores}, H., {Proust}, D., {Hammer}, F.,
  {Lehnert}, M., {Rawat}, A., {Amram}, P., {Balkowski}, C., {Burgarella}, D.,
  {Cassata}, P., {Cesarsky}, C., {Cimatti}, A., {Combes}, F., {Daddi}, E.,
  {Dannerbauer}, H., {di Serego Alighieri}, S., {Elbaz}, D., {Guiderdoni}, B.,
  {Kembhavi}, A., {Liang}, Y.~C., {Pozzetti}, L., {Vergani}, D., {Vernet}, J.,
  {Wozniak}, H., \& {Zheng}, X.~Z. 2007, \aap, 465, 1099

\bibitem[{{Rieke} {et~al.}(2004){Rieke}, {Young}, {Engelbracht}, {Kelly},
  {Low}, {Haller}, {Beeman}, {Gordon}, {Stansberry}, {Misselt}, {Cadien},
  {Morrison}, {Rivlis}, {Latter}, {Noriega-Crespo}, {Padgett}, {Stapelfeldt},
  {Hines}, {Egami}, {Muzerolle}, {Alonso-Herrero}, {Blaylock}, {Dole}, {Hinz},
  {Le Floc'h}, {Papovich}, {P{\'e}rez-Gonz{\'a}lez}, {Smith}, {Su}, {Bennett},
  {Frayer}, {Henderson}, {Lu}, {Masci}, {Pesenson}, {Rebull}, {Rho}, {Keene},
  {Stolovy}, {Wachter}, {Wheaton}, {Werner}, \& {Richards}}]{Rieke2004}
{Rieke}, G.~H., {Young}, E.~T., {Engelbracht}, C.~W., {Kelly}, D.~M., {Low},
  F.~J., {Haller}, E.~E., {Beeman}, J.~W., {Gordon}, K.~D., {Stansberry},
  J.~A., {Misselt}, K.~A., {Cadien}, J., {Morrison}, J.~E., {Rivlis}, G.,
  {Latter}, W.~B., {Noriega-Crespo}, A., {Padgett}, D.~L., {Stapelfeldt},
  K.~R., {Hines}, D.~C., {Egami}, E., {Muzerolle}, J., {Alonso-Herrero}, A.,
  {Blaylock}, M., {Dole}, H., {Hinz}, J.~L., {Le Floc'h}, E., {Papovich}, C.,
  {P{\'e}rez-Gonz{\'a}lez}, P.~G., {Smith}, P.~S., {Su}, K.~Y.~L., {Bennett},
  L., {Frayer}, D.~T., {Henderson}, D., {Lu}, N., {Masci}, F., {Pesenson}, M.,
  {Rebull}, L., {Rho}, J., {Keene}, J., {Stolovy}, S., {Wachter}, S.,
  {Wheaton}, W., {Werner}, M.~W., \& {Richards}, P.~L. 2004, \apjs, 154, 25

\bibitem[{{Rodighiero} {et~al.}(2010){Rodighiero}, {Cimatti}, {Gruppioni},
  {Popesso}, {Andreani}, {Altieri}, {Aussel}, {Berta}, {Bongiovanni},
  {Brisbin}, {Cava}, {Cepa}, {Daddi}, {Dominguez-Sanchez}, {Elbaz}, {Fontana},
  {F{\"o}rster Schreiber}, {Franceschini}, {Genzel}, {Grazian}, {Lutz},
  {Magdis}, {Magliocchetti}, {Magnelli}, {Maiolino}, {Mancini}, {Nordon},
  {Perez Garcia}, {Poglitsch}, {Santini}, {Sanchez-Portal}, {Pozzi},
  {Riguccini}, {Saintonge}, {Shao}, {Sturm}, {Tacconi}, {Valtchanov},
  {Wetzstein}, \& {Wieprecht}}]{Rodighiero2010}
{Rodighiero}, G., {Cimatti}, A., {Gruppioni}, C., {Popesso}, P., {Andreani},
  P., {Altieri}, B., {Aussel}, H., {Berta}, S., {Bongiovanni}, A., {Brisbin},
  D., {Cava}, A., {Cepa}, J., {Daddi}, E., {Dominguez-Sanchez}, H., {Elbaz},
  D., {Fontana}, A., {F{\"o}rster Schreiber}, N., {Franceschini}, A., {Genzel},
  R., {Grazian}, A., {Lutz}, D., {Magdis}, G., {Magliocchetti}, M., {Magnelli},
  B., {Maiolino}, R., {Mancini}, C., {Nordon}, R., {Perez Garcia}, A.~M.,
  {Poglitsch}, A., {Santini}, P., {Sanchez-Portal}, M., {Pozzi}, F.,
  {Riguccini}, L., {Saintonge}, A., {Shao}, L., {Sturm}, E., {Tacconi}, L.,
  {Valtchanov}, I., {Wetzstein}, M., \& {Wieprecht}, E. 2010, \aap, 518, L25+

\bibitem[{{Rowan-Robinson}(2001)}]{RR01}
{Rowan-Robinson}, M. 2001, \apj, 549, 745

\bibitem[{{Rowan-Robinson} {et~al.}(2008){Rowan-Robinson}, {Babbedge},
  {Oliver}, {Trichas}, {Berta}, {Lonsdale}, {Smith}, {Shupe}, {Surace},
  {Arnouts}, {Ilbert}, {Le F{\'e}vre}, {Afonso-Luis}, {Perez-Fournon},
  {Hatziminaoglou}, {Polletta}, {Farrah}, \& {Vaccari}}]{RR08}
{Rowan-Robinson}, M., {Babbedge}, T., {Oliver}, S., {Trichas}, M., {Berta}, S.,
  {Lonsdale}, C., {Smith}, G., {Shupe}, D., {Surace}, J., {Arnouts}, S.,
  {Ilbert}, O., {Le F{\'e}vre}, O., {Afonso-Luis}, A., {Perez-Fournon}, I.,
  {Hatziminaoglou}, E., {Polletta}, M., {Farrah}, D., \& {Vaccari}, M. 2008,
  \mnras, 386, 697

\bibitem[{{Salpeter}(1955)}]{Salpeter1955}
{Salpeter}, E.~E. 1955, \apj, 121, 161

\bibitem[{{Santini} {et~al.}(2009){Santini}, {Fontana}, {Grazian}, {Salimbeni},
  {Fiore}, {Fontanot}, {Boutsia}, {Castellano}, {Cristiani}, {de Santis},
  {Gallozzi}, {Giallongo}, {Menci}, {Nonino}, {Paris}, {Pentericci}, \&
  {Vanzella}}]{Santini2009}
{Santini}, P., {Fontana}, A., {Grazian}, A., {Salimbeni}, S., {Fiore}, F.,
  {Fontanot}, F., {Boutsia}, K., {Castellano}, M., {Cristiani}, S., {de
  Santis}, C., {Gallozzi}, S., {Giallongo}, E., {Menci}, N., {Nonino}, M.,
  {Paris}, D., {Pentericci}, L., \& {Vanzella}, E. 2009, \aap, 504, 751

\bibitem[{{Schlegel} {et~al.}(1998){Schlegel}, {Finkbeiner}, \&
  {Davis}}]{Schlegel1998}
{Schlegel}, D.~J., {Finkbeiner}, D.~P., \& {Davis}, M. 1998, \apj, 500, 525

\bibitem[{{Schulz} {et~al.}(2010){Schulz}, {Pearson}, {Clements}, {Altieri},
  {Amblard}, {Arumugam}, {Auld}, {Aussel}, {Babbedge}, {Blain}, {Bock},
  {Boselli}, {Buat}, {Burgarella}, {Castro-Rodr{\'{\i}}guez}, {Cava},
  {Chanial}, {Conley}, {Conversi}, {Cooray}, {Dowell}, {Dwek}, {Eales},
  {Elbaz}, {Fox}, {Franceschini}, {Gear}, {Giovannoli}, {Glenn}, {Griffin},
  {Halpern}, {Hatziminaoglou}, {Ibar}, {Isaak}, {Ivison}, {Lagache},
  {Levenson}, {Lu}, {Madden}, {Maffei}, {Mainetti}, {Marchetti}, {Marsden},
  {Mortier}, {Nguyen}, {O'Halloran}, {Oliver}, {Omont}, {Page}, {Panuzzo},
  {Papageorgiou}, {P{\'e}rez-Fournon}, {Pohlen}, {Rangwala}, {Rawlings},
  {Raymond}, {Rigopoulou}, {Rizzo}, {Roseboom}, {Rowan-Robinson}, {S{\'a}nchez
  Portal}, {Scott}, {Seymour}, {Shupe}, {Smith}, {Stevens}, {Symeonidis},
  {Trichas}, {Tugwell}, {Vaccari}, {Valiante}, {Valtchanov}, {Vigroux}, {Wang},
  {Ward}, {Wright}, {Xu}, \& {Zemcov}}]{Schulz2010}
{Schulz}, B., {Pearson}, C.~P., {Clements}, D.~L., {Altieri}, B., {Amblard},
  A., {Arumugam}, V., {Auld}, R., {Aussel}, H., {Babbedge}, T., {Blain}, A.,
  {Bock}, J., {Boselli}, A., {Buat}, V., {Burgarella}, D.,
  {Castro-Rodr{\'{\i}}guez}, N., {Cava}, A., {Chanial}, P., {Conley}, A.,
  {Conversi}, L., {Cooray}, A., {Dowell}, C.~D., {Dwek}, E., {Eales}, S.,
  {Elbaz}, D., {Fox}, M., {Franceschini}, A., {Gear}, W., {Giovannoli}, E.,
  {Glenn}, J., {Griffin}, M., {Halpern}, M., {Hatziminaoglou}, E., {Ibar}, E.,
  {Isaak}, K., {Ivison}, R.~J., {Lagache}, G., {Levenson}, L., {Lu}, N.,
  {Madden}, S., {Maffei}, B., {Mainetti}, G., {Marchetti}, L., {Marsden}, G.,
  {Mortier}, A.~M.~J., {Nguyen}, H.~T., {O'Halloran}, B., {Oliver}, S.~J.,
  {Omont}, A., {Page}, M.~J., {Panuzzo}, P., {Papageorgiou}, A.,
  {P{\'e}rez-Fournon}, I., {Pohlen}, M., {Rangwala}, N., {Rawlings}, J.~I.,
  {Raymond}, G., {Rigopoulou}, D., {Rizzo}, D., {Roseboom}, I.~G.,
  {Rowan-Robinson}, M., {S{\'a}nchez Portal}, M., {Scott}, D., {Seymour}, N.,
  {Shupe}, D.~L., {Smith}, A.~J., {Stevens}, J.~A., {Symeonidis}, M.,
  {Trichas}, M., {Tugwell}, K.~E., {Vaccari}, M., {Valiante}, E., {Valtchanov},
  I., {Vigroux}, L., {Wang}, L., {Ward}, R., {Wright}, G., {Xu}, C.~K., \&
  {Zemcov}, M. 2010, \aap, 518, L32+

\bibitem[{{Serjeant} {et~al.}(2008){Serjeant}, {Dye}, {Mortier}, {Peacock},
  {Egami}, {Cirasuolo}, {Rieke}, {Borys}, {Chapman}, {Clements}, {Coppin},
  {Dunlop}, {Eales}, {Farrah}, {Halpern}, {Mauskopf}, {Pope}, {Rowan-Robinson},
  {Scott}, {Smail}, \& {Vaccari}}]{Serjeant2008}
{Serjeant}, S., {Dye}, S., {Mortier}, A., {Peacock}, J., {Egami}, E.,
  {Cirasuolo}, M., {Rieke}, G., {Borys}, C., {Chapman}, S., {Clements}, D.,
  {Coppin}, K., {Dunlop}, J., {Eales}, S., {Farrah}, D., {Halpern}, M.,
  {Mauskopf}, P., {Pope}, A., {Rowan-Robinson}, M., {Scott}, D., {Smail}, I.,
  \& {Vaccari}, M. 2008, \mnras, 386, 1907

\bibitem[{{Shao} {et~al.}(2010){Shao}, {Lutz}, {Nordon}, {Maiolino},
  {Alexander}, {Altieri}, {Andreani}, {Aussel}, {Bauer}, {Berta},
  {Bongiovanni}, {Brandt}, {Brusa}, {Cava}, {Cepa}, {Cimatti}, {Daddi},
  {Dominguez-Sanchez}, {Elbaz}, {F{\"o}rster Schreiber}, {Geis}, {Genzel},
  {Grazian}, {Gruppioni}, {Magdis}, {Magnelli}, {Mainieri}, {P{\'e}rez
  Garc{\'{\i}}a}, {Poglitsch}, {Popesso}, {Pozzi}, {Riguccini}, {Rodighiero},
  {Rovilos}, {Saintonge}, {Salvato}, {Sanchez Portal}, {Santini}, {Sturm},
  {Tacconi}, {Valtchanov}, {Wetzstein}, \& {Wieprecht}}]{Shao2010}
{Shao}, L., {Lutz}, D., {Nordon}, R., {Maiolino}, R., {Alexander}, D.~M.,
  {Altieri}, B., {Andreani}, P., {Aussel}, H., {Bauer}, F.~E., {Berta}, S.,
  {Bongiovanni}, A., {Brandt}, W.~N., {Brusa}, M., {Cava}, A., {Cepa}, J.,
  {Cimatti}, A., {Daddi}, E., {Dominguez-Sanchez}, H., {Elbaz}, D.,
  {F{\"o}rster Schreiber}, N.~M., {Geis}, N., {Genzel}, R., {Grazian}, A.,
  {Gruppioni}, C., {Magdis}, G., {Magnelli}, B., {Mainieri}, V., {P{\'e}rez
  Garc{\'{\i}}a}, A.~M., {Poglitsch}, A., {Popesso}, P., {Pozzi}, F.,
  {Riguccini}, L., {Rodighiero}, G., {Rovilos}, E., {Saintonge}, A., {Salvato},
  M., {Sanchez Portal}, M., {Santini}, P., {Sturm}, E., {Tacconi}, L.~J.,
  {Valtchanov}, I., {Wetzstein}, M., \& {Wieprecht}, E. 2010, \aap, 518, L26+

\bibitem[{{Sharp} {et~al.}(2006){Sharp}, {Saunders}, {Smith}, {Churilov},
  {Correll}, {Dawson}, {Farrel}, {Frost}, {Haynes}, {Heald}, {Lankshear},
  {Mayfield}, {Waller}, \& {Whittard}}]{Sharp2006}
{Sharp}, R., {Saunders}, W., {Smith}, G., {Churilov}, V., {Correll}, D.,
  {Dawson}, J., {Farrel}, T., {Frost}, G., {Haynes}, R., {Heald}, R.,
  {Lankshear}, A., {Mayfield}, D., {Waller}, L., \& {Whittard}, D. 2006, in
  Presented at the Society of Photo-Optical Instrumentation Engineers (SPIE)
  Conference, Vol. 6269, Society of Photo-Optical Instrumentation Engineers
  (SPIE) Conference Series

\bibitem[{{Silverman} {et~al.}(2009){Silverman}, {Lamareille}, {Maier},
  {Lilly}, {Mainieri}, {Brusa}, {Cappelluti}, {Hasinger}, {Zamorani},
  {Scodeggio}, {Bolzonella}, {Contini}, {Carollo}, {Jahnke}, {Kneib}, {Le
  F{\`e}vre}, {Merloni}, {Bardelli}, {Bongiorno}, {Brunner}, {Caputi},
  {Civano}, {Comastri}, {Coppa}, {Cucciati}, {de la Torre}, {de Ravel},
  {Elvis}, {Finoguenov}, {Fiore}, {Franzetti}, {Garilli}, {Gilli}, {Iovino},
  {Kampczyk}, {Knobel}, {Kova{\v c}}, {Le Borgne}, {Le Brun}, {Mignoli},
  {Pello}, {Peng}, {Montero}, {Ricciardelli}, {Tanaka}, {Tasca}, {Tresse},
  {Vergani}, {Vignali}, {Zucca}, {Bottini}, {Cappi}, {Cassata}, {Fumana},
  {Griffiths}, {Kartaltepe}, {Koekemoer}, {Marinoni}, {McCracken}, {Memeo},
  {Meneux}, {Oesch}, {Porciani}, \& {Salvato}}]{Silverman2009}
{Silverman}, J.~D., {Lamareille}, F., {Maier}, C., {Lilly}, S.~J., {Mainieri},
  V., {Brusa}, M., {Cappelluti}, N., {Hasinger}, G., {Zamorani}, G.,
  {Scodeggio}, M., {Bolzonella}, M., {Contini}, T., {Carollo}, C.~M., {Jahnke},
  K., {Kneib}, J., {Le F{\`e}vre}, O., {Merloni}, A., {Bardelli}, S.,
  {Bongiorno}, A., {Brunner}, H., {Caputi}, K., {Civano}, F., {Comastri}, A.,
  {Coppa}, G., {Cucciati}, O., {de la Torre}, S., {de Ravel}, L., {Elvis}, M.,
  {Finoguenov}, A., {Fiore}, F., {Franzetti}, P., {Garilli}, B., {Gilli}, R.,
  {Iovino}, A., {Kampczyk}, P., {Knobel}, C., {Kova{\v c}}, K., {Le Borgne},
  J., {Le Brun}, V., {Mignoli}, M., {Pello}, R., {Peng}, Y., {Montero}, E.~P.,
  {Ricciardelli}, E., {Tanaka}, M., {Tasca}, L., {Tresse}, L., {Vergani}, D.,
  {Vignali}, C., {Zucca}, E., {Bottini}, D., {Cappi}, A., {Cassata}, P.,
  {Fumana}, M., {Griffiths}, R., {Kartaltepe}, J., {Koekemoer}, A., {Marinoni},
  C., {McCracken}, H.~J., {Memeo}, P., {Meneux}, B., {Oesch}, P., {Porciani},
  C., \& {Salvato}, M. 2009, \apj, 696, 396

\bibitem[{{Smail} {et~al.}(1997){Smail}, {Ivison}, \& {Blain}}]{Smail1997}
{Smail}, I., {Ivison}, R.~J., \& {Blain}, A.~W. 1997, \apjl, 490, L5+

\bibitem[{{Stern} {et~al.}(2005){Stern}, {Eisenhardt}, {Gorjian}, {Kochanek},
  {Caldwell}, {Eisenstein}, {Brodwin}, {Brown}, {Cool}, {Dey}, {Green},
  {Jannuzi}, {Murray}, {Pahre}, \& {Willner}}]{Stern2005}
{Stern}, D., {Eisenhardt}, P., {Gorjian}, V., {Kochanek}, C.~S., {Caldwell},
  N., {Eisenstein}, D., {Brodwin}, M., {Brown}, M.~J.~I., {Cool}, R., {Dey},
  A., {Green}, P., {Jannuzi}, B.~T., {Murray}, S.~S., {Pahre}, M.~A., \&
  {Willner}, S.~P. 2005, \apj, 631, 163

\bibitem[{{Surace} {et~al.}(2005){Surace}, {Shupe}, {Fang}, {Evans}, {Alexov},
  {Frayer}, {Lonsdale}, \& {SWIRE Team}}]{Surace2005}
{Surace}, J.~A., {Shupe}, D.~L., {Fang}, F., {Evans}, T., {Alexov}, A.,
  {Frayer}, D., {Lonsdale}, C.~J., \& {SWIRE Team}. 2005, in Bulletin of the
  American Astronomical Society, Vol.~37, Bulletin of the American Astronomical
  Society, 1246--+

\bibitem[{{Takeuchi} {et~al.}(2009){Takeuchi}, {Buat}, {Heinis}, {Giovannoli},
  {Yuan}, {Iglesias-Paramo}, {Murata}, \& {Burgarella}}]{Takeuchi2009}
{Takeuchi}, T.~T., {Buat}, V., {Heinis}, S., {Giovannoli}, E., {Yuan}, F.,
  {Iglesias-Paramo}, J., {Murata}, K.~L., \& {Burgarella}, D. 2009, ArXiv
  e-prints

\bibitem[{{Taylor} {et~al.}(2009){Taylor}, {Franx}, {van Dokkum}, {Quadri},
  {Gawiser}, {Bell}, {Barrientos}, {Blanc}, {Castander}, {Damen},
  {Gonzalez-Perez}, {Hall}, {Herrera}, {Hildebrandt}, {Kriek}, {Labb{\'e}},
  {Lira}, {Maza}, {Rudnick}, {Treister}, {Urry}, {Willis}, \&
  {Wuyts}}]{Taylor2009}
{Taylor}, E.~N., {Franx}, M., {van Dokkum}, P.~G., {Quadri}, R.~F., {Gawiser},
  E., {Bell}, E.~F., {Barrientos}, L.~F., {Blanc}, G.~A., {Castander}, F.~J.,
  {Damen}, M., {Gonzalez-Perez}, V., {Hall}, P.~B., {Herrera}, D.,
  {Hildebrandt}, H., {Kriek}, M., {Labb{\'e}}, I., {Lira}, P., {Maza}, J.,
  {Rudnick}, G., {Treister}, E., {Urry}, C.~M., {Willis}, J.~P., \& {Wuyts}, S.
  2009, \apjs, 183, 295

\bibitem[{{Truch} {et~al.}(2009){Truch}, {Ade}, {Bock}, {Chapin}, {Devlin},
  {Dicker}, {Griffin}, {Gundersen}, {Halpern}, {Hargrave}, {Hughes}, {Klein},
  {Marsden}, {Martin}, {Mauskopf}, {Moncelsi}, {Barth Netterfield}, {Olmi},
  {Pascale}, {Patanchon}, {Rex}, {Scott}, {Semisch}, {Thomas}, {Tucker},
  {Tucker}, {Viero}, \& {Wiebe}}]{Truch2009}
{Truch}, M.~D.~P., {Ade}, P.~A.~R., {Bock}, J.~J., {Chapin}, E.~L., {Devlin},
  M.~J., {Dicker}, S.~R., {Griffin}, M., {Gundersen}, J.~O., {Halpern}, M.,
  {Hargrave}, P.~C., {Hughes}, D.~H., {Klein}, J., {Marsden}, G., {Martin},
  P.~G., {Mauskopf}, P., {Moncelsi}, L., {Barth Netterfield}, C., {Olmi}, L.,
  {Pascale}, E., {Patanchon}, G., {Rex}, M., {Scott}, D., {Semisch}, C.,
  {Thomas}, N.~E., {Tucker}, C., {Tucker}, G.~S., {Viero}, M.~P., \& {Wiebe},
  D.~V. 2009, \apj, 707, 1723

\bibitem[{{Viero} {et~al.}(2010){Viero}, {Moncelsi}, {Buitrago}, {Marsden},
  {Bauer}, {Trujillo}, {Conselice}, {P{\'e}rez-Gonz{\'a}lez}, {Chapin},
  {Devlin}, {Halpern}, {Mentuch}, {Netterfield}, {Pascale}, {Rex}, {Scott},
  {Truch}, \& {Wiebe}}]{Viero2010}
{Viero}, M.~P., {Moncelsi}, L., {Buitrago}, F., {Marsden}, G., {Bauer}, A.~E.,
  {Trujillo}, I., {Conselice}, C.~J., {P{\'e}rez-Gonz{\'a}lez}, P.~G.,
  {Chapin}, E.~L., {Devlin}, M.~J., {Halpern}, M., {Mentuch}, E.,
  {Netterfield}, C.~B., {Pascale}, E., {Rex}, M., {Scott}, D., {Truch},
  M.~D.~P., \& {Wiebe}, D.~V. 2010, ArXiv e-prints

\bibitem[{{Vlahakis} {et~al.}(2005){Vlahakis}, {Dunne}, \&
  {Eales}}]{Vlahakis2005}
{Vlahakis}, C., {Dunne}, L., \& {Eales}, S. 2005, \mnras, 364, 1253

\bibitem[{{Vollmann} \& {Eversberg}(2006)}]{Vollmann2006}
{Vollmann}, K. \& {Eversberg}, T. 2006, Astronomische Nachrichten, 327, 862

\bibitem[{{Wei{\ss}} {et~al.}(2009){Wei{\ss}}, {Kov{\'a}cs}, {Coppin}, {Greve},
  {Walter}, {Smail}, {Dunlop}, {Knudsen}, {Alexander}, {Bertoldi}, {Brandt},
  {Chapman}, {Cox}, {Dannerbauer}, {De Breuck}, {Gawiser}, {Ivison}, {Lutz},
  {Menten}, {Koekemoer}, {Kreysa}, {Kurczynski}, {Rix}, {Schinnerer}, \& {van
  der Werf}}]{Weiss2009}
{Wei{\ss}}, A., {Kov{\'a}cs}, A., {Coppin}, K., {Greve}, T.~R., {Walter}, F.,
  {Smail}, I., {Dunlop}, J.~S., {Knudsen}, K.~K., {Alexander}, D.~M.,
  {Bertoldi}, F., {Brandt}, W.~N., {Chapman}, S.~C., {Cox}, P., {Dannerbauer},
  H., {De Breuck}, C., {Gawiser}, E., {Ivison}, R.~J., {Lutz}, D., {Menten},
  K.~M., {Koekemoer}, A.~M., {Kreysa}, E., {Kurczynski}, P., {Rix}, H.,
  {Schinnerer}, E., \& {van der Werf}, P.~P. 2009, \apj, 707, 1201

\bibitem[{{Wiebe} {et~al.}(2009){Wiebe}, {Ade}, {Bock}, {Chapin}, {Devlin},
  {Dicker}, {Griffin}, {Gundersen}, {Halpern}, {Hargrave}, {Hughes}, {Klein},
  {Marsden}, {Martin}, {Mauskopf}, {Netterfield}, {Olmi}, {Pascale},
  {Patanchon}, {Rex}, {Scott}, {Semisch}, {Thomas}, {Truch}, {Tucker},
  {Tucker}, \& {Viero}}]{Wiebe2009}
{Wiebe}, D.~V., {Ade}, P.~A.~R., {Bock}, J.~J., {Chapin}, E.~L., {Devlin},
  M.~J., {Dicker}, S., {Griffin}, M., {Gundersen}, J.~O., {Halpern}, M.,
  {Hargrave}, P.~C., {Hughes}, D.~H., {Klein}, J., {Marsden}, G., {Martin},
  P.~G., {Mauskopf}, P., {Netterfield}, C.~B., {Olmi}, L., {Pascale}, E.,
  {Patanchon}, G., {Rex}, M., {Scott}, D., {Semisch}, C., {Thomas}, N.,
  {Truch}, M.~D.~P., {Tucker}, C., {Tucker}, G.~S., \& {Viero}, M.~P. 2009,
  \apj, 707, 1809

\bibitem[{{Wolf} {et~al.}(2008){Wolf}, {Hildebrandt}, {Taylor}, \&
  {Meisenheimer}}]{wolf08}
{Wolf}, C., {Hildebrandt}, H., {Taylor}, E.~N., \& {Meisenheimer}, K. 2008,
  ArXiv e-prints

\bibitem[{{Wolf} {et~al.}(2004){Wolf}, {Meisenheimer}, {Kleinheinrich},
  {Borch}, {Dye}, {Gray}, {Wisotzki}, {Bell}, {Rix}, {Cimatti}, {Hasinger}, \&
  {Szokoly}}]{wolf04}
{Wolf}, C., {Meisenheimer}, K., {Kleinheinrich}, M., {Borch}, A., {Dye}, S.,
  {Gray}, M., {Wisotzki}, L., {Bell}, E.~F., {Rix}, H.-W., {Cimatti}, A.,
  {Hasinger}, G., \& {Szokoly}, G. 2004, \aap, 421, 913

\bibitem[{{Wyder} {et~al.}(2007){Wyder}, {Martin}, {Schiminovich}, {Seibert},
  {Budav{\'a}ri}, {Treyer}, {Barlow}, {Forster}, {Friedman}, {Morrissey},
  {Neff}, {Small}, {Bianchi}, {Donas}, {Heckman}, {Lee}, {Madore}, {Milliard},
  {Rich}, {Szalay}, {Welsh}, \& {Yi}}]{Wyder2007}
{Wyder}, T.~K., {Martin}, D.~C., {Schiminovich}, D., {Seibert}, M.,
  {Budav{\'a}ri}, T., {Treyer}, M.~A., {Barlow}, T.~A., {Forster}, K.,
  {Friedman}, P.~G., {Morrissey}, P., {Neff}, S.~G., {Small}, T., {Bianchi},
  L., {Donas}, J., {Heckman}, T.~M., {Lee}, Y., {Madore}, B.~F., {Milliard},
  B., {Rich}, R.~M., {Szalay}, A.~S., {Welsh}, B.~Y., \& {Yi}, S.~K. 2007,
  \apjs, 173, 293

\end{thebibliography}

\end{document}